# A Unique Mathematical Derivation of the Fundamental Laws of Nature Based on a New Algebraic-Axiomatic (Matrix) Approach †,‡

**Ramin Zahedi**

Logic and Philosophy of Science Research Group, Hokkaido University, Hokkaido 060-0808, Japan; zahedi@let.hokudai.ac.jp or zahedi.r@gmail.com

† Invited and presented article at The 21st International Conference on General Relativity and Gravitation, Columbia University, New York, NY, USA, 2016.

‡ This work has been done and published during my research fellowship, 2007–2016.

**Abstract:** In this article, as a new mathematical approach to origin of the laws of nature, using a new basic algebraic axiomatic (matrix) formalism based on the ring theory and Clifford algebras (presented in Section 2), "it is shown that certain mathematical forms of fundamental laws of nature, including laws governing the fundamental forces of nature (represented by a set of two definite classes of general covariant massive field equations, with new matrix formalisms), are derived uniquely from only a very few axioms." In agreement with the rational Lorentz group, it is also basically assumed that the components of relativistic energy-momentum can only take rational values. In essence, the main scheme of this new mathematical axiomatic approach to the fundamental laws of nature is as follows: First, based on the assumption of the rationality of *D*-momentum and by linearization (along with a parameterization procedure) of the Lorentz invariant energy-momentum quadratic relation, a unique set of Lorentz invariant systems of homogeneous linear equations (with matrix formalisms compatible with certain Clifford and symmetric algebras) is derived. Then by an initial quantization (followed by a basic procedure of minimal coupling to space-time geometry) of these determined systems of linear equations, a set of two classes of general covariant massive (tensor) field equations (with matrix formalisms compatible with certain Clifford, and Weyl algebras) is derived uniquely as well.

Each class of the derived general covariant field equations also includes a definite form of torsion field appearing as the generator of the corresponding field' invariant mass. In addition, it is shown that the (1 + 3)-dimensional cases of two classes of derived field equations represent a new general covariant massive formalism of bispinor fields of spin-2, and spin-1 particles, respectively. In fact, these uniquely determined bispinor fields represent a unique set of new generalized massive forms of the laws governing the fundamental forces of nature, including the Einstein (gravitational), Maxwell (electromagnetic) and Yang-Mills (nuclear) field equations. Moreover, it is also shown that the (1 + 2)-dimensional cases of two classes of these field equations represent (asymptotically) a new general covariant massive formalism of bispinor fields of spin-3/2 and spin-1/2 particles, corresponding to the Dirac and Rarita–Schwinger equations.

As a particular consequence, it is shown that a certain massive formalism of general relativity—with a definite form of torsion field appeared originally as the generator of gravitational field's invariant mass—is obtained only by first quantization (followed by a basic procedure of minimal coupling to space-time geometry) of a certain set of special relativistic algebraic matrix equations. It has been also proved that Lagrangian densities specified for the originally derived new massive forms of the Maxwell, Yang-Mills and Dirac field equations, are also gauge invariant, where the invariant mass of each field is generated solely by the corresponding torsion field. In addition, in agreement with recent astronomical data, a new particular form of massive boson is identified (corresponding to the U(1) gauge symmetry group) with invariant mass: $m_\gamma \approx 4.90571 \times 10^{-50}$ kg, generated by a coupled torsion field of the background space-time geometry.

Moreover, based on the definite mathematical formalism of this axiomatic approach, along with the C, P and T symmetries (represented basically by the corresponding quantum operators) of the fundamentally derived field equations, it is concluded that the universe could be realized solely by the (1 + 2) and (1 + 3)-dimensional space-times (where this conclusion, in particular, is based on the T-symmetry). It is proved that 'CPT' is the only (unique) combination of C, P, and T symmetries that could be defined as a symmetry for interacting fields. In addition, on the basis of these discrete symmetries of derived field equations, it has also been shown that only left-handed particle fields (along with their complementary right-handed fields) could be coupled with the corresponding (any) source currents. Furthermore, it has been shown that the metric of background space-time is diagonalized for the uniquely derived fermion field equations (defined and expressed solely in (1 + 2)-dimensional space-time), where this property generates a certain set of additional symmetries corresponding uniquely to the $SU(2)_L \otimes U(2)_R$ symmetry group for spin-1/2 fermion fields (representing "1 + 3" generations of four fermions, including a group of eight leptons and a group of eight quarks), and also the $SU(2)_L \otimes U(2)_R$ and SU(3) gauge symmetry groups for spin-1 boson fields coupled to the spin-1/2 fermionic source currents. Hence, along with the known elementary particles, eight new elementary particles including four new charge-less right-handed spin-1/2 fermions (two leptons and two quarks), a spin-3/2 fermion, and also three new spin-1 (massive) bosons, are predicted uniquely by this mathematical axiomatic approach. As a particular result, based on the definite formulation of derived Maxwell (and Yang-Mills) field equations, it has also been concluded that magnetic monopoles cannot exist in nature.[1]

---

## 1. Introduction and Summary

Why do the fundamental forces of nature (i.e., the forces that appear to cause all the movements and interactions in the universe) manifest in the way, shape, and form that they do? This is one of the greatest ontological questions that science can investigate. In this article, we'll consider this basic and crucial question (and a number of relevant issues) via a new axiomatic mathematical formalism. By definition, a basic law of physics (or a scientific law in general) is: "A theoretical principle deduced from particular facts, applicable to a defined group or class of phenomena, and expressible by the statement that a particular phenomenon always occurs if certain conditions be present" [1]. Eugene Wigner's foundational paper—"On the Unreasonable Effectiveness of Mathematics in the Natural Sciences"—famously observed that purely mathematical structures and formalisms often lead to deep physical insights, in turn serving as the basis for highly successful physical theories [2]. However, all the known fundamental laws of physics (and corresponding mathematical formalisms which are used for their representations), are generally the conclusions of a number of repeated experiments and observations over years and have become accepted universally within the scientific community [3,4]. It should be noted that all the current fundamental field equations that comprise the general formulatory framework of physics are formulated and obtained from the results of experiments (such as the Einstein, Maxwell, Dirac, Schrödinger equations and so on, using in all modern classical and quantum field equations). The extent of their validity is merely the extent to which they correctly predict and agree with experimental results. From a historical point of view, these field equations can be also equivalently written and formulated on the basis of Lagrangian, Hamiltonian, and similar field theoretic formulations in connection with the principle of stationary action [5–10].

---

1 https://cds.cern.ch/record/1980381, https://ui.adsabs.harvard.edu/#abs/2015arXiv150101373Z, https://Inspirehep.net/record/1387680.

This article is based on my earlier publications (Refs. [11–14], Springer, 1996–1998). In this article, as a new mathematical approach to origin of the laws of nature, using a new basic algebraic axiomatic (matrix) formalism based on the ring theory and Clifford algebras (presented in Section 2), "it is shown that certain mathematical forms of fundamental laws of nature, including laws governing the fundamental forces of nature (represented by a set of two definite classes of general covariant massive field equations, with new matrix formalisms), are derived uniquely from only a very few axioms"and, in agreement with the rational Lorentz group, it is also basically assumed that the components of relativistic energy-momentum can only take rational values.

Concerning the basic assumption of the rationality of relativistic energy-momentum, it is necessary to note that the rational Lorentz symmetry group is not only dense in the general form of Lorentz group, but also is compatible with the necessary conditions required basically for the formalism of a consistent relativistic quantum theory [15]. In essence, the main scheme of this new mathematical axiomatic approach to fundamental laws of nature is as follows. First based on the assumption of rationality of $D$-momentum, by linearization (along with a parameterization procedure) of the Lorentz invariant energy-momentum quadratic relation, a unique set of Lorentz invariant systems of homogeneous linear equations (with matrix formalisms compatible with certain Clifford, and symmetric algebras) is derived. Then by initial quantization (followed by a basic procedure of minimal coupling to space-time geometry) of these determined systems of linear equations, a set of two classes of general covariant massive (tensor) field equations (with matrix formalisms compatible with certain Clifford, and Weyl algebras) is derived uniquely as well. Each class of the derived general covariant field equations also includes a definite form of torsion field appearing as the generator of the corresponding field's invariant mass. In addition, it is shown that the (1 + 3)-dimensional cases of two classes of derived field equations represent a new general covariant massive formalism of bispinor fields of spin-2, and spin-1 particles, respectively. In fact, these uniquely determined bispinor fields represent a unique set of new generalized massive forms of the laws governing the fundamental forces of nature, including the Einstein (gravitational), Maxwell (electromagnetic) and Yang-Mills (nuclear) field equations. Moreover, it is also shown that the (1 + 2)-dimensional cases of two classes of these field equations represent (asymptotically) a new general covariant massive formalism of bispinor fields of spin-3/2 and spin-1/2 particles, respectively, corresponding to the Dirac and Rarita–Schwinger equations.

Of particular consequence, it is shown that a certain massive formalism of general relativity—with a definite form of torsion field appeared originally as the generator of gravitational field's invariant mass—is obtained only by initial quantization (followed by a basic procedure of minimal coupling to space-time geometry) of a certain set of special relativistic algebraic matrix equations. It has been also proved that Lagrangian densities specified for the originally derived new massive forms of the Maxwell, Yang-Mills and Dirac field equations, are also gauge invariant, where the invariant mass of each field is generated solely by the corresponding torsion field. In addition, in agreement with recent astronomical data, a new particular form of massive boson is identified (corresponding to U(1) gauge group) with invariant mass: $m_\gamma \approx 4.90571 \times 10^{-50}$ kg, generated by a coupled torsion field of the background space-time geometry.

Moreover, based on the definite mathematical formalism of this axiomatic approach, along with the C, P and T symmetries (represented basically by the corresponding quantum operators) of the fundamentally derived field equations, it has been concluded that the universe could be realized solely with the (1 + 2) and (1 + 3)-dimensional space-times (where this conclusion, in particular, is based on the T-symmetry). It is proved that 'CPT' is the only (unique) combination of C, P, and T symmetries that could be defined as a symmetry for interacting fields. In addition, on the basis of these discrete symmetries of derived field equations, it has been also shown that only left-handed particle fields (along with their complementary right-handed fields) could be coupled to the corresponding (any) source currents. Furthermore, it has been shown that the metric of background space-time is diagonalized for the uniquely derived fermion field equations (defined and expressed solely in (1 + 2)-dimensional space-time), where this property generates a certain set of additional symmetries corresponding uniquely to the $SU(2)_L \otimes U(2)_R$ symmetry group for spin-1/2 fermion fields (representing "1 + 3" generations of four fermions, including a group of eight leptons and a group of eight quarks), and also the $SU(2)_L \otimes U(2)_R$ and SU(3) gauge symmetry groups for spin-1 boson fields coupled to the spin-1/2 fermionic source currents. Hence, along with the known elementary particles, eight new elementary particles, including: four new charge-less right-handed spin-1/2 fermions (two leptons and two quarks, represented by "$z_e$, $z_n$ and $z_u$, $z_d$"), a spin-3/2 fermion, and also three new spin-1 massive bosons (represented by $\tilde{W}^+, \tilde{W}^-, \vec{Z}$, where in particular, the new boson $\vec{Z}$ is complementary right-handed particle of ordinary $Z$ boson), have been predicted uniquely and expressly by this new mathematical axiomatic approach.

As a particular result, presented in Section 3.6, based on the definite and unique formulation of the derived Maxwell's equations (and also determined Yang-Mills equations, represented uniquely with two specific forms of gauge symmetries, in Section 3.15), it has also been concluded generally that magnetic monopoles cannot exist in nature.

### *1.1. The Main Results Obtained in This Article are Based on the Following Three Basic Assumptions (as Postulates)*

**(1)-** "A new definite axiomatic generalization of the axiom of "no zero divisors" of integral domains (including the ring of integers $\mathbb{Z}$);"

This algebraic postulate (as a new mathematical concept) is formulated as follows:

**"**Let $A=[a_{ij}]$ be a $n\times n$ matrix with entries expressed by the following linear homogeneous polynomials in $s$ variables over the integral domain $\mathbb{Z}$: $a_{ij} = a_{ij}(b_1, b_2, b_3, \ldots, b_s) = \sum_{k=1}^{s} H_{ijk} b_k$; suppose also "$\exists r \in \mathbb{N}$: $A^r = F(b_1, b_2, b_3, \ldots, b_s) I_n$", where $F(b_1, b_2, b_3, \ldots, b_s)$ is a homogeneous polynomial of degree $r \geq 2$, and $I_n$ is $n\times n$ identity matrix; Then the following axiom is assumed (as a new axiomatic generalization of the ordinary axiom of "no zero divisors" of integral domain $\mathbb{Z}$):

$$(A^r = 0) \Leftrightarrow (A\times M = 0,\ M \neq 0) \quad (1)$$

where $M$ is a non-zero arbitrary $n\times 1$ column matrix".

The axiomatic relation (1) is a logical biconditional, where $(A^r = 0)$ and $(A\times M = 0,\ M \neq 0)$ are respectively the antecedent and consequent of this biconditional. In addition, based on the initial assumption $\exists r \in \mathbb{N}$: $A^r = F(b_1, b_2, b_3, \ldots, b_s) I_n$, the axiomatic biconditional (1) could be also represented as follows:

$$[F(b_1, b_2, b_3, \ldots, b_s) = 0] \Leftrightarrow (A\times M = 0,\ M \neq 0) \quad (1\text{-}1)$$

where the homogeneous equation $F(b_1, b_2, b_3, \ldots, b_s) = 0$, and system of linear equations $(A\times M = 0,\ M \neq 0)$ are respectively the antecedent and consequent of biconditional (1-1). The axiomatic biconditional (1-1), defines a system of linear equations of the type $A\times M = 0$ $(M \neq 0)$, as the algebraic equivalent representation of $r^{th}$ degree homogeneous equation $F(b_1, b_2, b_3, \ldots, b_s) = 0$ (over the integral domain $\mathbb{Z}$). In addition, according to the Ref. [16], since $F(b_1, b_2, b_3, \ldots, b_s) = 0$ is a homogeneous equation over $\mathbb{Z}$, it is also concluded that homogeneous equations defined over the field of rational numbers $\mathbb{Q}$, obey the axiomatic relations (1) and (1-1) as well. As particular outcome of this new mathematical axiomatic formalism (based on the axiomatic relations (1) and (1-1), including their basic algebraic properties), in Section 3.4, it is shown that using a unique set of general covariant massive (tensor) field equations (with new matrix formalism compatible with Clifford, and Weyl algebras), corresponding with the fundamental field equations of physics, are derived—where, in agreement with the rational Lorentz symmetry group, it has been basically assumed that the components of relativistic energy-momentum can only take the rational values. In Sections 3.2–3.15, we present in detail the main applications of this basic algebraic assumption (along with the following basic assumptions (2) and (3)) in fundamental physics.

**(2)-** "In agreement with the rational Lorentz symmetry group, we assume basically that the components of relativistic energy-momentum (*D*-momentum) can only take the rational values."

Concerning this assumption, it is necessary to note that the rational Lorentz symmetry group is not only dense in the general form of the Lorentz group, but is also compatible with the necessary basic conditions required for the formalism of a consistent relativistic quantum theory [15]. Moreover, this assumption is clearly also compatible with any quantum circumstance in which the energy-momentum of a relativistic particle is transferred as integer multiples of the quantum of action "*h*" (Planck constant).

Before defining the next basic assumption, it should be noted that from basic assumptions (1) and (2), it follows directly that the Lorentz invariant energy-momentum quadratic relation (represented by formula (52), in Section 3.1) is a particular form of homogeneous quadratic equation (represented by formula (18-2) in Section 2.2). Hence, using the set of systems of linear equations that are determined uniquely as equivalent algebraic representations of the corresponding set of quadratic homogeneous equations (given by Equation (18-2) in various numbers of unknown variables, respectively), a unique set of the Lorentz invariant systems of homogeneous linear equations (with matrix formalisms compatible with certain Clifford, and symmetric algebras) are also determined, representing equivalent algebraic forms of the energy-momentum quadratic relation in various space-time dimensions, respectively. Subsequently, we have shown that by an initial quantization (followed by a basic procedure of minimal coupling to space-time geometry) of these determined systems of linear equations, a unique set of two definite classes of general covariant massive (tensor) field equations (with matrix formalisms compatible with certain Clifford, and Weyl algebras) is also derived, corresponding with various space-time dimensions, respectively. In addition, it is also shown that this derived set of two classes of general covariant field equations represent new tensor massive (matrix) formalism of the fundamental field equations of physics, corresponding with fundamental laws of nature (including the laws governing the fundamental forces of nature). Following these essential results, in addition to the basic assumptions (1) and (2), it would also be assumed that:

**(3)-** "We assume that the mathematical formalism of the fundamental laws of nature, are defined solely by the axiomatic matrix constitution formulated uniquely on the basis of postulates (1) and (2)".

In addition to this basic assumption, in Section 3.11, the C, P and T symmetries of the uniquely derived general covariant field equations (that are field Equations (3) and (4) in Section 1.2), would be represented by their corresponding quantum matrix operators.

### *1.2. A Summary of the Main Consequences of Basic Assumptions (1)–(3)*

In the following, we present a summary description of the main consequences of basic assumptions (1)–(3) (mentioned in Section 1.1) in fundamental physics. In this article, the metric signature (+ − ... −), the geometrized units [17] and also the following sign conventions have been used in the representations of the Riemann curvature tensor $R^{\rho}{}_{\sigma\mu\nu}$, Ricci tensor $R_{\mu\nu}$ and Einstein tensor $G_{\mu\nu}$:

$$R^{\rho}_{\sigma\mu\nu}=(\partial_{\nu}\Gamma^{\rho}_{\sigma\mu}+\Gamma^{\rho}_{\lambda\nu}\Gamma^{\lambda}_{\sigma\mu})-(\partial_{\mu}\Gamma^{\rho}_{\sigma\nu}+\Gamma^{\rho}_{\lambda\mu}\Gamma^{\lambda}_{\sigma\nu}),\ \nabla_{\sigma}R_{\mu\nu\rho}{}^{\sigma}=\nabla_{\nu}R_{\mu\rho}-\nabla_{\mu}R_{\nu\rho},\ G_{\mu\nu}=-8\pi T_{\mu\nu}+\ldots \quad (2)$$

On the basis of assumptions (1)–(3), two sets of the general covariant field equations (compatible with the Clifford algebras) are derived solely as follows:

$$(i\hbar\alpha^{\mu}\nabla_{\mu}-m_0^{(R)}\tilde{\alpha}^{\mu}k_{\mu})\Psi_R=0 \quad (3)$$

$$(i\hbar\alpha^{\mu}D_{\mu}-m_0^{(F)}\tilde{\alpha}^{\mu}k_{\mu})\Psi_F=0 \quad (4)$$

where

$$\alpha^{\mu}=\beta^{\mu}+\beta'^{\mu},\ \tilde{\alpha}^{\mu}=\beta^{\mu}-\beta'^{\mu} \quad (5)$$

$i\hbar\nabla_{\mu}$ and $i\hbar D_{\mu}$ are the general relativistic forms of energy-momentum quantum operator (where $\nabla_{\mu}$ is the general covariant derivative and $D_{\mu}$ is gauge covariant derivative, defining in Sections 3.4–3.6), $m_0^{(R)}$ and $m_0^{(F)}$ are the fields' invariant masses, $k_{\mu}=(c/\sqrt{g^{00}},0,\ldots,0)$ is the general covariant velocity in stationary reference frame (that is a time-like covariant vector), $\beta^{\mu}$ and $\beta'^{\mu}$ are two contravariant square matrices (given by formulas (6) and (7)), $\Psi_R$ is a column matrix given by formulas (6) and (7), which contains the components of field strength tensor $R_{\mu\nu\rho\sigma}$ (equivalent to the Riemann curvature tensor), and also the components of a covariant quantity which defines the corresponding source current (by relations (6) and (7)), $\Psi_F$ is also a column matrix given by formulas (6) and (7), which contains the components of tensor field $F_{\mu\nu}$ (defined as the gauge field strength tensor), and also the components of a covariant quantity that defines the corresponding source current (by relations (6) and (7)). In Section 3.11, based on a basic class of discrete symmetries of general covariant field Equations (3) and (4), it would be concluded that these equations could be defined solely in (1 + 2) and (1 + 3) space-time dimensions, where the (1 + 2) and (1 + 3)-dimensional cases these field equations are given uniquely as follows (in terms of the above-mentioned quantities), respectively:

For (1 + 2)-dimensional space-time we have:

$$\beta^0=\begin{bmatrix}0 & 0\\ 0 & -(\sigma^0+\sigma^1)\end{bmatrix},\ \beta'_0=\begin{bmatrix}\sigma^0+\sigma^1 & 0\\ 0 & 0\end{bmatrix},\ \beta^1=\begin{bmatrix}0 & \sigma^2\\ -\sigma^2 & 0\end{bmatrix},\ \beta'_1=\begin{bmatrix}0 & \sigma^3\\ -\sigma^3 & 0\end{bmatrix},$$

$$\beta^2=\begin{bmatrix}0 & -\sigma^1\\ -\sigma^0 & 0\end{bmatrix},\ \beta'_2=\begin{bmatrix}0 & -\sigma^0\\ -\sigma^1 & 0\end{bmatrix},\ \sigma^0=\begin{bmatrix}1 & 0\\ 0 & 0\end{bmatrix},\ \sigma^1=\begin{bmatrix}0 & 0\\ 0 & 1\end{bmatrix},\ \sigma^2=\begin{bmatrix}0 & 1\\ 0 & 0\end{bmatrix},\ \sigma^3=\begin{bmatrix}0 & 0\\ -1 & 0\end{bmatrix},$$

$$\Psi_R=\begin{bmatrix}R_{\rho\sigma10}\\ 0\\ R_{\rho\sigma21}\\ \phi^{(R)}_{\rho\sigma}\end{bmatrix},\quad \Psi_F=\begin{bmatrix}F_{10}\\ 0\\ F_{21}\\ \phi^{(F)}\end{bmatrix},\quad \begin{array}{l} J^{(R)}_{\rho\sigma\nu}=-(\breve{\nabla}_{\nu}+\dfrac{im_0^{(R)}}{\hbar}k_{\nu})\phi^{(R)}_{\rho\sigma},\\ J^{(F)}_{\nu}=-(\breve{D}_{\nu}+\dfrac{im_0^{(F)}}{\hbar}k_{\nu})\phi^{(F)};\end{array} \quad (6)$$

For (1 + 3)-dimensional space-time of we get:

$$\beta^0=\begin{bmatrix}0 & 0\\ 0 & -(\gamma^0+\gamma^1)\end{bmatrix},\ \beta'_0=\begin{bmatrix}(\gamma^0+\gamma^1) & 0\\ 0 & 0\end{bmatrix},\ \beta^1=\begin{bmatrix}0 & \gamma^2\\ -\gamma^3 & 0\end{bmatrix},\ \beta'_1=\begin{bmatrix}0 & \gamma^3\\ -\gamma^2 & 0\end{bmatrix},$$

$$\beta^2=\begin{bmatrix}0&\gamma^4\\ \gamma^5&0\end{bmatrix},\ \beta_2'=\begin{bmatrix}0&-\gamma^5\\ -\gamma^4&0\end{bmatrix},\ \beta^3=\begin{bmatrix}0&\gamma^6\\ -\gamma^7&0\end{bmatrix},\ \beta_3'=\begin{bmatrix}0&\gamma^7\\ -\gamma^6&0\end{bmatrix},$$

$$\gamma^0=\begin{bmatrix}1&0&0&0\\0&1&0&0\\0&0&0&0\\0&0&0&0\end{bmatrix},\ \gamma^1=\begin{bmatrix}0&0&0&0\\0&0&0&0\\0&0&1&0\\0&0&0&1\end{bmatrix},\ \gamma^2=\begin{bmatrix}0&0&0&1\\0&0&0&0\\0&0&0&0\\-1&0&0&0\end{bmatrix},\ \gamma^3=\begin{bmatrix}0&0&0&0\\0&0&-1&0\\0&1&0&0\\0&0&0&0\end{bmatrix},$$

$$\gamma^4=\begin{bmatrix}0&0&0&0\\0&0&0&1\\0&0&0&0\\0&-1&0&0\end{bmatrix},\ \gamma^5=\begin{bmatrix}0&0&-1&0\\0&0&0&0\\1&0&0&0\\0&0&0&0\end{bmatrix},\ \gamma^6=\begin{bmatrix}0&0&0&0\\0&0&0&0\\0&0&0&1\\0&0&-1&0\end{bmatrix},\ \gamma^7=\begin{bmatrix}0&-1&0&0\\1&0&0&0\\0&0&0&0\\0&0&0&0\end{bmatrix},$$

$$\Psi_R=\begin{bmatrix}R_{\rho\sigma10}\\R_{\rho\sigma20}\\R_{\rho\sigma30}\\0\\R_{\rho\sigma23}\\R_{\rho\sigma31}\\R_{\rho\sigma12}\\\phi^{(R)}_{\rho\sigma}\end{bmatrix},\quad \Psi_F=\begin{bmatrix}F_{10}\\F_{20}\\F_{30}\\0\\F_{23}\\F_{31}\\F_{12}\\\phi^{(F)}\end{bmatrix},\quad \begin{aligned}J^{(R)}_{\rho\sigma\nu}&=-(\breve{\nabla}_\nu+\frac{im_0^{(R)}}{\hbar}k_\nu)\phi^{(R)}_{\rho\sigma},\\ J^{(F)}_\nu&=-(\breve{D}_\nu+\frac{im_0^{(F)}}{\hbar}k_\nu)\phi^{(F)};\end{aligned}\tag{7}$$

In formulas (6) and (7), $J^{(R)}_{\nu\rho\sigma}$ and $J^{(F)}_\nu$ are the covariant source currents expressed necessarily in terms of the covariant quantities $\phi^{(R)}_{\rho\sigma}$ and $\phi^{(F)}$ (as initially given quantities). Moreover, in Sections 3.4–3.15, it has been shown that the field equations in (1 + 2) dimensions, are compatible with the matrix representation of Clifford algebra $C\ell_{1,2}$, and represent (asymptotically) a new general covariant massive formalism of bispinor fields of spin-3/2 and spin-1/2 particles, respectively. It has also been shown that these field equations in (1 + 3) dimensions are compatible with the matrix representation of Clifford algebra $C\ell_{1,3}$, and represent solely a new general covariant massive formalism of bispinor fields of spin-2 and spin-1 particles, respectively.

In addition, from the field Equations (3) and (4), the following field equations (with ordinary tensor formulations) could be also obtained, respectively:

$$\breve{\nabla}_\lambda R_{\rho\sigma\mu\nu}+\breve{\nabla}_\mu R_{\rho\sigma\nu\lambda}+\breve{\nabla}_\nu R_{\rho\sigma\lambda\mu}=T^{\tau}{}_{\lambda\mu}R_{\rho\sigma\tau\nu}+T^{\tau}{}_{\mu\nu}R_{\rho\sigma\tau\lambda}+T^{\tau}{}_{\nu\lambda}R_{\rho\sigma\tau\mu},\tag{3-1}$$

$$\breve{\nabla}_\mu R_{\rho\sigma}{}^{\mu\nu}-(im_0^{(R)}/\hbar)k_\mu R_{\rho\sigma}{}^{\mu\nu}=-J^{(R)\nu}_{\rho\sigma};\tag{3-2}$$

$$R^\rho{}_{\sigma\mu\nu}=(\partial_\nu\Gamma^\rho{}_{\sigma\mu}+\Gamma^\rho{}_{\lambda\nu}\Gamma^\lambda{}_{\sigma\mu})-(\partial_\mu\Gamma^\rho{}_{\sigma\nu}+\Gamma^\rho{}_{\lambda\mu}\Gamma^\lambda{}_{\sigma\nu}),$$

$$J^{(R)}_{\rho\sigma\nu}=-(\breve{\nabla}_\nu+\frac{im_0^{(R)}}{\hbar}k_\nu)\phi^{(R)}_{\rho\sigma},\ T_{\tau\mu\nu}=\frac{im_0^{(R)}}{2\hbar}(g_{\tau\mu}k_\nu-g_{\tau\nu}k_\mu).\tag{3-3}$$

and

$$\breve{D}_\lambda F_{\mu\nu}+\breve{D}_\mu F_{\nu\lambda}+\breve{D}_\nu F_{\lambda\mu}=0,\tag{4-1}$$

$$\breve{D}_\mu F^{\mu\nu}=-J^{\nu(F)};\tag{4-2}$$

$$F_{\mu\nu}=\breve{D}_\nu A_\mu-\breve{D}_\mu A_\nu,\tag{4-3}$$

$$J_{\nu}^{(F)} = -(\breve{D}_{\nu} + \frac{im_0^{(F)}}{\hbar} k_{\nu})\phi^{(F)}, \;\; Z_{\tau\mu\nu} = \frac{im_0^{(F)}}{2\hbar}(g_{\tau\mu}k_{\nu} - g_{\tau\nu}k_{\mu}).$$

where in Equations (3-1)–(3-2), $\Gamma^{\rho}{}_{\sigma\mu}$ is the affine connection given by: $\Gamma^{\rho}{}_{\sigma\mu} = \bar{\Gamma}^{\rho}{}_{\sigma\mu} - K^{\rho}{}_{\sigma\mu}$, $\bar{\Gamma}^{\rho}{}_{\sigma\mu}$ is the Christoffel symbol (or the torsion-free connection), $K^{\rho}{}_{\sigma\mu}$ is a contorsion tensor defined by: $K_{\rho\sigma\mu} = (im_0^{(R)}/2\hbar)g_{\rho\mu}k_{\sigma}$ (that is anti-symmetric in the first and last indices), $T_{\rho\sigma\mu}$ is its corresponding torsion tensor given by: $T_{\rho\sigma\mu} = K_{\rho\mu\sigma} - K_{\rho\sigma\mu}$ (as the generator of the gravitational field's invariant mass), $\breve{\nabla}_{\mu}$ is general covariant derivative defined with torsion $T_{\rho\sigma\mu}$. In Equations (4-1)–(4-3), $\breve{D}_{\mu}$ is the general relativistic form of gauge covariant derivative defined with torsion field $Z_{\tau\mu\nu}$ (which generates the gauge field's invariant mass), and $A_{\mu}$ denotes the corresponding gauge (potential) field.

In Section 3.11, on the basis of definite mathematical formalism of this axiomatic approach, along with the C, P and T symmetries (represented basically by the corresponding quantum operators, in Section 3.11) of the fundamentally derived field equations, it has been concluded that the universe could be realized solely with the (1 + 2) and (1 + 3)-dimensional space-times (where this conclusion, in particular, is based on the T-symmetry). It is proved that 'CPT' is the only (unique) combination of C, P, and T symmetries that could be defined as a symmetry for interacting fields. In addition, on the basis of these discrete symmetries of derived field equations, it has been also shown that only left-handed particle fields (along with their complementary right-handed fields) could be coupled to the corresponding (any) source currents. Furthermore, it has been shown that the metric of background space-time is diagonalized for the uniquely derived fermion field equations (defined and expressed solely in (1 + 2)-dimensional space-time), where this property generates a certain set of additional symmetries corresponding uniquely to the $SU(2)_L \otimes U(2)_R$ symmetry group for spin-1/2 fermion fields (representing "1 + 3" generations of four fermions, including a group of eight leptons and a group of eight quarks), and also the $SU(2)_L \otimes U(2)_R$ and SU(3) gauge symmetry groups for spin-1 boson fields coupled to the spin-1/2 fermionic source currents. Hence, along with the known elementary particles, eight new elementary particles, including four new charge-less right-handed spin-1/2 fermions (two leptons and two quarks, represented by "$z_e$, $z_n$ and $z_u$, $z_d$"), a spin-3/2 fermion, and also three new spin-1 massive bosons (represented by $\tilde{W}^{+}, \tilde{W}^{-}, \vec{Z}$, where in particular the new boson $\vec{Z}$ is complementary right-handed particle of ordinary $Z$ boson), have been predicted uniquely by this new mathematical axiomatic approach (as shown in Section 3.15).

As a consequence, in Section 3.6 it is shown that a certain massive formalism of the general theory of relativity—with a definite torsion field which generates the gravitational field's mass—is obtained only by first quantization (followed by a basic procedure of minimal coupling to space-time geometry) of a set of special relativistic algebraic matrix relations. In Section 3.9, it is also proved that Lagrangian densities specified for the derived unique massive forms of Maxwell, Yang-Mills and Dirac equations, are gauge-invariant as well, where the invariant mass of each field is generated by the corresponding torsion field. In addition, in Section 3.10, in agreement with recent astronomical data, a new massive boson is identified (corresponding to U(1) gauge group) with invariant mass: $m_{\gamma} \approx 4.90571 \times 10^{-50}$ kg, generated by a coupling torsion field of the background space-time geometry. Furthermore, in Section 3.6, based on the definite and unique formulation of the derived Maxwell's equations (and also determined Yang-Mills equations, represented unqiely with two specific forms of gauge symmetries), it is also concluded that magnetic monopoles cannot exist in nature.

As shown in Section 3.7, if the Ricci curvature tensor $R_{\mu\nu}$ is defined by the following relation in terms of Riemann curvature tensor (which is determined by field Equations (3-1)–(3-3)):

$$(\breve{\nabla}_{\sigma}+\frac{im_0^{(R)}}{\hbar}k_{\sigma})R_{\mu\nu\rho}{}^{\sigma}=(\breve{\nabla}_{\nu}+\frac{im_0^{(R)}}{\hbar}k_{\nu})R_{\mu\rho}-(\breve{\nabla}_{\mu}+\frac{im_0^{(R)}}{\hbar}k_{\mu})R_{\nu\rho}, \quad (8\text{-}1)$$

then from this expression for the current in terms of the stress-energy tensor $T_{\mu\nu}$ :

$$J_{\rho\sigma\nu}^{(R)}=-8\pi[(\nabla_{\sigma}+\frac{im_0^{(R)}}{\hbar}k_{\sigma})T_{\rho\nu}-(\nabla_{\rho}+\frac{im_0^{(R)}}{\hbar}k_{\rho})T_{\sigma\nu}]+8\pi B[(\nabla_{\sigma}+\frac{im_0^{(R)}}{\hbar}k_{\sigma})Tg_{\rho\nu}-(\nabla_{\rho}+\frac{im_0^{(R)}}{\hbar}k_{\rho})Tg_{\sigma\nu}] \quad (8\text{-}2)$$

where $T=T^{\mu}{}_{\mu}$, the gravitational field equations (including a cosmological constant $\Lambda$ emerged naturally in the course of derivation process) could be equivalently derived from the massless case of tensor field Equations (3-1)–(3-3) in (1 + 3) space-time dimensions, as follows:

$$R_{\mu\nu}=-8\pi T_{\mu\nu}+4\pi Tg_{\mu\nu}-\Lambda g_{\mu\nu} \quad (9)$$

Let us emphasize again that the results obtained in this article are the direct outcome of a new algebraic-axiomatic approach,[2] presented in Section 2. This algebraic approach—in the form of a basic linearization theory—has been constructed on the basis of a new single axiom (that is the axiom (17) in Section 2.1) proposed to replace with the ordinary axiom of "no zero divisors" of integral domains (that is the axiom (16) in Section 2). In fact, as noted in Sections 1.1 and 2.1, the new proposed axiom is a definite generalized form of ordinary axiom (16), which has been formulated in terms of square matrices (using basically as primary objects for representing the elements of underlying algebra, i.e., integral domains including the ring of integers). In Section 3, based on this new algebraic axiomatic formalism, as a new mathematical approach to origin of the laws of nature, "it is shown that certain mathematical forms of fundamental laws of nature, including laws governing the fundamental forces of nature (represented by a set of two definite classes of general covariant massive field equations, with new matrix formalisms), are derived uniquely from only a very few axioms" in which, in agreement with the rational Lorentz group, it is also assumed that the components of relativistic energy-momentum can only take rational values.

## 2. Theory of Linearization: A New Algebraic-Axiomatic (Matrix) Formalism Based on the Ring Theory and Clifford Algebras

In this Section, a new algebraic theory of linearization (including the simultaneous parameterization) of the homogeneous equations is presented that is formulated on the basis of ring theory and matrix representation of the generalized Clifford algebras (associated with homogeneous forms of degree $r \geq 2$ defined over the integral domain $\mathbb{Z}$).

Mathematical models of physical processes include certain classes of mathematical objects and the relations between these objects. The models of this type, which are most commonly used, are groups, rings, vector spaces, and linear algebras. A group is a set $G$ with a single operation (multiplication) $a \times b = c$ ; $a, b, c \in G$ which obeys the known conditions [18,19]. A ring is a set of elements $R$, where two binary operations, namely, addition and multiplication, are defined. With respect to addition this set is a group, and multiplication is connected with addition by the

2 Besides, we may argue that our presented axiomatic matrix approach (for a direct derivation and formulating the fundamental laws of nature uniquely) is not subject to the Gödel's incompleteness theorems [18]. As in our axiomatic approach, firstly, we have basically changed (i.e., replaced and generalized) one of the main Peano axioms (when these axioms algebraically are augmented with the operations of addition and multiplication [19–21]) for integers, which is the algebraic axiom of "no zero divisors". Secondly, based on our approach, one of the axiomatic properties of integers (i.e., axiom of "no zero divisors") could be accomplished solely by the arbitrary square matrices (with integer components). This axiomatic reformulation of algebraic properties of integers thoroughly has been presented in Section 2 of this article.

distributivity laws: $a\times(b+c)=(a\times b)+(a\times c)$, $(b+c)\times a=(b\times a)+(c\times a)$; $a,b,c\in R$. The rings reflect the structural properties of the set *R*. As distinct from the group models, those connected with rings are not frequently applied, although in physics various algebras of matrices, algebras of hyper-complex numbers, Grassman and Clifford algebras are widely used. This is due to the intricacy of finding a connection between the binary relations of addition and multiplication and the element of the rings [12,22,23]. This Section is devoted to the development of a rather simple approach of establishing such a connection and an analysis of concrete problems on this basis.

I have found that if the algebraic axiom of "no zero divisors" of integral domains is generalized expressing in terms of the square matrices (as it has been formulated by the axiomatic relation (17)), fruitful new results hold. In this Section, first on the basis of the matrix representation of the generalized Clifford algebras (associated with homogeneous polynomials of degree $r \geq 2$ over the integral domain $\mathbb{Z}$), we have presented a new generalized formulation of the algebraic axiom of "no zero divisors" of integral domains. Subsequently, a linearization theory has been constructed axiomatically that implies (necessarily and sufficiently) any homogeneous equation of degree $r \geq 2$ over the integral domain $\mathbb{Z}$, should be linearized (and parameterized simultaneously), and its solution investigated systematically via its equivalent linearized-parameterized formulation (representing as a certain type of system of linear homogeneous equations). In Sections 2.2 and 2.4, by this axiomatic approach a class of homogeneous quadratic equations (in various numbers of variables) over $\mathbb{Z}$ has been considered explicitly.

### *2.1. The Basic Properties of the Integral Domain $\mathbb{Z}$*

The ordinary basic properties of the integral domain $\mathbb{Z}$ with binary operations $(+,\times)$ are represented as follows, respectively [22,23] ($\forall a_i, a_j, a_k, \ldots \in \mathbb{Z}$):

Closure:

$$a_k + a_l \in \mathbb{Z},\; a_k \times a_l \in \mathbb{Z} \tag{10}$$

Associativity:

$$a_k + (a_l + a_p) = (a_k + a_l) + a_p,\; a_k \times (a_l \times a_p) = (a_k \times a_l) \times a_p \tag{11}$$

Commutativity:

$$a_k + a_l = a_l + a_k,\; a_k \times a_l = a_l \times a_k \tag{12}$$

Existence of identity elements:

$$a_k + 0 = a_k,\; a_k \times 1 = a_k \tag{13}$$

Existence of inverse element (for operator of addition):

$$a_k + (-a_k) = 0 \tag{14}$$

Distributivity:

$$a_k \times (a_l + a_p) = (a_k \times a_l) + (a_k \times a_p),\; (a_k + a_l) \times a_p = (a_k \times a_p) + (a_l \times a_p) \tag{15}$$

No zero divisors (as a logical bi-conditional for operator of multiplication):

$$a_k = 0 \Leftrightarrow (a_k \times a_l = 0,\; a_l \neq 0) \tag{16}$$

Axiom (16), equivalently, could be also expressed as follows,

$$(a_k = 0 \vee a_l = 0) \Leftrightarrow a_k \times a_l = 0 \quad (16\text{-}1)$$

In this article, as a new basic algebraic property of the domain of integers, we present the following new axiomatic generalization of the ordinary axiom of "no zero divisors" (16), which particularly has been formulated on the basis of matrix formalism of Clifford algebras (associated with homogeneous polynomials of degree $r \geq 2$, over the integral domain ℤ):

"Let $A = [a_{ij}]$ be a $n \times n$ matrix with entries expressed by the following linear homogeneous polynomials in $s$ variables over the integral domain ℤ: $a_{ij} = a_{ij}(b_1, b_2, b_3, \ldots, b_s) = \sum_{k=1}^{s} H_{ijk} b_k$; suppose also " $\exists\, r \in$ ℕ: $A^r = F(b_1, b_2, b_3, \ldots, b_s) I_n$ ", where $F(b_1, b_2, b_3, \ldots, b_s)$ is a homogeneous polynomial of degree $r \geq 2$, and $I_n$ is $n \times n$ identity matrix; Then the following axiom is assumed (as a new axiomatic generalization of the ordinary axiom of "no zero divisors" of integral domain ℤ):

$$(A^r = 0) \Leftrightarrow (A \times M = 0,\ M \neq 0) \quad (17)$$

where $M$ is a non-zero arbitrary $n \times 1$ column matrix".

The axiomatic relation (17) is a logical biconditional, where $(A^r = 0)$ and $(A \times M = 0,\ M \neq 0)$ are respectively the antecedent and consequent of this biconditional. In addition, based on the initial assumption $\exists\, r \in$ ℕ: $A^r = F(b_1, b_2, b_3, \ldots, b_s) I_n$, the axiomatic biconditional (17) could be also represented as follows:

$$[F(b_1, b_2, b_3, \ldots, b_s) = 0] \Leftrightarrow (A \times M = 0,\ M \neq 0) \quad (17\text{-}1)$$

where the homogeneous equation $F(b_1, b_2, b_3, \ldots, b_s) = 0$, and system of linear equations $(A \times M = 0,\ M \neq 0)$ are respectively the antecedent and consequent of biconditional (17-1). The axiomatic biconditional (17-1), defines a system of linear equations of the type $A \times M = 0$ $(M \neq 0)$, as the algebraic equivalent representation of $r^{th}$ degree homogeneous equation $F(b_1, b_2, b_3, \ldots, b_s) = 0$ (over the integral domain ℤ). The axiom (17) (or (17-1)) for $n = 1$, is equivalent to the ordinary axiom of "no zero divisors" (16). In fact, the axiom (16), as a particular case, can be obtained from the axiom (17) (or (17-1), but not vice versa.

Moreover, according to the Ref. [16], since $F(b_1, b_2, b_3, \ldots, b_s) = 0$ is a *homogeneous* equation over ℤ, it is also concluded that homogeneous equations defined over the field of rational numbers ℚ, obey the axiomatic relations (17) and (17-1).

As a crucial additional issue concerning the axiom (17), it should be noted that the condition " $\exists\, r \in$ ℕ: $A^r = F(b_1, b_2, b_3, \ldots, b_s) I_n$" which is assumed initially in the axiom (17), is also compatible with matrix representation of the generalized Clifford algebras [24–33] associated with the $r^{th}$ degree homogeneous polynomials $F(b_1, b_2, b_3, \ldots, b_s)$. In fact, we may represent uniquely the square matrix $A$ (with assumed properties in the axiom (17)) by this homogeneous linear form: $A = \sum_{k=1}^{s} b_k E_k$, then the relation: $A^r = F(b_1, b_2, b_3, \ldots, b_s) I_n$ implies that the square matrices $E_k$ (which their entries are independent from the variables $b_k$) would be generators of the corresponding generalized Clifford algebra associated with the $r^{th}$ degree homogeneous polynomial $F(b_1, b_2, b_3, \ldots, b_s)$. However, in some particular cases and applications, we may also assume some additional conditions for the generators $E_k$, such as the Hermiticity or anti-Hermiticity (see Sections 2.2, 2.4 and 3.1). In Section 3, we use these algebraic properties of the square matrix $A$ (corresponding with the homogeneous quadratic equations), where we present explicitly the main applications of the axiomatic relations (17) and (17-1) in foundations of physics (where we also use the basic assumptions (2) and (3) mentioned in Section 1.1).

It is noteworthy that since the axiom (17) has been formulated solely in terms of square matrices, in Ref. [34][3] we have shown that all the ordinary algebraic axioms (10)–(15) of the integral domain ⊚ (except the axiom of "no zero divisors" (16)) in addition to the new axiom (17) could be also reformulated uniformly in terms of the set of square matrices. Hence, we may conclude that the square matrices, logically, are the most elemental algebraic objects for representing the basic properties of set of integers (as the most fundamental set of mathematics).

In the following, based on the axiomatic relation (17-1), we have constructed a corresponding basic algebraic linearization (including a parameterization procedure) approach applicable to the all classes of homogeneous equation. Hence, it could be also shown that for any given homogeneous equation of degree $r \geq 2$ over the ring $\mathbb{Z}$ (or field $\mathbb{Q}$), a square matrix $A$ exists that obeys the relation (17-1). In this regard, for various classes of homogeneous equations, their equivalent systems of linear equations would be derived systematically. As a particular crucial case, in Sections 2.2 and 2.4, by derivation of the systems of linear equations equivalent to a class of quadratic homogeneous equations (in various number of unknown variables) over the integral domain $\mathbb{Z}$ (or field $\mathbb{Q}$), these equations have been analyzed (and solved) thoroughly by this axiomatic approach. In the following, the basic schemes of this axiomatic linearization-parameterization approach are described.

First, it should be noted that since the entries $a_{ij}$ of square matrix $A$ are linear homogeneous forms expressed in terms of the integral variables $b_p$, i.e., $a_{ij} = \sum_{k=1}^{s} H_{ijk} b_k$, we may also represent the square matrix $A$ by this linear matrix form: $A = \sum_{k=1}^{s} b_k E_k$, then (as noted above) the relation: $A^r = F(b_1, b_2, b_3, \ldots, b_s) I_n$ implies that the square matrices $E_k$ (the entries of which are independent from the variables $b_k$) would be generators of the corresponding generalized Clifford algebra associated with the $r^{th}$ degree homogeneous polynomial $F(b_1, b_2, b_3, \ldots, b_s)$ [28–33]. However, for some particular cases of the $r^{th}$ degree homogeneous forms $F(b_1, b_2, b_3, \ldots, b_s)$ (for $r \geq 2$, such as the standard quadratic forms defined in the quadratic Equation (18) in Section 2.2), without any restriction in the existence and procedure of derivation of their corresponding square matrices $A = \sum_{k=1}^{s} b_k E_k$ (with the algebraic properties assuming in axiom (17)) obeying the Clifford algebraic relation: $A^r = F(b_1, b_2, b_3, \ldots, b_s) I_n$, we may also assume certain additional conditions for the matrix generators $E_k$, such as the Hermiticity (or anti-Hermiticity), and so on (see Sections 2.2, 2.4 and 3.1). In fact, these conditions could be required, for example if a homogeneous invariant relation (of physics) is represented by a homogeneous algebraic equation of the type: $F(b_1, b_2, b_3, \ldots, b_s) = 0$, with the algebraic properties as assumed in the axiom (17), where the variables $b_k$ denote the components of corresponding physical quantity (such as the relativistic energy-momentum, as it has been assumed in Section 3.1 of this article based on the basic assumption (2) noted in Sections 1.1 and 3.1).

---

[3] Some of recent citations to this paper include: Gudder, S. Reconditioning in Discrete Quantum Field Theory. *Int. J. Theor. Phys.* **2017**, 1–14; McKenzie, A. The Level I Multiverse is not the same as the Level III Multiverse. *arXiv* **2017**, arXiv:1702.02019; McKenzie, A. Some Remarks on the Mathematical Structure of the Multiverse. *PhilSci-Archive, University of Pittsburgh Publication, Pittsburgh, USA,* 2016. *Paper ID*: 11921, URL: http://philsci-archive.pitt.edu/11921/, *arXiv* **2016**, arXiv:1602.04247.

In Section 2.2, as one of the main applications of the axiomatic relations (17) and (17-1), we derive a unique set the square matrices $A_{n\times n}$ (by assuming a minimum value for $n$, i.e., the size of the corresponding matrix $A_{n\times n}$) corresponding to the quadratic homogeneous equations of the type: $\sum_{i=0}^{s} e_i f_i = 0$, for $s$ = 0, 1, 2, 3, 4, ..., respectively. Subsequently, in Section 2.4, by solving the corresponding systems of linear equations $A \times M = 0$, we obtain the general parametric solutions of the quadratic homogeneous equations $\sum_{i=0}^{s} e_i f_i = 0$, for $s$ = 0, 1, 2, 3, 4,.., respectively. In addition, in Section 2.3 using this systematic axiomatic approach, for some particular forms of homogeneous equations of degrees 3, 4 and 5, their equivalent systems of linear equations have been derived as well. It is noteworthy that using this general axiomatic approach (on the basis of the logical biconditional relations (17) and (17-1)), for any given $r^{th}$ degree homogeneous equation in $s$ unknown variables over the integral domain over $\mathbb{Z}$, its equivalent system(s) of linear equations $A \times M = 0$ is derivable (with a unique size, if in the course of the derivation, we also assume a minimum value for $n$, i.e., the size of corresponding square matrix $A_{n\times n}$). Furthermore, for a given homogeneous equation of degree $r$ in $s$ unknown variables, the minimum value for $n$, i.e., the size of the corresponding square matrix $A_{n\times n}$ in its equivalent matrix equation: $A \times M = 0$, is: $n_{\min} = r^{s-1} \times r^{s-1}$ for $r = 2$, and $n_{\min} = r^{s} \times r^{s}$ for $r > 2$. For additional detail concerning the general methodology of the derivation of square matrix $A_{n\times n}$ and the matrix equation $A \times M = 0$ equivalent to a given homogeneous equation of degree $r$ in $s$ unknown variables—on the basis of the axiomatic relations (17) and (17-1)—see also the preprint versions of this article in Ref. [34].

### 2.2. The Applications of Axiom (17-1) to Homogeneous Quadratic Equations

In this section, on the basis of axiomatic relation (17-1) and the general methodological notes (mentioned above), for the following general form of homogeneous quadratic equations their equivalent systems of linear equations are derived (uniquely):

$$Q(e_0, f_0, e_1, f_1, \ldots, e_s, f_s) = \sum_{i=0}^{s} e_i f_i = 0 \tag{18}$$

Equation (18) for $s$ = 0, 1, 2, 3 , 4, … is represented by, respectively:

$$\sum_{i=0}^{0} e_i f_i = e_0 f_0 = 0, \tag{19}$$

$$\sum_{i=0}^{1} e_i f_i = e_0 f_0 + e_1 f_1 = 0, \tag{20}$$

$$\sum_{i=0}^{2} e_i f_i = e_0 f_0 + e_1 f_1 + e_2 f_2 = 0, \tag{21}$$

$$\sum_{i=0}^{3} e_i f_i = e_0 f_0 + e_1 f_1 + e_2 f_2 + e_3 f_3 = 0, \tag{22}$$

$$\sum_{i=0}^{4} e_i f_i = e_0 f_0 + e_1 f_1 + e_2 f_2 + e_3 f_3 + e_4 f_4 = 0. \tag{23}$$

It is necessary to note that quadratic Equation (18) is isomorphic to the following ordinary representations of homogeneous quadratic equations:

$$\sum_{i,j=0}^{s} G_{ij} c_i c_j = 0, \tag{18-1}$$

$$\sum_{i,j=0}^{s} G_{ij} c_i c_j = \sum_{i,j=0}^{s} G_{ij} d_i d_j \ , \tag{18-2}$$

using the linear transformations:

$$\begin{bmatrix} e_0 \\ e_1 \\ e_3 \\ . \\ . \\ . \\ e_s \end{bmatrix} = \begin{bmatrix} G_{00} & G_{01} & G_{02} & . & . & . & G_{0s} \\ G_{10} & G_{11} & G_{12} & . & . & . & G_{1s} \\ G_{20} & G_{21} & G_{22} & . & . & . & G_{2s} \\ . & & & & & & \\ . & & & & & & \\ . & & & & & & \\ G_{s0} & G_{s1} & G_{s2} & . & . & . & G_{ss} \end{bmatrix} \begin{bmatrix} c_0 + d_0 \\ c_1 + d_1 \\ c_2 + d_2 \\ . \\ . \\ . \\ c_s + d_s \end{bmatrix}, \quad \begin{bmatrix} f_0 \\ f_1 \\ f_3 \\ . \\ . \\ . \\ f_s \end{bmatrix} = \begin{bmatrix} c_0 - d_0 \\ c_1 - d_1 \\ c_2 - d_2 \\ . \\ . \\ . \\ c_s - d_s \end{bmatrix} \tag{18-3}$$

where $[G_{ij}]$ is a symmetric and invertible square matrix, i.e, $G_{ij} = G_{ji}$ and $\det[G_{ij}] \neq 0$, and the quadratic form $\sum_{i,j=0}^{s} G_{ij} c_i c_j$ in Equation (18-1) could be obtained via transformations (18-3), only by taking $d_i = 0$.

We note here, and show below, that the reason for choosing Equation (18) as the standard general form for representing the homogeneous quadratic equations (that could also be transformed to the ordinary representations of homogeneous quadratic Equations (18-1) and (18-2), by linear transformations (18-3)) is not only its very simple algebraic structure, but also the simple linear homogeneous forms of the entries of square matrices $A$ (expressed in terms of variables $e_i, f_i$) in the corresponding systems of linear equations $A \times M = 0$ obtained as the equivalent form of quadratic Equation (18) in various number of unknown variables.

Moreover, as shown in the following, we may also assume certain Hermiticity and anti-Hermiticity conditions for the deriving square matrices $A$ (in the corresponding systems of linear equations $A \times M = 0$ equivalent to the quadratic Equation (18)), without any restriction in the existence and procedure of derivation of these matrices. By adding these particular conditions, for a specific number of variables in Equation (18), its equivalent matrix equation $A \times M = 0$ could be determined uniquely. In Section 3, where we use the algebraic results obtained in Sections 2.2 and 2-4 on the basis of axiomatic relations (17) and (17-1) in fact the assumption of these Hermiticity and anti-Hermiticity properties is a necessary issue. These Hermiticity and anti-Hermiticity additional conditions are defined as follows:

"First, by supposing: $e_0 = f_0$ and $e_i = -f_i$ (for $i = 1, 2, \ldots, s$ ), the quadratic Equation (18) would be represented as: $e_0^2 - \sum_{i=1}^{s} e_i^2 = 0$, and consequently the corresponding square matrix $A$ in the deriving system of linear equations $A \times M = 0$ (which equivalently represent the quadratic Equation (18), based on the axiomatic relation (17-1)) could be also expressed by the homogeneous linear matrix form: $A=\sum_{i=0}^{s} e_i E_i$ , where the real matrices $E_i$ are generators of the corresponding Clifford algebra associated with the standard quadratic form $e_0^2-\sum_{i=1}^{s} e_i^2$ .

Now for defining the relevant Hermiticity and anti-Hermiticity conditions, we assume that any square matrix $A$ in the deriving matrix equation: $A\times M = 0$ (as the equivalent representation of quadratic Equation (18)), should also has this additional property that by supposing: $e_0=f_0$ and $e_i=-f_i$ by which the square matrix $A$ could be represented as: $A=\sum_{i=0}^{s} e_i E_i$ , the matrix generator $E_0$ be Hermitian: $E_0=E_0^*$ , and matrix generators $E_i$ (for $i = 1, 2, \ldots, s$ ) be anti-Hermitian: $E_i=-E_i^*$

As noted previously and shown below, by assuming the above additional Hermiticity and anti-Hermiticity conditions, the system of linear equations $A\times M = 0$ corresponding to quadratic Equation (18), is determined uniquely for any specific number of variables $e_i, f_i$. Hence, starting from the simplest (or most trivial) case of quadratic Equation (18)—i.e., Equation (19— its equivalent system of linear equations is given uniquely as follows:

$$A\times M = \begin{bmatrix} 0 & e_0 \\ f_0 & 0 \end{bmatrix} \begin{bmatrix} m_1 \\ m_2 \end{bmatrix} = 0 \tag{24}$$

where it is assumed $M \neq 0$ , and in agreement with (17-1) we also have:

$$A^2 = \begin{bmatrix} 0 & e_0 \\ f_0 & 0 \end{bmatrix} \times \begin{bmatrix} 0 & e_0 \\ f_0 & 0 \end{bmatrix} = (e_0 f_0) I_2 \tag{24-1}$$

For Equation (20), the corresponding equivalent system of linear equations is determined as:

$$A\times M = \begin{bmatrix} 0 & A' \\ A'' & 0 \end{bmatrix} \begin{bmatrix} M'' \\ M' \end{bmatrix} = \begin{bmatrix} 0 & 0 & e_0 & f_1 \\ 0 & 0 & e_1 & -f_0 \\ f_0 & f_1 & 0 & 0 \\ e_1 & -e_0 & 0 & 0 \end{bmatrix} \begin{bmatrix} m_1 \\ m_2 \\ m_3 \\ m \end{bmatrix} = 0 \tag{25}$$

where we have:

$$M = \begin{bmatrix} M'' \\ M' \end{bmatrix} \neq 0,\ A^2 = \begin{bmatrix} 0 & 0 & e_0 & f_1 \\ 0 & 0 & e_1 & -f_0 \\ f_0 & f_1 & 0 & 0 \\ e_1 & -e_0 & 0 & 0 \end{bmatrix} \times \begin{bmatrix} 0 & 0 & e_0 & f_1 \\ 0 & 0 & e_1 & -f_0 \\ f_0 & f_1 & 0 & 0 \\ e_1 & -e_0 & 0 & 0 \end{bmatrix} = (e_0 f_0 + e_1 f_1) I_4 \tag{25-1}$$

Notice that matrix Equation (25) could be represented by two matrix equations, as follows:

$$A'\times M' = \begin{bmatrix} e_0 & f_1 \\ e_1 & -f_0 \end{bmatrix} \begin{bmatrix} m_3 \\ m \end{bmatrix} = 0, \tag{25-2}$$

$$A''\times M'' = \begin{bmatrix} f_0 & f_1 \\ e_1 & -e_0 \end{bmatrix} \begin{bmatrix} m_1 \\ m_2 \end{bmatrix} = 0 \tag{25-3}$$

The matrix Equations (25-2) and (25-3) are equivalent (due to the assumption of arbitrariness of parameters $m_1, m_2, m_3, m$), so we may choose the matrix Equation (25-2) as the system of linear equations equivalent to the quadratic Equation (20)—where for simplicity in the indices of parameters $m_i$, we may simply replace arbitrary parameter $m_3$ with arbitrary parameter $m_1$, as follows (for $\begin{bmatrix} m_1 \\ m \end{bmatrix} \neq 0$):

$$\begin{bmatrix} e_0 & f_1 \\ e_1 & -f_0 \end{bmatrix} \begin{bmatrix} m_1 \\ m \end{bmatrix} = 0. \tag{26}$$

The system of linear equations corresponding to the quadratic Equation (21) is obtained as:

$$A \times M = \begin{bmatrix} 0 & A' \\ A'' & 0 \end{bmatrix} \begin{bmatrix} M'' \\ M' \end{bmatrix} = \begin{bmatrix} 0 & 0 & 0 & 0 & e_0 & 0 & -e_2 & f_1 \\ 0 & 0 & 0 & 0 & 0 & e_0 & -e_1 & -f_2 \\ 0 & 0 & 0 & 0 & -f_2 & -f_1 & -f_0 & 0 \\ 0 & 0 & 0 & 0 & e_1 & -e_2 & 0 & -f_0 \\ f_0 & 0 & -e_2 & f_1 & 0 & 0 & 0 & 0 \\ 0 & f_0 & -e_1 & -f_2 & 0 & 0 & 0 & 0 \\ -f_2 & -f_1 & -e_0 & 0 & 0 & 0 & 0 & 0 \\ e_1 & -e_2 & 0 & -e_0 & 0 & 0 & 0 & 0 \end{bmatrix} \begin{bmatrix} m_1 \\ m_2 \\ m_3 \\ m_4 \\ m_5 \\ m_6 \\ m_7 \\ m \end{bmatrix} = 0, \tag{27}$$

where in agreement with (17) we have:

$$A^2 = \begin{bmatrix} 0 & 0 & 0 & 0 & e_0 & 0 & -e_2 & f_1 \\ 0 & 0 & 0 & 0 & 0 & e_0 & -e_1 & -f_2 \\ 0 & 0 & 0 & 0 & -f_2 & -f_1 & -f_0 & 0 \\ 0 & 0 & 0 & 0 & e_1 & -e_2 & 0 & -f_0 \\ f_0 & 0 & -e_2 & f_1 & 0 & 0 & 0 & 0 \\ 0 & f_0 & -e_1 & -f_2 & 0 & 0 & 0 & 0 \\ -f_2 & -f_1 & -e_0 & 0 & 0 & 0 & 0 & 0 \\ e_1 & -e_2 & 0 & -e_0 & 0 & 0 & 0 & 0 \end{bmatrix} \times \begin{bmatrix} 0 & 0 & 0 & 0 & e_0 & 0 & -e_2 & f_1 \\ 0 & 0 & 0 & 0 & 0 & e_0 & -e_1 & -f_2 \\ 0 & 0 & 0 & 0 & -f_2 & -f_1 & -f_0 & 0 \\ 0 & 0 & 0 & 0 & e_1 & -e_2 & 0 & -f_0 \\ f_0 & 0 & -e_2 & f_1 & 0 & 0 & 0 & 0 \\ 0 & f_0 & -e_1 & -f_2 & 0 & 0 & 0 & 0 \\ -f_2 & -f_1 & -e_0 & 0 & 0 & 0 & 0 & 0 \\ e_1 & -e_2 & 0 & -e_0 & 0 & 0 & 0 & 0 \end{bmatrix} = (e_0 f_0 + e_1 f_1 + e_2 f_2) I_8 \tag{27-1}$$

In addition, similar to Equation (25), the obtained matrix Equation (27) is equivalent to a system of two matrix equations, as follows:

$$A' \times M' = \begin{bmatrix} e_0 & 0 & -e_2 & f_1 \\ 0 & e_0 & -e_1 & -f_2 \\ -f_2 & -f_1 & -f_0 & 0 \\ e_1 & -e_2 & 0 & -f_0 \end{bmatrix} \begin{bmatrix} m_5 \\ m_6 \\ m_7 \\ m \end{bmatrix} = 0, \tag{27-2}$$

$$A'' \times M'' = \begin{bmatrix} f_0 & 0 & -e_2 & f_1 \\ 0 & f_0 & -e_1 & -f_2 \\ -f_2 & -f_1 & -e_0 & 0 \\ e_1 & -e_2 & 0 & -e_0 \end{bmatrix} \begin{bmatrix} m_1 \\ m_2 \\ m_3 \\ m_4 \end{bmatrix} = 0 \tag{27-3}$$

The matrix Equations (27-2) and (27-3) are equivalent (due to the assumption of arbitrariness of parameters $m_1, m_2, \ldots, m_7, m$), so we may choose the Equation (27-2) as the system of linear equations corresponding to the quadratic Equation (21)—where for simplicity in the indices of parameters $m_i$, we may simply replace the arbitrary parameters $m_5$, $m_6$, $m_7$ with parameters $m_1$, $m_2$, $m_3$, as follows:

$$\begin{bmatrix} e_0 & 0 & -e_2 & f_1 \\ 0 & e_0 & -e_1 & -f_2 \\ -f_2 & -f_1 & -f_0 & 0 \\ e_1 & -e_2 & 0 & -f_0 \end{bmatrix}\begin{bmatrix} m_1 \\ m_2 \\ m_3 \\ m \end{bmatrix} = 0, \quad \begin{bmatrix} m_1 \\ m_2 \\ m_3 \\ m \end{bmatrix} \neq 0. \tag{28}$$

Similarly, for the quadratic Equations (22) the corresponding system of linear equations is obtained uniquely as follows:

$$\begin{bmatrix} e_0 & 0 & 0 & 0 & 0 & -e_3 & e_2 & f_1 \\ 0 & e_0 & 0 & 0 & e_3 & 0 & -e_1 & f_2 \\ 0 & 0 & e_0 & 0 & -e_2 & e_1 & 0 & f_3 \\ 0 & 0 & 0 & e_0 & -f_1 & -f_2 & -f_3 & 0 \\ 0 & f_3 & -f_2 & -e_1 & -f_0 & 0 & 0 & 0 \\ -f_3 & 0 & f_1 & -e_2 & 0 & -f_0 & 0 & 0 \\ f_2 & -f_1 & 0 & -e_3 & 0 & 0 & -f_0 & 0 \\ e_1 & e_2 & e_3 & 0 & 0 & 0 & 0 & -f_0 \end{bmatrix}\begin{bmatrix} m_1 \\ m_2 \\ m_3 \\ m_4 \\ m_5 \\ m_6 \\ m_7 \\ m \end{bmatrix} = 0 \tag{29}$$

where the column parametric matrix $M$ in (29) is non-zero $M \neq 0$.

In a similar manner, the uniquely obtained system of linear equations corresponding to the quadratic Equation (23), is given by:

$$\begin{bmatrix} e_0 & 0 & 0 & 0 & 0 & 0 & 0 & 0 & 0 & 0 & 0 & -e_4 & 0 & -e_3 & -e_2 & f_1 \\ 0 & e_0 & 0 & 0 & 0 & 0 & 0 & 0 & 0 & 0 & e_4 & 0 & e_3 & 0 & -e_1 & -f_2 \\ 0 & 0 & e_0 & 0 & 0 & 0 & 0 & 0 & 0 & -e_4 & 0 & 0 & e_2 & e_1 & 0 & f_3 \\ 0 & 0 & 0 & e_0 & 0 & 0 & 0 & 0 & e_4 & 0 & 0 & 0 & -f_1 & f_2 & -f_3 & 0 \\ 0 & 0 & 0 & 0 & e_0 & 0 & 0 & 0 & 0 & -e_3 & -e_2 & -e_1 & 0 & 0 & 0 & -f_4 \\ 0 & 0 & 0 & 0 & 0 & e_0 & 0 & 0 & e_3 & 0 & f_1 & -f_2 & 0 & 0 & f_4 & 0 \\ 0 & 0 & 0 & 0 & 0 & 0 & e_0 & 0 & e_2 & -f_1 & 0 & f_3 & 0 & -f_4 & 0 & 0 \\ 0 & 0 & 0 & 0 & 0 & 0 & 0 & e_0 & e_1 & f_2 & -f_3 & 0 & f_4 & 0 & 0 & 0 \\ 0 & 0 & 0 & f_4 & 0 & f_3 & f_2 & f_1 & -f_0 & 0 & 0 & 0 & 0 & 0 & 0 & 0 \\ 0 & 0 & -f_4 & 0 & -f_3 & 0 & -e_1 & e_2 & 0 & -f_0 & 0 & 0 & 0 & 0 & 0 & 0 \\ 0 & f_4 & 0 & 0 & -f_2 & e_1 & 0 & -e_3 & 0 & 0 & -f_0 & 0 & 0 & 0 & 0 & 0 \\ -f_4 & 0 & 0 & 0 & -f_1 & e_2 & e_3 & 0 & 0 & 0 & 0 & -f_0 & 0 & 0 & 0 & 0 \\ 0 & f_3 & f_2 & -e_1 & 0 & 0 & 0 & e_4 & 0 & 0 & 0 & 0 & -f_0 & 0 & 0 & 0 \\ -f_3 & 0 & f_1 & e_2 & 0 & 0 & -e_4 & 0 & 0 & 0 & 0 & 0 & 0 & -f_0 & 0 & 0 \\ -f_2 & -f_1 & 0 & -e_3 & 0 & e_4 & 0 & 0 & 0 & 0 & 0 & 0 & 0 & 0 & -f_0 & 0 \\ e_1 & -e_2 & e_3 & 0 & -e_4 & 0 & 0 & 0 & 0 & 0 & 0 & 0 & 0 & 0 & 0 & -f_0 \end{bmatrix}\begin{bmatrix} m_1 \\ m_2 \\ m_3 \\ m_4 \\ m_5 \\ m_6 \\ m_7 \\ m_8 \\ m_9 \\ m_{10} \\ m_{11} \\ m_{12} \\ m_{13} \\ m_{14} \\ m_{15} \\ m \end{bmatrix} = 0 \tag{30}$$

where we have assumed the parametric column matrix $M$ in (30) is non-zero, $M \neq 0$.

In a similar manner, the systems of linear equations (written in matrix forms similar to the matrix Equations (24), (26), (28)–(30)) with larger sizes are obtained for the quadratic Equation (18) in more variables (i.e., for $s = 6, 7, 8, \ldots$), where the size of the square matrices of the corresponding matrix equations is $2^s \times 2^s$ (which could be reduced to $2^{s-1} \times 2^{s-1}$ for $s \geq 2$). In general (as it has been also mentioned in Section 2.1), the size of the $n \times n$ square matrices $A$ (with the minimum value for $n$) in the matrix equations $A \times M = 0$ corresponding to the homogeneous polynomials $F(b_1, b_2, b_3, \ldots, b_s)$ of degree $r$ defined in axiom (17) is $r^s \times r^s$ (which for $r = 2$ this size, in particular, could be reduced to $2^{s-1} \times 2^{s-1}$). Moreover, based on the axiom (17), by solving the obtained system of linear equations corresponding to a homogeneous equation of degree $r$ we may systematically show (and decide) whether this equation has the integral solution.

*2.3. The Applications of Axiom (17-1) to Higher Degree Homogeneous Equations*

Similar to the uniquely obtained systems of linear equations corresponding to the homogeneous quadratic equations (in Section 2.2), in this section in agreement with the axiom (17), we present the obtained systems of linear equations, i.e., $A \times M = 0$ (by assuming the minimum value for $n$, i.e., the size of square matrix $A_{n \times n}$), corresponding to some homogeneous equations of degrees 3, 4 and 5, respectively. For the homogeneous equation of degree three of the type:

$$F(e_0,f_0,e_1,f_1,e_2,f_2)=e_0^2 f_0 - e_0 f_0^2 + e_2^2 f_2 - e_2 f_2^2 + e_1 f_1 g_1 = 0, \tag{31}$$

the corresponding system of linear equations is given as follows:

$$A \times M = \begin{bmatrix} 0 & 0 & A_1 \\ A_2 & 0 & 0 \\ 0 & A_3 & 0 \end{bmatrix} \begin{bmatrix} m_1 \\ m_2 \\ \vdots \\ m_{27} \end{bmatrix} = 0, \tag{32}$$

where $A$ is a 27 × 27 square matrix written in terms of the square 9 × 9 matrices $A_1, A_2$ and $A_3$, given by:

$$A_1 = \begin{bmatrix} -e_2+f_2 & 0 & 0 & 0 & 0 & 0 & -e_0+f_0 & e_1 & 0 \\ 0 & -e_2+f_2 & 0 & 0 & 0 & 0 & 0 & e_0 & g_1 \\ 0 & 0 & -e_2+f_2 & 0 & 0 & 0 & f_1 & 0 & -f_0 \\ -f_0 & e_1 & 0 & e_2 & 0 & 0 & 0 & 0 & 0 \\ 0 & -e_0+f_0 & g_1 & 0 & e_2 & 0 & 0 & 0 & 0 \\ f_1 & 0 & e_0 & 0 & 0 & e_2 & 0 & 0 & 0 \\ 0 & 0 & 0 & e_0 & e_1 & 0 & -f_2 & 0 & 0 \\ 0 & 0 & 0 & 0 & -f_0 & g_1 & 0 & -f_2 & 0 \\ 0 & 0 & 0 & f_1 & 0 & -e_0+f_0 & 0 & 0 & -f_2 \end{bmatrix},$$
$$A_2 = \begin{bmatrix} -f_2 & 0 & 0 & 0 & 0 & 0 & -e_0+f_0 & e_1 & 0 \\ 0 & -f_2 & 0 & 0 & 0 & 0 & 0 & e_0 & g_1 \\ 0 & 0 & -f_2 & 0 & 0 & 0 & f_1 & 0 & -f_0 \\ -f_0 & e_1 & 0 & -e_2+f_2 & 0 & 0 & 0 & 0 & 0 \\ 0 & -e_0+f_0 & g_1 & 0 & -e_2+f_2 & 0 & 0 & 0 & 0 \\ f_1 & 0 & e_0 & 0 & 0 & -e_2+f_2 & 0 & 0 & 0 \\ 0 & 0 & 0 & e_0 & e_1 & 0 & e_2 & 0 & 0 \\ 0 & 0 & 0 & 0 & -f_0 & g_1 & 0 & e_2 & 0 \\ 0 & 0 & 0 & f_1 & 0 & -e_0+f_0 & 0 & 0 & e_2 \end{bmatrix}, \tag{33}$$

$$A_3 = \begin{bmatrix} e_2 & 0 & 0 & 0 & 0 & 0 & -e_0+f_0 & e_1 & 0 \\ 0 & e_2 & 0 & 0 & 0 & 0 & 0 & e_0 & g_1 \\ 0 & 0 & e_2 & 0 & 0 & 0 & f_1 & 0 & -f_0 \\ -f_0 & e_1 & 0 & -f_2 & 0 & 0 & 0 & 0 & 0 \\ 0 & -e_0+f_0 & g_1 & 0 & -f_2 & 0 & 0 & 0 & 0 \\ f_1 & 0 & e_0 & 0 & 0 & -f_2 & 0 & 0 & 0 \\ 0 & 0 & 0 & e_0 & e_1 & 0 & -e_2+f_2 & 0 & 0 \\ 0 & 0 & 0 & 0 & -f_0 & g_1 & 0 & -e_2+f_2 & 0 \\ 0 & 0 & 0 & f_1 & 0 & -e_0+f_0 & 0 & 0 & -e_2+f_2 \end{bmatrix}$$

The uniquely obtained system of linear equations (i.e., $A \times M = 0$, by assuming the minimum size for the square matrix $A_{n \times n}$) corresponds to the well-known homogeneous equation of degree three:

$$F(a,b,c) = 2(a^3 - c^3 + Bb^3) = 0 \tag{34}$$

and has the following form (in compatible with the new axiom (17)):

$$A \times M = \begin{bmatrix} 0 & 0 & A_1 \\ A_2 & 0 & 0 \\ 0 & A_3 & 0 \end{bmatrix} \begin{bmatrix} m_1 \\ m_2 \\ \vdots \\ m_{27} \end{bmatrix} = 0, \tag{35}$$

where $A$ is a 27 × 27 square matrix written in terms of the 9 × 9 matrices $A_1$, $A_2$ and $A_3$ given by:

$$A_1 = \begin{bmatrix} -a & 0 & 0 & 0 & 0 & 0 & -2c & b & 0 \\ 0 & -a & 0 & 0 & 0 & 0 & 0 & c & 2b \\ 0 & 0 & -a & 0 & 0 & 0 & b & 0 & c \\ c & b & 0 & -a & 0 & 0 & 0 & 0 & 0 \\ 0 & -2c & 2b & 0 & -a & 0 & 0 & 0 & 0 \\ b & 0 & c & 0 & 0 & -a & 0 & 0 & 0 \\ 0 & 0 & 0 & c & b & 0 & 2a & 0 & 0 \\ 0 & 0 & 0 & 0 & c & 2b & 0 & 2a & 0 \\ 0 & 0 & 0 & b & 0 & -2c & 0 & 0 & 2a \end{bmatrix},$$

$$A_2 = \begin{bmatrix} 2a & 0 & 0 & 0 & 0 & 0 & -2c & b & 0 \\ 0 & 2a & 0 & 0 & 0 & 0 & 0 & c & 2b \\ 0 & 0 & 2a & 0 & 0 & 0 & b & 0 & c \\ c & b & 0 & -a & 0 & 0 & 0 & 0 & 0 \\ 0 & -2c & 2b & 0 & -a & 0 & 0 & 0 & 0 \\ b & 0 & c & 0 & 0 & -a & 0 & 0 & 0 \\ 0 & 0 & 0 & c & b & 0 & -a & 0 & 0 \\ 0 & 0 & 0 & 0 & c & 2b & 0 & -a & 0 \\ 0 & 0 & 0 & b & 0 & -2c & 0 & 0 & -a \end{bmatrix}, \tag{36}$$

$$A_3 = \begin{bmatrix} -a & 0 & 0 & 0 & 0 & 0 & -2c & b & 0 \\ 0 & -a & 0 & 0 & 0 & 0 & 0 & c & 2b \\ 0 & 0 & -a & 0 & 0 & 0 & b & 0 & c \\ c & b & 0 & 2a & 0 & 0 & 0 & 0 & 0 \\ 0 & -2c & 2b & 0 & 2a & 0 & 0 & 0 & 0 \\ b & 0 & c & 0 & 0 & 2a & 0 & 0 & 0 \\ 0 & 0 & 0 & c & b & 0 & -a & 0 & 0 \\ 0 & 0 & 0 & 0 & c & 2b & 0 & -a & 0 \\ 0 & 0 & 0 & b & 0 & -2c & 0 & 0 & -a \end{bmatrix}.$$

For the 4th degree homogeneous equation of the type:

$$F(e_1, e_2, f_1, f_2, f_3, f_4) = -e_1 e_2^3 + e_1^3 e_2 + f_1 f_2 f_3 f_4 = 0 \tag{37}$$

the corresponding system of linear equations is given as:

$$A \times M = \begin{bmatrix} 0 & 0 & 0 & A_1 \\ -A_2 & 0 & 0 & 0 \\ 0 & A_3 & 0 & 0 \\ 0 & 0 & A_4 & 0 \end{bmatrix} \begin{bmatrix} m_1 \\ m_2 \\ ::: \\ m_{16} \end{bmatrix} - 0, \tag{38}$$

where $A$ is a 16 × 16 square matrix represented in terms of the 4 × 4 matrices $A_1, A_2, A_3, A_4$:

$$A_1 = \begin{bmatrix} e_1 + e_2 & 0 & 0 & f_1 \\ -f_2 & e_1 & 0 & 0 \\ 0 & f_3 & -e_1 + e_2 & 0 \\ 0 & 0 & f_4 & -e_2 \end{bmatrix}, \quad A_2 = \begin{bmatrix} -e_2 & 0 & 0 & f_1 \\ f_2 & e_1 + e_2 & 0 & 0 \\ 0 & -f_3 & e_1 & 0 \\ 0 & 0 & f_4 & -e_1 + e_2 \end{bmatrix},$$
$$A_3 = \begin{bmatrix} -e_1 + e_2 & 0 & 0 & f_1 \\ f_2 & -e_2 & 0 & 0 \\ 0 & f_3 & e_1 + e_2 & 0 \\ 0 & 0 & -f_4 & e_1 \end{bmatrix}, \quad A_4 = \begin{bmatrix} e_1 & 0 & 0 & -f_1 \\ f_2 & -e_1 + e_2 & 0 & 0 \\ 0 & f_3 & -e_2 & 0 \\ 0 & 0 & f_4 & e_1 + e_2 \end{bmatrix}. \tag{39}$$

In addition, the system of linear equations corresponding to 5th degree homogeneous equation of the type,

$$F(e_1, e_2, f_1, f_2, f_3, f_4, f_5) = e_1^4 e_2 - e_1^3 e_2^2 - e_1^2 e_2^3 + e_1 e_2^4 + f_1 f_2 f_3 f_4 f_5 = 0 \tag{40}$$

is determined as:

$$A \times M = \begin{bmatrix} 0 & 0 & 0 & 0 & A_1 \\ A_2 & 0 & 0 & 0 & 0 \\ 0 & A_3 & 0 & 0 & 0 \\ 0 & 0 & A_4 & 0 & 0 \\ 0 & 0 & 0 & A_5 & 0 \end{bmatrix} \begin{bmatrix} m_1 \\ m_2 \\ m_3 \\ ::: \\ m_{25} \end{bmatrix} = 0, \tag{41}$$

where $A$ is a 25 × 25 square matrix expressed in terms of the following 5 × 5 matrices $A_1, A_2, A_3, A_4, A_5$:

$$A_1 = \begin{bmatrix} e_1 - e_2 & 0 & 0 & 0 & f_1 \\ f_2 & e_1 & 0 & 0 & 0 \\ 0 & f_3 & e_2 & 0 & 0 \\ 0 & 0 & f_4 & -e_1 + e_2 & 0 \\ 0 & 0 & 0 & f_5 & -e_1 - e_2 \end{bmatrix},$$
$$A_2 = \begin{bmatrix} -e_1 - e_2 & 0 & 0 & 0 & f_1 \\ f_2 & e_1 - e_2 & 0 & 0 & 0 \\ 0 & f_3 & e_1 & 0 & 0 \\ 0 & 0 & f_4 & e_2 & 0 \\ 0 & 0 & 0 & f_5 & -e_1 + e_2 \end{bmatrix},$$
$$A_3 = \begin{bmatrix} -e_1 + e_2 & 0 & 0 & 0 & f_1 \\ f_2 & -e_1 - e_2 & 0 & 0 & 0 \\ 0 & f_3 & e_1 - e_2 & 0 & 0 \\ 0 & 0 & f_4 & e_1 & 0 \\ 0 & 0 & 0 & f_5 & e_2 \end{bmatrix}, \quad (42)$$
$$A_4 = \begin{bmatrix} e_2 & 0 & 0 & 0 & f_1 \\ f_2 & -e_1 + e_2 & 0 & 0 & 0 \\ 0 & f_3 & -e_1 - e_2 & 0 & 0 \\ 0 & 0 & f_4 & e_1 - e_2 & 0 \\ 0 & 0 & 0 & f_5 & e_1 \end{bmatrix},$$
$$A_5 = \begin{bmatrix} e_1 & 0 & 0 & 0 & f_1 \\ f_2 & e_2 & 0 & 0 & 0 \\ 0 & f_3 & -e_1 + e_2 & 0 & 0 \\ 0 & 0 & f_4 & -e_1 - e_2 & 0 \\ 0 & 0 & 0 & f_5 & e_1 - e_2 \end{bmatrix}.$$

### *2.4. Obtaining the General Parametric Solutions of Homogeneous Quadratic Equations by Solving their Corresponding Systems of the Linear Equations (Derived in Section 2.2)*

In this section by solving the derived systems of the linear Equations (26), (28)–(30) that correspond with the quadratic homogeneous Equations (20)–(23) in Section 2.2, the general parametric solutions of these equations are obtained for unknowns $e_i$ and $f_i$. There are the standard methods for obtaining the general solutions of the systems of homogeneous linear equations in integers [35,36]. Using these methods, for the system of linear Equation (26) (and consequently, its corresponding quadratic Equation (20)) we get directly the following general parametric solutions for unknowns $e_0, e_1$ and $f_0, f_1$:

$$e_0 = lk_0 m, \quad f_0 = lk_1 m_1, \quad e_1 = lk_1 m, \quad f_1 = -lk_0 m_1 \quad (43)$$

where $k_0, k_1$, $m_1, m, l$ are arbitrary parameters. In the matrix representation, the general parametric solution (43) has the following form:

$$\begin{bmatrix} e_0 \\ e_1 \end{bmatrix} = lM_eK = lm\begin{bmatrix} 1 & 0 \\ 0 & 1 \end{bmatrix}\begin{bmatrix} k_0 \\ k_1 \end{bmatrix}, \begin{bmatrix} f_0 \\ f_1 \end{bmatrix} = lM_fK = l\begin{bmatrix} 0 & m_1 \\ -m_1 & 0 \end{bmatrix}\begin{bmatrix} k_0 \\ k_1 \end{bmatrix} \quad (43\text{-}1)$$

where $M_e = mI_2$, $K$ is a column parametric matrix and $M_f$ is also a parametric anti-symmetric matrix.

For the system of linear Equation (28) (and, consequently, for its corresponding quadratic homogeneous Equation (21)), the following general parametric solution is obtained directly:

$$e_0 = lk_0m,\ f_0 = l(k_1m_1 - k_2m_2),\ e_1 = lk_1m,\ f_1 = l(k_2m_3 - k_0m_1),\ e_2 = lk_2m,\ f_2 = l(k_0m_2 - k_1m_3) \quad (44)$$

where $k_0, k_1, k_2, m_1, m_2, m_3, m, l$ are arbitrary parameters. In matrix representation the general parametric solution (44) could be also written as follows:

$$\begin{bmatrix} e_0 \\ e_1 \\ e_2 \end{bmatrix} = lM_eK = lm\begin{bmatrix} 1 & 0 & 0 \\ 0 & 1 & 0 \\ 0 & 0 & 1 \end{bmatrix}\begin{bmatrix} k_0 \\ k_1 \\ k_2 \end{bmatrix}, \begin{bmatrix} f_0 \\ f_1 \\ f_2 \end{bmatrix} = lM_fK = l\begin{bmatrix} 0 & m_1 & -m_2 \\ -m_1 & 0 & m_3 \\ m_2 & -m_3 & 0 \end{bmatrix}\begin{bmatrix} k_0 \\ k_1 \\ k_2 \end{bmatrix} \quad (44\text{-}1)$$

where $M_e = mI_3$, $K$ is a column parametric matrix and $M_f$ is also a parametric anti-symmetric matrix.

In addition, it could be simply shown that by adding two particular solutions of the types $\{e_i, f_i\}$ and $\{e'_i, f_i\}$ of homogeneous quadratic Equation (18), the new solution $\{e_i + e'_i, f_i\}$ is also obtained, as follows:

$$\left(\sum_{i=0}^{s} e_i f_i = 0,\ \sum_{i=0}^{s} e'_i f_i = 0\right) \Rightarrow \sum_{i=0}^{s}(e_i f_i + e'_i f_i) = 0 \Rightarrow \sum_{i=0}^{s}(e_i + e'_i) f_i = 0 \quad (44\text{-}2)$$

Using the general basic property (44-2) in addition to the general parametric solution (44) of quadratic Equation (21) (which has been obtained directly from the system of linear Equation (28) corresponding to quadratic Equation (21)), exceptionally, the following equivalent general parametric solution is also obtained for quadratic Equation (21):

$$\begin{gathered} e_0 = l(k_0m - km_3),\ f_0 = l(k_1m_1 - k_2m_2),\ e_1 = l(k_1m - km_2), \\ f_1 = l(k_2m_3 - k_0m_1),\ e_2 = l(k_2m - km_1),\ f_2 = l(k_0m_2 - k_1m_3) \end{gathered} \quad (45)$$

where $k_0, k_1, k_2, k$, $m_1, m_2, m_3, m, l$ are arbitrary parameters.

Moreover, the parametric solution (45) by the direct bijective replacements of six unknown variables $(e_i, f_i)$ (where $i$ = 0, 1, 2) with the six new variables of the type $h_{\mu\nu}$, is given by: $e_0 \to h_{23}$, $e_1 \to h_{20}$, $e_2 \to h_{21}$, $f_0 \to h_{10}$, $f_1 \to h_{31}, f_2 \to h_{03}$, in addition to the replacements of nine arbitrary parameters $u_0, u_1, u_2, u_3$, $v_0, v_1, v_2, v_3, w$, with new nine parameters of the types $u_0, u_1, u_2, u_3$, $v_0, v_1, v_2, v_3, w$, given as: $k_0 \to u_3$, $k_1 \to u_0$, $k_2 \to u_1$, $k \to u_2$, $k \to u_2$, $m_1 \to v_1,\ m_2 \to v_0,\ m_3 \to v_3,\ m \to v_2,\ l \to w$, exceptionally, could be also represented as follows:

$$\begin{aligned} h_{23} &= w(u_3 v_2 - u_2 v_3),\ h_{10} = w(u_0 v_1 - u_1 v_0),\ h_{20} = w(u_0 v_2 - u_2 v_0), \\ h_{31} &= w(u_1 v_3 - u_3 v_1),\ h_{21} = w(u_1 v_2 - u_2 v_1),\ h_{03} = w(u_3 v_0 - u_0 v_3); \end{aligned} \quad (45\text{-}1)$$

where it could be expressed by a single uniform formula as well (for $\mu, \nu = 0,1,2,3$):

$$h_{\mu\nu} = w(u_\nu v_\mu - u_\mu v_\nu) \quad (45\text{-}2)$$

A crucial and important issue concerning the algebraic representation (45-2) (as the differences of products of two parametric variables $u_\mu$ and $v_\nu$) for the general parametric solution (45), is that it generates a symmetric algebra Sym(*V*) on the vector space $V$, where $(u_\mu, v_\nu) \in V$ [37]. This essential property of the form (45-2) could be used for various purposes in the following and also in Section 3, in which we show the applications of this axiomatic linearization-parameterization approach and the results obtained in this Section and Section 2.4, to the foundations of physics.

In addition, as it is also shown in the following, it should be mentioned again that the algebraic form (45-2) (representing the symmetric algebra Sym(*V*)), exceptionally, is determined solely by the parametric solution (44-1) (obtained from the system of Equation (28)) by using the identity (44-2). In fact, from the parametric solutions obtained directly from the subsequent systems of linear equations. i.e., Equations (29), (30) and so on (corresponding to the quadratic Equations (22) and (23), …, and subsequent equations, i.e., $\sum_{i=0}^{s} e_i f_i = 0$ for $s \geq 3$ ), the expanded parametric solutions of the type (45) (equivalent to the algebraic form (45-2)) are not derived.

In the following (also see Ref. [34]), we present the parametric solutions that are obtained directly from the systems of linear Equations (29) and (30) and so on, which also would be the parametric solutions of their corresponding quadratic Equation (18) in various number of unknowns (on the basis of axiom (17)). Meanwhile, the following obtained parametric solutions for the systems of linear Equations (29) and (30) and so on, similar to the parametric solutions (43) and (44), include one parametric term for each of unknowns $e_i$, and the sum of $s$ parametric terms for each of unknowns $f_i$ (where $i = 0,1,2,3,\ldots,s$ ).

Hence, the following parametric solution is derived directly from the system of linear Equation (29) (that would be also the solution of its corresponding quadratic Equation (22)):

$$\begin{aligned} e_0 &= lk_0 m,\ f_0 = l(k_1 m_1 + k_2 m_2 + k_3 m_3),\ e_1 = lk_1 m,\ f_1 = l(-k_0 m_1 + k_3 m_6 - k_2 m_7), \\ e_2 &= lk_2 m,\ f_2 = l(-k_0 m_2 - k_3 m_5 + k_1 m_7),\ e_3 = lk_3 m,\ f_3 = l(-k_0 m_3 + k_2 m_5 - k_1 m_6). \end{aligned} \quad (46)$$

where $k_0, k_1, k_2, k_3, l$ are arbitrary parameters. In the matrix representation, the parametric solution (46) is represented as follows:

$$\begin{bmatrix} e_0 \\ e_1 \\ e_2 \\ e_3 \end{bmatrix} = lM_e K = lm \begin{bmatrix} 1 & 0 & 0 & 0 \\ 0 & 1 & 0 & 0 \\ 0 & 0 & 1 & 0 \\ 0 & 0 & 0 & 1 \end{bmatrix} \begin{bmatrix} k_0 \\ k_1 \\ k_2 \\ k_3 \end{bmatrix},\ \begin{bmatrix} f_0 \\ f_1 \\ f_2 \\ f_3 \end{bmatrix} = lM_f K = l \begin{bmatrix} o & m_1 & m_2 & m_3 \\ -m_1 & 0 & -m_7 & m_6 \\ -m_2 & m_7 & 0 & -m_5 \\ -m_3 & -m_6 & m_5 & 0 \end{bmatrix} \begin{bmatrix} k_0 \\ k_1 \\ k_2 \\ k_3 \end{bmatrix} \quad (46\text{-}1)$$

where $M_e = mI_4$, $K$ is a column parametric matrix and $M_f$ is also a parametric anti-symmetric matrix.

However, in solutions (46) or (46-1) the parameters $m_1, m_2, m_3, m_4, m_5, m_6, m_7, m$ are not arbitrary and in fact, in the course of obtaining the solution (46) from the system of linear Equation (29), a condition appears for these parameters as follows:

$$m_4 m + m_1 m_5 + m_2 m_6 + m_3 m_7 = 0 \tag{47}$$

The condition (47) is also a homogeneous quadratic equation that should be solved first, in order to obtain a general parametric solution for the system of linear Equation (29). Since the parameter $m_4$ has not appeared in the solution (46), it could be assumed that $m_4 = 0$, and the condition (47) is reduced to the following homogeneous quadratic equation, which is equivalent to the quadratic Equation (20) (corresponding to the system of linear Equation (28)):

$$m_4 = 0, \; m_1 m_5 + m_2 m_6 + m_3 m_7 = 0 \tag{47-1}$$

where the parameter $m$ would be arbitrary. The condition (47-1) is equivalent to the quadratic Equation (21). Hence by using the general parametric solution (45-1) (as the most symmetric solution obtained for quadratic Equation (21) by solving its corresponding system of linear Equation (28)), the following general parametric solution for the condition (47-1) is obtained:

$$\begin{aligned} &m_1 = w(u_0 v_1 - u_1 v_0), \; m_2 = w(u_0 v_2 - u_2 v_0), \; m_3 = w(u_0 v_3 - u_3 v_0), \\ &m_5 = w(u_3 v_2 - u_2 v_3), \; m_6 = w(u_1 v_3 - u_3 v_1), \; m_7 = w(u_2 v_1 - u_1 v_2) \end{aligned} \tag{48}$$

where $u_0, u_1, u_2, u_3, \; v_0, v_1, v_2, v_3, w, \; m$ are arbitrary parameters. Now by replacing the solutions (48) (obtained for $m_1, m_2, m_3, m_5, m_6, m_7$ in terms of the new parameters $u_0, u_1, u_2, u_3$, $v_0, v_1, v_2, v_3, w$) in the relations (46), the general parametric solution of the system of linear Equation (29) (and its corresponding quadratic Equation (22)) is obtained in terms of the arbitrary parameters $k_0, k_1, k_2, k_3, \; u_0, u_1, u_2, u_3, \; v_0, v_1, v_2, v_3, w, \; m, l$.

For the system of linear Equation (30) and its corresponding quadratic Equation (23), the following parametric solution is obtained:

$$\begin{aligned} &e_0 = lk_0 m, \; f_0 = l(k_1 m_1 - k_2 m_2 + k_3 m_3 - k_4 m_5), \; e_1 = lk_1 m, \; f_1 = l(-k_0 m_1 + k_4 m_{12} + k_3 m_{14} + k_2 m_{15}), \\ &e_2 = lk_2 m, \; f_2 = l(k_0 m_2 + k_4 m_{11} + k_3 m_{13} - k_1 m_{15}), \; e_3 = lk_3 m, \; f_3 = l(-k_0 m_3 + k_4 m_{10} - k_2 m_{13} - k_1 m_{14}), \\ &e_4 = lk_4 m, \; f_4 = l(k_0 m_5 - k_3 m_{10} - k_2 m_{11} - k_1 m_{12}) \end{aligned} \tag{49}$$

where $k_0, k_1, k_2, k_3, k_4, l$ are arbitrary parameters. In the matrix representation, the solution (49) can be also written as follows:

$$\begin{bmatrix} e_0 \\ e_1 \\ e_2 \\ e_3 \\ e_4 \end{bmatrix} = lM_e K = lm \begin{bmatrix} 1 & 0 & 0 & 0 & 0 \\ 0 & 1 & 0 & 0 & 0 \\ 0 & 0 & 1 & 0 & 0 \\ 0 & 0 & 0 & 1 & 0 \\ 0 & 0 & 0 & 0 & 1 \end{bmatrix} \begin{bmatrix} k_0 \\ k_1 \\ k_2 \\ k_3 \\ k_4 \end{bmatrix}, \; \begin{bmatrix} f_0 \\ f_1 \\ f_2 \\ f_3 \\ f_4 \end{bmatrix} = lM_f K = l \begin{bmatrix} 0 & m_1 & -m_2 & m_3 & -m_5 \\ -m_1 & 0 & m_{15} & m_{14} & m_{12} \\ m_2 & -m_{15} & 0 & m_{13} & m_{11} \\ -m_3 & -m_{14} & -m_{13} & 0 & m_{10} \\ m_5 & -m_{12} & -m_{11} & -m_{10} & 0 \end{bmatrix} \begin{bmatrix} k_0 \\ k_1 \\ k_2 \\ k_3 \\ k_4 \end{bmatrix} \tag{49-1}$$

where $M_e = mI_5$, $K$ is a column parametric matrix and $M_f$ is also a parametric anti-symmetric matrix.

However, similar to the system of Equation (29), in the course of obtaining the solutions (49) or (49–1) from the system of linear Equation (30), the following conditions appear for parameters $m_1, m_2, m_3, m_5, m_{10}, m_{11}, m_{12}, m_{13}, m_{14}, m_{15}$:

$$\begin{aligned} &m_4 m = -m_1 m_{13} - m_2 m_{14} - m_3 m_{15}, \\ &m_6 m = m_1 m_{11} + m_2 m_{12} - m_5 m_{15}, \\ &m_7 m = m_1 m_{10} - m_3 m_{12} - m_5 m_{14}, \\ &m_8 m = m_2 m_{10} + m_3 m_{11} + m_5 m_{13}, \\ &m_9 m = m_{10} m_{15} - m_{11} m_{14} + m_{12} m_{13}. \end{aligned} \tag{50}$$

that are similar to condition (47). Here also by the same approach, since the parameters $m_4, m_6, m_7, m_8, m_9$ have not appeared in the solution (49), it can be assumed that $m_4 = m_6 = m_7 = m_8 = m_9 = 0$, and the set of conditions (50) are reduced to the following system of homogeneous quadratic equations which are similar to the quadratic Equation (20) (corresponding to the system of linear Equation (28)):

$$\begin{aligned} &m_4 = m_6 = m_7 = m_8 = m_9 = 0, \\ &m_1 m_{10} - m_3 m_{12} - m_5 m_{14} = 0, \\ &m_1 m_{11} + m_2 m_{12} - m_5 m_{15} = 0, \\ &m_1 m_{13} + m_2 m_{14} + m_3 m_{15} = 0, \\ &m_2 m_{10} + m_5 m_{13} + m_3 m_{11} = 0, \\ &m_{10} m_{15} + m_{12} m_{13} - m_{11} m_{14} = 0; \end{aligned} \tag{50-1}$$

The conditions (50-1) are also similar to the quadratic Equation (21). Hence using again the general parametric solution (45-1), the following general parametric solutions for the system of homogeneous quadratic Equation (50-1) are obtained directly:

$$\begin{aligned} &m_1 = w(u_0 v_1 - u_1 v_0),\ m_2 = w(u_2 v_0 - u_0 v_2), \\ &m_3 = w(u_0 v_3 - u_3 v_0),\ m_4 = 0, \\ &m_5 = w(u_4 v_0 - u_0 v_4),\ m_6 = 0, \\ &m_7 = 0, m_8 = 0, m_9 = 0, \\ &m_{10} = w(u_3 v_4 - u_4 v_3),\ m_{11} = w(u_2 v_4 - u_4 v_2), \\ &m_{12} = w(u_1 v_4 - u_4 v_1),\ m_{13} = w(u_2 v_3 - u_3 v_2), \\ &m_{14} = w(u_1 v_3 - u_3 v_1),\ m_{15} = w(u_1 v_2 - u_2 v_1). \end{aligned} \tag{50-2}$$

where $u_0, u_1, u_2, u_3, u_4, v_0, v_1, v_2, v_3, v_4$ and $w$ are arbitrary parameters. Now by replacing the solution (51) (that has been obtained for $m_1, m_2, m_3, m_5, m_{10}, m_{11}, m_{12}, m_{13}, m_{14}, m_{15}$ in terms of the new parameters $u_0, u_1, u_2, u_3, u_4, v_0, v_1, v_2, v_3, v_4, w$) in the relations (49), the general parametric solution of the system of linear Equation (30) (and its corresponding quadratic Equation (23)) is obtained in terms of the arbitrary parameters $k_0, k_1, k_2, k_3, k_4, u_0, u_1, u_2, u_3, u_4, v_0, v_1, v_2, v_3, v_4, w,$ $m, l$.

Meanwhile, similar to relations (48) and (51), it should be noted that arbitrary parameter $m_1$ in the general parametric solution (43) and the arbitrary parameters $m_1, m_2, m_3$ in the general parametric solution (44) (which have been obtained as the solutions of quadratic Equations (20) and (21), respectively, by solving their equivalent systems of linear Equations (26) and (28)), by keeping their arbitrariness property, could be expressed in terms of new arbitrary parameters $u_0, u_1, v_0, v_1$ and $u_0, u_1, u_2, v_0, v_1, v_2$, as follows, respectively:

$$m_1 = w(u_0 v_1 - u_1 v_0); \tag{43-2}$$

$$m_1 = w(u_0 v_1 - u_1 v_0),\ m_2 = w(u_0 v_2 - u_2 v_0),\ m_3 = w(u_0 v_3 - u_3 v_0). \tag{44-3}$$

In fact, as a particular common algebraic property of both parametric relations (43-2) and (44-2), it could be shown directly that by choosing appropriate integer values for parameters $u_0, u_1, v_0, v_1, w$ in the relation (43-2), the parameter $m_1$ (defined in terms of arbitrary parameters $u_0, u_1, v_0, v_1, w)$ could take any given integer value, and similarly, by choosing appropriate integer values for parameters $u_0, u_1, u_2, v_0, v_1, v_2, w$ in the relation (43-2), the parameters $m_1, m_2, m_3$ (defined in terms of arbitrary parameters $u_0, u_1, u_2, v_0, v_1, v_2, w)$ could also take any given integer values.

Therefore, using this common algebraic property of the parametric relations (43-2) and (44-2), the arbitrary parameter $m_1$ in general parametric solutions (43), and arbitrary parameters $m_1, m_2, m_3$ in general parametric solutions (44), could be equivalently replaced by new arbitrary parameters $u_0, u_1, v_0, v_1, w$ and $u_0, u_1, u_2, v_0, v_1, v_2, w$, respectively. In addition, for the general quadratic homogeneous Equation (18) with a larger number of unknowns, the general parametric solutions could be obtained by the same approaches used above for quadratic Equations (20)–(23), i.e., by solving their corresponding systems of linear equations (defined on the basis of axiom (17)). Moreover, using the isomorphic transformations (18-3) and the above general parametric solutions obtained for quadratic Equations (20)–(23), …, (via solving their corresponding systems of linear Equations (26), (28)–(30), …), the general parametric solutions of quadratic equations of the regular type (18-2) (in various number of unknown) are also obtained straightforwardly. All the parametric solutions that are obtained by this new systematic matrix approach for the homogeneous quadratic equations and also higher degree homogeneous equations of the type $F(x_1, x_2, x_3, \ldots, x_s) = 0$ (defined in the axiom (17)), are fully compatible with the solutions and conclusions that have been obtained previously for various homogeneous equations by different and miscellaneous methods and approaches [16,35,36]. In Section 3, we have used the uniquely specified systems of homogeneous linear equations (and also their general parametric solutions) corresponding with the homogeneous quadratic equations. It has been assumed that the components of the relativistic energy-momentum vector (as one of the most basic physical quantities) in the Lorentz invariant energy-momentum (homogeneous) quadratic relation, can only take the rational values.

## 3. A Unique Mathematical Derivation of the Laws Governing the Fundamental Forces of Nature: Based on a New Algebraic-Axiomatic (Matrix) Approach

In this Section, as a new mathematical approach to the origin of the laws of nature, using the new basic algebraic axiomatic (matrix) formalism as presented in Section 2 "it is shown that certain mathematical forms of fundamental laws of nature, including laws governing the fundamental forces of nature (represented by a set of two definite classes of general covariant massive field equations, with new matrix formalisms), are derived uniquely from only a very few axioms" where in agreement with the rational Lorentz group, it is also basically assumed that the components of relativistic energy-momentum can only take rational values. Concerning the basic assumption of the rationality of relativistic energy-momentum, it is necessary to note that the rational Lorentz symmetry group is not only dense in the general form of Lorentz group, but also is compatible with the necessary conditions required basically for the formalism of a consistent relativistic quantum theory [15]. In essence, the main scheme of this new mathematical axiomatic approach to fundamental laws of nature is as follows. First in Section 3.1, based on the assumption of rationality of *D*-momentum, by linearization (along with a parameterization procedure) of the Lorentz invariant energy-momentum quadratic relation, a unique set of Lorentz invariant systems of homogeneous linear equations (with matrix formalisms compatible with certain Clifford, and symmetric algebras) is derived.

Then in Section 3.4, by initial quantization (followed by a basic procedure of minimal coupling to space-time geometry) of these determined systems of linear equations, a set of two classes of general covariant massive (tensor) field equations (with matrix formalisms compatible with certain Clifford, and Weyl algebras) is derived. Each class of the derived general covariant field equations also includes a definite form of torsion field appeared as generator of the corresponding field' invariant mass. In addition, in Sections 3.4 and 3.11, it is shown that the (1 + 3)-dimensional cases of two classes of derived field equations represent a new general covariant massive formalism of bispinor fields of spin-2, and spin-1 particles, respectively. In fact, these uniquely determined bispinor fields represent a unique set of new generalized massive forms of the laws governing the fundamental forces of nature, including the Einstein (gravitational), Maxwell (electromagnetic) and Yang-Mills (nuclear) field equations. Moreover, it is also shown that the (1 + 2)-dimensional cases of two classes of these field equations represent (asymptotically) a new general covariant massive formalism of bispinor fields of spin-3/2 and spin-1/2 particles, respectively, corresponding with the Dirac and Rarita–Schwinger equations.

As a particular consequence, in Section 3.6, it is shown that a certain massive formalism of general relativity—with a definite form of torsion field appearing originally as the generator of the gravitational field's invariant mass—is obtained only by the initial quantization (followed by a basic procedure of minimal coupling to space-time geometry) of a certain set of special relativistic algebraic matrix equations. In Section 3.9, it has been proven that Lagrangian densities specified for the originally derived new massive forms of the Maxwell, Yang-Mills and Dirac field equations, are also gauge-invariant, where the invariant mass of each field is generated solely by the corresponding torsion field. In addition, in Section 3.10, in agreement with recent astronomical data, a particular new form of massive boson is identified (corresponding to U(1) gauge group) with invariant mass: $m_\gamma \approx 4.90571 \times 10^{-50}$ kg, generated by a coupled torsion field of the background space-time geometry.

Moreover, in Section 3.12, based on the definite mathematical formalism of this axiomatic approach, along with the C, P and T symmetries (represented basically by the corresponding quantum operators) of the fundamentally derived field equations, it has been concluded that the universe could be realized solely with the (1 + 2) and (1 + 3)-dimensional space-times (where this conclusion, in particular, is based on the time-reversal symmetry). In Sections 3.13 and 3.14, it is proved that 'CPT' is the only (unique) combination of C, P, and T symmetries that could be defined as a symmetry for interacting fields. In addition, in Section 3.14, on the basis of these discrete symmetries of derived field equations, it has been also shown that only left-handed particle fields (along with their complementary right-handed fields) could be coupled to the corresponding (any) source currents. Furthermore, in Section 3.15, it has been shown that metric of the background space-time is diagonalized for the uniquely derived fermion field equations (defined and expressed solely in (1 + 2)-dimensional space-time), where this property generates a certain set of additional symmetries corresponding uniquely to the $SU(2)_L \otimes U(2)_R$ symmetry group for spin-1/2 fermion fields (representing "1 + 3" generations of four fermions, including a group of eight leptons and a group of eight quarks), and also the $SU(2)_L \otimes U(2)_R$ and SU(3) gauge symmetry groups for spin-1 boson fields coupled to the spin-1/2 fermionic source currents. Hence, along with the known elementary particles, eight new elementary particles, including: four new charge-less right-handed spin-1/2 fermions (two leptons and two quarks, represented by "$z_e$, $z_n$ and $z_u$, $z_d$"), a spin-3/2 fermion, and also three new spin-1 massive bosons (represented by $\tilde{W}^+, \tilde{W}^-, \vec{Z}$, where in particular, the new boson $\vec{Z}$ is complementary right-handed particle of ordinary $Z$ boson), are uniquely predicted by this new mathematical axiomatic approach.

Furthermore, as a particular result, it is generally concluded in Section 3.6—based on the definite and unique formulation of the derived Maxwell's equations (and also determined Yang-Mills equations, represented uniquely with two specific forms of gauge symmetries, in Section 3.15, formulas (114-4)–(114-9)— that magnetic monopoles cannot exist in nature.

### *3.1. The Main Results Obtained in This Article are Based on the Following Three Postulates*

As noted in Section 1.1, the main results obtained in this article are based on the following three basic assumptions (as postulates):

**(1)-** "A new definite axiomatic generalization of the axiom of "no zero divisors" of integral domains (including the integr ring $\mathbb{Z}$) is assumed (represented by formula (17), in Section 2.1);"

This basic assumption (as a postulate) is a new mathematical concept. In Section 2.1, based on this new axiom, a general algebraic axiomatic (matrix) approach (in the form of a basic linearization-parameterization theory) to homogeneous equations of degree $r \geq 2$ (over the integer domain, extendable to field of rational numbers), has been formulated. A summary of the main results obtained from this axiomatic approach have been presented in Section 1.1. As particular outcome of this new mathematical axiomatic formalism (based on the axiomatic relations (17) and (17-1), including their basic algebraic properties presented in detail, in Sections 2.1–2.4), in Section 3.4, it is shown that using, a unique set of general covariant massive (tensor) field equations (with new matrix formalism compatible with Clifford, and Weyl algebras), corresponding with the fundamental field equations of physics, are derived—where, in agreement with the rational Lorentz symmetry group, it has been assumed that the components of relativistic energy-momentum can only take the rational values. In Sections 3.2–3.15, we present in detail the main applications of this basic algebraic assumption (along with the following basic assumptions (2) and (3)) to fundamental physics.

**(2)-** "In agreement with the rational Lorentz symmetry group, we assume basically that the components of relativistic energy-momentum (*D*-momentum) can only take the rational values;"

Concerning this assumption, it is necessary to note that the rational Lorentz symmetry group is not only dense in the general form of Lorentz group, but also is compatible with the necessary conditions required basically for the formalism of a consistent relativistic quantum theory [15]. Moreover, this assumption is clearly also compatible with any quantum circumstance in which the energy-momentum of a relativistic particle is transferred as integer multiples of the quantum of action "*h*" (Planck constant).

Before defining the next basic assumption, it should be noted that from basic assumptions (1) and (2), it follows directly that the Lorentz invariant energy-momentum quadratic relation (represented by formula (52)) is a particular form of homogeneous quadratic equation (18-2). Hence, using the set of systems of linear equations that have been determined uniquely as equivalent algebraic representations of the corresponding set of quadratic homogeneous equations (given by equation (18-2) in various number of unknown variables, respectively), a unique set of the Lorentz invariant systems of homogeneous linear equations (with matrix formalisms compatible with certain Clifford, and symmetric algebras) are also determined, representing equivalent algebraic forms of the energy-momentum quadratic relation in various space-time dimensions, respectively. Subsequently, we have shown that by first quantization (followed by a basic procedure of minimal coupling to space-time geometry) of these determined systems of linear equations, a unique set of two definite classes of general covariant massive (tensor) field equations (with matrix formalisms compatible with certain Clifford, and Weyl algebras) is also derived, corresponding to various space-time dimensions, respectively. In addition, it is also shown that this derived set of two classes of general covariant field equations represent new tensor massive (matrix) formalism of the fundamental field equations of physics, corresponding to fundamental laws of nature (including the laws governing the fundamental forces of nature). Following these essential results, in addition to the basic assumptions (1) and (2), it would be also basically assumed that:

**(3)-** "We assume that the mathematical formalism of the fundamental laws of nature, are defined solely by the axiomatic matrix constitution formulated uniquely on the basis of postulates (1) and (2)".

In addition to this basic assumption, in Section 3.11, the C, P and T symmetries of uniquely derived general covariant field equations (that are Equations (71) and (72), in Section 3.4), are also represented basically by their corresponding quantum matrix operators.

As the next step, in the following, based on the basic assumption (2), i.e., the assumption of rationality of the relativistic energy-momentum, the following Lorentz invariant quadratic relations (expressed in terms of the components of *D*-momentums $p_\mu$, $p'_\mu$ of a relativistic massive particle (given bytwo reference frames), and also components of quantity $p^{st}_\mu = m_0 k_\mu$, where $m_0$ is the invariant mass of particle and $k_\mu$ is its covariant velocity in the stationary reference frame):

$$g^{\mu\nu} p_\mu p_\nu = g^{\mu\nu} p'_\mu p'_\nu, \tag{51}$$

$$g^{\mu\nu} p_\mu p_\nu = g^{\mu\nu} p^{st}_\mu p^{st}_\nu = g^{\mu\nu}(m_0 k_\mu)(m_0 k_\nu) = g^{00}(m_0 k_0)^2 = (m_0 c)^2. \tag{52}$$

would be particular cases of homogeneous quadratic Equation (18-2) in Section 2.2, and hence, they would be necessarily subject to the process of linearization (along with a parameterization procedure) using the systematic axiomatic approach presented Sections 2, 2.2 and 2.4 (formulated based on the basic assumption (1)).

The Lorentz invariant relations (51) and (52) (as the norm of the relativistic energy-momentum) have been defined in the *D*-dimensional space-time, where $m_0$ is the invariant mass of the particle, $p_\mu$ and $p'_\mu$ are its relativistic energy-momentums (i.e., *D*-momentums) given respectively in two reference frames, $k_\mu$ is a time-like covariant vector given by: $k_\mu = (k_0, 0, \ldots, 0) = (c/\sqrt{g^{00}}, 0, \ldots, 0)$, "*c*" is the speed of light, and the components of metric have constant values. As noted in Section 1.2, in this article the sign conventions (2) (including the metric signature (+ − − … − )) and geometric units would be used (where in particular "*c* = 1"). However, for the clarity, in some of relativistic formulas (such as the relativistic matrix relations), the speed of light "*c*" is indicated formally.

As a crucial issue here, it should be noted that in the invariant quadratic relations (51) and (52), the components of metric which have the constant values (as assumed), have necessarily been written by their general representations $g^{\mu\nu}$ (and not by the Minkowski metric $\eta^{\mu\nu}$, and so on). This follows from the fact that by axiomatic approach of linearization-parameterization (presented in Sections 2.1–2.4) of quadratic relations (51) and (52) (as particular forms of homogeneous quadratic Equation (18-2) which could be expressed equivalently by quadratic equations of the types (18) via the linear transformations (18-3)), their corresponding algebraic equivalent systems of linear equations could be determined uniquely. In fact, based on the formulations of systems of linear equations obtained uniquely for the quadratic Equation (18) in Sections 2.2–2.4, it is concluded directly that the algebraic equivalent systems of linear equations corresponding to the relations (51) and (52), are determined uniquely if and *only* if these quadratic relations be expressed in terms of the components $g^{\mu\nu}$ represented by their general forms (and not in terms of any special presentation of the metric's components, such as the Minkowski metric, and so on). However, after the derivation of corresponding systems of linear equations (representing uniquely the equivalent algebraic matrix forms of the quadratic relations (51) and (52) in various space-time dimensions), the Minkowski metric could be used in these equations (and the subsequent relativistic equations and relations as well).

Hence, using the systems of linear Equations (24), (26), (28)–(30), …, obtained uniquely on the basis of the axiom (17) by linearization (along with a parameterization procedure) of the homogeneous quadratic Equations (19)–(23), … (which could be transformed directly to the general quadratic Equation (18-2), by the isomorphic linear transformations (18-3)) and also using the parametric relations (43-2), (44-3), (48) and (52) (expressed in terms of the arbitrary parameters $u_\mu$ and $v_\mu$), as the result of linearization (along with a parameterization procedure) of the invariant quadratic relations (51) and (52), the following systems of linear equations are also derived uniquely corresponding with various space-time dimensions, respectively:

For (1 + 0)-dimensional case of the invariant relation (51), we obtain:

$$[g^{0\nu}(p_\nu + p'_\nu)][s] = 0 \tag{53}$$

where $\nu = 0$ and parameter $s$ is arbitrary;

For (1 + 1)-dimensional space-time we have:

$$\begin{bmatrix} g^{0\nu}(p_\nu + p'_\nu) & p_1 - p'_1 \\ g^{1\nu}(p_\nu + p'_\nu) & -(p_0 - p'_0) \end{bmatrix} \begin{bmatrix} (u_0 v_1 - u_1 v_0)w \\ s \end{bmatrix} = 0 \tag{54}$$

where $\nu = 0,1$ and $u_0, u_1, v_0, v_1, w, s$ are arbitrary parameters;

For (1 + 2)-dimensional space-time we have:

$$\begin{bmatrix} g^{0\nu}(p_\nu + p'_\nu) & 0 & -g^{2\nu}(p_\nu + p'_\nu) & p_1 - p'_1 \\ 0 & g^{0\nu}(p_\nu + p'_\nu) & -g^{1\nu}(p_\nu + p'_\nu) & -(p_2 - p'_2) \\ -(p_2 - p'_2) & -(p_1 - p'_1) & -(p_0 - p'_0) & 0 \\ g^{1\nu}(p_\nu + p'_\nu) & -g^{2\nu}(p_\nu + p'_\nu) & 0 & -(p_0 - p'_0) \end{bmatrix} \begin{bmatrix} (u_0 v_1 - u_1 v_0)w \\ (u_2 v_0 - u_0 v_2)w \\ (u_1 v_2 - u_2 v_1)w \\ s \end{bmatrix} = 0 \tag{55}$$

where $\nu = 0, 1, 2$ and $u_0, u_1, u_2,\ v_0, v_1, v_2, w;\ s$ are arbitrary parameters;

For (1 + 3) -dimensional space-time we obtain:

$$\begin{bmatrix} e_0 & 0 & 0 & 0 & 0 & -e_3 & e_2 & f_1 \\ 0 & e_0 & 0 & 0 & e_3 & 0 & -e_1 & f_2 \\ 0 & 0 & e_0 & 0 & -e_2 & e_1 & 0 & f_3 \\ 0 & 0 & 0 & e_0 & -f_1 & -f_2 & -f_3 & 0 \\ 0 & f_3 & -f_2 & -e_1 & -f_0 & 0 & 0 & 0 \\ -f_3 & 0 & f_1 & -e_2 & 0 & -f_0 & 0 & 0 \\ f_2 & -f_1 & 0 & -e_3 & 0 & 0 & -f_0 & 0 \\ e_1 & e_2 & e_3 & 0 & 0 & 0 & 0 & -f_0 \end{bmatrix} \begin{bmatrix} (u_0 v_1 - u_1 v_0)w \\ (u_0 v_2 - u_2 v_0)w \\ (u_0 v_3 - u_3 v_0)w \\ 0 \\ (u_3 v_2 - u_2 v_3)w \\ (u_1 v_3 - u_3 v_1)w \\ (u_2 v_1 - u_1 v_2)w \\ s \end{bmatrix} = 0 \tag{56}$$

where $\nu = 0, 1, 2, 3$ and $w, u_0, u_1, u_2, u_3,\ v_0, v_1, v_2, v_3,\ s$ are arbitrary parameters, and we also have:

$$\begin{aligned} e_0 &= g^{0\nu}(p_\nu + p'_\nu),\ f_0 = -(p_0 - p'_0)_0, \\ e_1 &= g^{1\nu}(p_\nu + p'_\nu),\ f_1 = -(p_1 - p'_1), \\ e_2 &= g^{2\nu}(p_\nu + p'_\nu),\ f_2 = -(p_2 - p'_2), \\ e_3 &= g^{3\nu}(p_\nu + p'_\nu),\ f_3 = -(p_3 - p'_3)\ ; \end{aligned} \tag{56-1}$$

For (1 + 4)-dimensional case, the system of linear equations corresponding to the invariant quadratic relation (51) is specified as follows:

$$\begin{bmatrix} e_0 & 0 & 0 & 0 & 0 & 0 & 0 & 0 & 0 & 0 & 0 & -e_4 & 0 & -e_3 & -e_2 & f_1 \\ 0 & e_0 & 0 & 0 & 0 & 0 & 0 & 0 & 0 & 0 & e_4 & 0 & e_3 & 0 & -e_1 & -f_2 \\ 0 & 0 & e_0 & 0 & 0 & 0 & 0 & 0 & 0 & -e_4 & 0 & 0 & e_2 & e_1 & 0 & f_3 \\ 0 & 0 & 0 & e_0 & 0 & 0 & 0 & 0 & e_4 & 0 & 0 & 0 & -f_1 & f_2 & -f_3 & 0 \\ 0 & 0 & 0 & 0 & e_0 & 0 & 0 & 0 & 0 & -e_3 & -e_2 & -e_1 & 0 & 0 & 0 & -f_4 \\ 0 & 0 & 0 & 0 & 0 & e_0 & 0 & 0 & e_3 & 0 & f_1 & -f_2 & 0 & 0 & f_4 & 0 \\ 0 & 0 & 0 & 0 & 0 & 0 & e_0 & 0 & e_2 & -f_1 & 0 & f_3 & 0 & -f_4 & 0 & 0 \\ 0 & 0 & 0 & 0 & 0 & 0 & 0 & e_0 & e_1 & f_2 & -f_3 & 0 & f_4 & 0 & 0 & 0 \\ 0 & 0 & 0 & f_4 & 0 & f_3 & f_2 & f_1 & -f_0 & 0 & 0 & 0 & 0 & 0 & 0 & 0 \\ 0 & 0 & -f_4 & 0 & -f_3 & 0 & -e_1 & e_2 & 0 & -f_0 & 0 & 0 & 0 & 0 & 0 & 0 \\ 0 & f_4 & 0 & 0 & -f_2 & e_1 & 0 & -e_3 & 0 & 0 & -f_0 & 0 & 0 & 0 & 0 & 0 \\ -f_4 & 0 & 0 & 0 & -f_1 & e_2 & e_3 & 0 & 0 & 0 & 0 & -f_0 & 0 & 0 & 0 & 0 \\ 0 & f_3 & f_2 & -e_1 & 0 & 0 & 0 & e_4 & 0 & 0 & 0 & 0 & -f_0 & 0 & 0 & 0 \\ -f_3 & 0 & f_1 & e_2 & 0 & 0 & -e_4 & 0 & 0 & 0 & 0 & 0 & 0 & -f_0 & 0 & 0 \\ -f_2 & -f_1 & 0 & -e_3 & 0 & e_4 & 0 & 0 & 0 & 0 & 0 & 0 & 0 & 0 & -f_0 & 0 \\ e_1 & -e_2 & e_3 & 0 & -e_4 & 0 & 0 & 0 & 0 & 0 & 0 & 0 & 0 & 0 & 0 & -f_0 \end{bmatrix} \begin{bmatrix} (u_0v_1-u_1v_0)w \\ (u_2v_0-u_0v_2)w \\ (u_0v_3-u_3v_0)w \\ 0 \\ (u_4v_0-u_0v_4)w \\ 0 \\ 0 \\ 0 \\ 0 \\ (u_3v_4-u_4v_3)w \\ (u_2v_4-u_4v_2)w \\ (u_1v_4-u_4v_1)w \\ (u_2v_3-u_3v_2)w \\ (u_1v_3-u_3v_1)w \\ (u_1v_2-u_2v_1)w \\ s \end{bmatrix} = 0 \quad (57)$$

where $\nu = 0, 1, 2, 3, 4$, $u_0, u_1, u_2, u_3, u_4$, $v_0, v_1, v_2, v_3, v_4, w$, $s$ are arbitrary parameters, and we have:

$$\begin{aligned} e_0 &= g^{0\nu}(p_\nu + p'_\nu), \; f_0 = p_0 - p'_0, \\ e_1 &= g^{1\nu}(p_\nu + p'_\nu), \; f_1 = p_1 - p'_1, \\ e_2 &= g^{2\nu}(p_\nu + p'_\nu), \; f_2 = p_2 - p'_2, \\ e_3 &= g^{3\nu}(p_\nu + p'_\nu), \; f_3 = p_3 - p'_3, \\ e_4 &= g^{4\nu}(p_\nu + p'_\nu), \; f_4 = p_4 - p'_4; \end{aligned} \quad (57\text{-}1)$$

The systems of linear equations that are obtained for (1 + 5) and higher dimensional cases of the invariant quadratic relation (51), have also the formulations similar to the obtained systems of linear Equations (53)–(57), and would be expressed by the matrix product of a $2^N \times 2^N$ square matrix and a $2^N \times 1$ column matrix in (1 + *N*)-dimensional space-time. For (1 + 5)-dimensional case of the invariant relation (51), the column matrix of the corresponding system of linear equations (expressed by the matrix product of a $32 \times 32$ square matrix and a $32 \times 1$ column matrix) are given by (where $u_0, u_1, u_2, u_3, u_4, u_5$, $v_0, v_1, v_2, v_3, v_4, v_5, w, s$ are arbitrary parameters):

$$S=\begin{bmatrix}S'\\S''\end{bmatrix},\quad S'=\begin{bmatrix}(u_0v_1-u_1v_0)w\\(u_0v_2-u_2v_0)w\\(u_0v_3-u_3v_0)w\\0\\(u_0v_4-u_4v_0)w\\0\\0\\0\\(u_0v_5-u_5v_0)w\\0\\0\\0\\0\\0\\0\\0\end{bmatrix},\quad S''=\begin{bmatrix}0\\0\\0\\(u_5v_4-u_4v_5)w\\0\\(u_3v_5-u_5v_3)w\\(u_5v_2-u_2v_5)w\\(u_1v_5-u_5v_1)w\\0\\(u_4v_3-u_3v_4)w\\(u_2v_4-u_4v_2)w\\(u_4v_1-u_1v_4)w\\(u_2v_3-u_3v_2)w\\(u_3v_1-u_1v_3)w\\(u_1v_2-u_2v_1)w\\s\end{bmatrix}. \tag{57-2}$$

In a similar manner, using the axiomatic approach presented in Section 2, the systems of linear equations corresponding to the energy-momentum invariant relation (52) in various space-time dimensions are obtained uniquely as follows, respectively (note that by using the geometric units, we would take $c=1$):

For (1 + 0)-dimensional space-time we obtain:

$$\begin{bmatrix}g^{0\nu}p_\nu-g^{00}\dfrac{m_0c}{\sqrt{g^{00}}}\end{bmatrix}[s]=0 \tag{58}$$

where $\nu = 0$ and parameter $s$ is arbitrary;

For (1 + 1)-dimensional space-time we have:

$$\begin{bmatrix}g^{0\nu}p_\nu-g^{00}(\dfrac{m_0c}{\sqrt{g^{00}}}) & p_1\\ g^{1\nu}p_\nu+g^{10}(\dfrac{m_0c}{\sqrt{g^{00}}}) & -(p_0+(\dfrac{m_0c}{\sqrt{g^{00}}}))\end{bmatrix}\begin{bmatrix}(u_0v_1-u_1v_0)w\\s\end{bmatrix}=0 \tag{59}$$

where $\nu = 0, 1$ and $u_0, u_1, v_0, v_1, w, s$ are arbitrary parameters;

For (1 + 2) dimensions we have (where $\nu = 0, 1, 2$ and $u_0, u_1, u_2, v_0, v_1, v_2, w, s$ are arbitrary parameters):

$$\begin{bmatrix} g^{0\nu}p_\nu - g^{00}(\frac{m_0c}{\sqrt{g^{00}}}) & 0 & -g^{2\nu}p_\nu + g^{20}(\frac{m_0c}{\sqrt{g^{00}}}) & p_1 \\ 0 & g^{0\nu}p_\nu - g^{00}(\frac{m_0c}{\sqrt{g^{00}}}) & -g^{1\nu}p_\nu + g^{10}(\frac{m_0c}{\sqrt{g^{00}}}) & -p_2 \\ -p_2 & -p_1 & -(p_0 + (\frac{m_0c}{\sqrt{g^{00}}})) & 0 \\ g^{1\nu}p_\nu + g^{10}(\frac{m_0c}{\sqrt{g^{00}}}) & -g^{2\nu}p_\nu - g^{20}(\frac{m_0c}{\sqrt{g^{00}}}) & 0 & -(p_0 + (\frac{m_0c}{\sqrt{g^{00}}})) \end{bmatrix} \begin{bmatrix} (u_0v_1 - u_1v_0)w \\ (u_2v_0 - u_0v_2)w \\ (u_1v_2 - u_2v_1)w \\ s \end{bmatrix} = 0 \quad (60)$$

For (1 + 3)-dimensional space-time we obtain:

$$\begin{bmatrix} e_0 & 0 & 0 & 0 & 0 & -e_3 & e_2 & f_1 \\ 0 & e_0 & 0 & 0 & e_3 & 0 & -e_1 & f_2 \\ 0 & 0 & e_0 & 0 & -e_2 & e_1 & 0 & f_3 \\ 0 & 0 & 0 & e_0 & -f_1 & -f_2 & -f_3 & 0 \\ 0 & f_3 & -f_2 & -e_1 & -f_0 & 0 & 0 & 0 \\ -f_3 & 0 & f_1 & -e_2 & 0 & -f_0 & 0 & 0 \\ f_2 & -f_1 & 0 & -e_3 & 0 & 0 & -f_0 & 0 \\ e_1 & e_2 & e_3 & 0 & 0 & 0 & 0 & -f_0 \end{bmatrix} \begin{bmatrix} (u_0v_1 - u_1v_0)w \\ (u_0v_2 - u_2v_0)w \\ (u_0v_3 - u_3v_0)w \\ 0 \\ (u_3v_2 - u_2v_3)w \\ (u_1v_3 - u_3v_1)w \\ (u_2v_1 - u_1v_2)w \\ s \end{bmatrix} = 0 \quad (61)$$

where $\nu$ = 0, 1, 2, 3 and $u_0, u_1, u_2, u_3,$ $v_0, v_1, v_2, v_3, w,$ $s$ are arbitrary parameters, and we also have:

$$\begin{aligned} & e_0 = g^{0\nu}p_\nu - g^{00}(m_0c/\sqrt{g^{00}}),\ f_0 = p_0 + (m_0c/\sqrt{g^{00}}), \\ & e_1 = g^{1\nu}p_\nu - g^{10}(m_0c/\sqrt{g^{00}}),\ f_1 = p_1, \\ & e_2 = g^{2\nu}p_\nu - g^{20}(m_0c/\sqrt{g^{00}}),\ f_2 = p_2, \\ & e_3 = g^{3\nu}p_\nu - g^{30}(m_0c/\sqrt{g^{00}}),\ f_3 = p_3\ ; \end{aligned} \quad (61\text{-}1)$$

For (1 + 4)-dimensional space-time, the system of linear equations corresponding to the invariant quadratic relation (52) is derived as follows:

$$\begin{bmatrix} e_0 & 0 & 0 & 0 & 0 & 0 & 0 & 0 & 0 & 0 & 0 & -e_4 & 0 & -e_3 & -e_2 & f_1 \\ 0 & e_0 & 0 & 0 & 0 & 0 & 0 & 0 & 0 & 0 & e_4 & 0 & e_3 & 0 & -e_1 & -f_2 \\ 0 & 0 & e_0 & 0 & 0 & 0 & 0 & 0 & 0 & -e_4 & 0 & 0 & e_2 & e_1 & 0 & f_3 \\ 0 & 0 & 0 & e_0 & 0 & 0 & 0 & 0 & e_4 & 0 & 0 & 0 & -f_1 & f_2 & -f_3 & 0 \\ 0 & 0 & 0 & 0 & e_0 & 0 & 0 & 0 & 0 & -e_3 & -e_2 & -e_1 & 0 & 0 & 0 & -f_4 \\ 0 & 0 & 0 & 0 & 0 & e_0 & 0 & 0 & e_3 & 0 & f_1 & -f_2 & 0 & 0 & f_4 & 0 \\ 0 & 0 & 0 & 0 & 0 & 0 & e_0 & 0 & e_2 & -f_1 & 0 & f_3 & 0 & -f_4 & 0 & 0 \\ 0 & 0 & 0 & 0 & 0 & 0 & 0 & e_0 & e_1 & f_2 & -f_3 & 0 & f_4 & 0 & 0 & 0 \\ 0 & 0 & 0 & f_4 & 0 & f_3 & f_2 & f_1 & -f_0 & 0 & 0 & 0 & 0 & 0 & 0 & 0 \\ 0 & 0 & -f_4 & 0 & -f_3 & 0 & -e_1 & e_2 & 0 & -f_0 & 0 & 0 & 0 & 0 & 0 & 0 \\ 0 & f_4 & 0 & 0 & -f_2 & e_1 & 0 & -e_3 & 0 & 0 & -f_0 & 0 & 0 & 0 & 0 & 0 \\ -f_4 & 0 & 0 & 0 & -f_1 & e_2 & e_3 & 0 & 0 & 0 & 0 & -f_0 & 0 & 0 & 0 & 0 \\ 0 & f_3 & f_2 & -e_1 & 0 & 0 & 0 & e_4 & 0 & 0 & 0 & 0 & -f_0 & 0 & 0 & 0 \\ -f_3 & 0 & f_1 & e_2 & 0 & 0 & -e_4 & 0 & 0 & 0 & 0 & 0 & 0 & -f_0 & 0 & 0 \\ -f_2 & -f_1 & 0 & -e_3 & 0 & e_4 & 0 & 0 & 0 & 0 & 0 & 0 & 0 & 0 & -f_0 & 0 \\ e_1 & -e_2 & e_3 & 0 & -e_4 & 0 & 0 & 0 & 0 & 0 & 0 & 0 & 0 & 0 & 0 & -f_0 \end{bmatrix} \begin{bmatrix} (u_0v_1 - u_1v_0)w \\ (u_2v_0 - u_0v_2)w \\ (u_0v_3 - u_3v_0)w \\ 0 \\ (u_4v_0 - u_0v_4)w \\ 0 \\ 0 \\ 0 \\ 0 \\ (u_3v_4 - u_4v_3)w \\ (u_2v_4 - u_4v_2)w \\ (u_1v_4 - u_4v_1)w \\ (u_2v_3 - u_3v_2)w \\ (u_1v_3 - u_3v_1)w \\ (u_1v_2 - u_2v_1)w \\ s \end{bmatrix} = 0 \quad (62)$$

where $\nu$ = 0,1,2,3,4 and $u_0, u_1, u_2, u_3, u_4,\ v_0, v_1, v_2, v_3, v_4, w,\ s$ are arbitrary parameters, and we have:

$$\begin{aligned}
&e_0 = g^{0\nu} p_\nu - g^{00}(m_0 c/\sqrt{g^{00}}),\ f_0 = p_0 + (m_0 c/\sqrt{g^{00}}),\\
&e_1 = g^{1\nu} p_\nu - g^{10}(m_0 c/\sqrt{g^{00}}),\ f_1 = p_1,\\
&e_2 = g^{2\nu} p_\nu - g^{20}(m_0 c/\sqrt{g^{00}}),\ f_2 = p_2,\\
&e_3 = g^{3\nu} p_\nu - g^{30}(m_0 c/\sqrt{g^{00}}),\ f_3 = p_3,\\
&e_4 = g^{4\nu} p_\nu - g^{40}(m_0 c/\sqrt{g^{00}}),\ f_4 = p_4.
\end{aligned} \tag{62-1}$$

The systems of linear equations that are obtained for (1 + 5) and higher dimensional cases of the energy-momentum quadratic relation (52), have also the formulations similar to the obtained systems of linear Equations (58)–(62), and would be expressed by the matrix product of a $2^N \times 2^N$ square matrix and a $2^N \times 1$ column matrix in (1 + *N*)-dimensional space-time. For the (1 + 5)-dimensional case of energy-momentum relation (52), the column matrix of the corresponding system of linear equations (expressed by the matrix product of a $32 \times 32$ square matrix and a $32 \times 1$ column matrix, similar to (57-2)) is given by:

$$S = \begin{bmatrix} S' \\ S'' \end{bmatrix},\ S' = \begin{bmatrix} (u_0 v_1 - u_1 v_0) w \\ (u_0 v_2 - u_2 v_0) w \\ (u_0 v_3 - u_3 v_0) w \\ 0 \\ (u_0 v_4 - u_4 v_0) w \\ 0 \\ 0 \\ 0 \\ (u_0 v_5 - u_5 v_0) w \\ 0 \\ 0 \\ 0 \\ 0 \\ 0 \\ 0 \\ 0 \end{bmatrix},\ S'' = \begin{bmatrix} 0 \\ 0 \\ 0 \\ (u_5 v_4 - u_4 v_5) w \\ 0 \\ (u_3 v_5 - u_5 v_3) w \\ (u_5 v_2 - u_2 v_5) w \\ (u_1 v_5 - u_5 v_1) w \\ 0 \\ (u_4 v_3 - u_3 v_4) w \\ (u_2 v_4 - u_4 v_2) w \\ (u_4 v_1 - u_1 v_4) w \\ (u_2 v_3 - u_3 v_2) w \\ (u_3 v_1 - u_1 v_3) w \\ (u_1 v_2 - u_2 v_1) w \\ s \end{bmatrix}. \tag{62-2}$$

where $u_0, u_1, u_2, u_3, u_4, u_5,\ v_0, v_1, v_2, v_3, v_4, v_5, w, s$ are arbitrary parameters.

*3.2. Deivation of the Rational Lorentz Transformations*

From the derived systems of linear Equations (54)–(57) corresponding to the (1 + 1)–(1 + 4)-dimensional cases of the invariant relation (51), and also using the general parametric solutions (43)–(51) (obtained for systems of linear Equations (26)–(30)), the rational Lorentz transformations (which are completely dense in the standard group of Lorentz transformations [15], as noted in Section 3.1) are derived for momentums $p_\mu$ and $p'_\mu$. For instance, assuming the Minkowski metric from the system of linear Equation (55), a parametric form of rational Lorentz transformations for three-momentums $p_\mu$ and $p'_\mu$ in (1 + 2) dimensional space-time is derived as follows:

$$\begin{bmatrix} \frac{1+z_0^2+z_1^2+z_2^2}{1-z_0^2-z_1^2+z_2^2} & \frac{2(z_0+z_1z_2)}{1-z_0^2-z_1^2+z_2^2} & \frac{-2(z_1-z_0z_2)}{1-z_0^2-z_1^2+z_2^2} \\ \frac{2(z_0-z_1z_2)}{1-z_0^2-z_1^2+z_2^2} & \frac{1+z_0^2-z_1^2-z_2^2}{1-z_0^2-z_1^2+z_2^2} & \frac{2(z_2-z_0z_1)}{1-z_0^2-z_1^2+z_2^2} \\ \frac{-2(z_1+z_0z_2)}{1-z_0^2-z_1^2+z_2^2} & \frac{2(z_2+z_0z_1)}{1-z_0^2-z_1^2+z_2^2} & \frac{1-z_0^2+z_1^2-z_2^2}{1-z_0^2-z_1^2+z_2^2} \end{bmatrix} \begin{bmatrix} p_0 \\ p_1 \\ p_2 \end{bmatrix} = \begin{bmatrix} p_0' \\ p_1' \\ p_2' \end{bmatrix} \tag{63}$$

where the parameters $s_\mu$ in (63) are given by the formulas: $z_0=(u_0v_1-u_1v_0)w$, $z_2=(u_2v_0-u_0v_2)w$, $z_3=(u_1v_2-u_2v_1)w$; that are expressed in terms of the arbitrary parameters $u_0,u_1,u_2,v_0,v_1,v_2,w$. These parameters would be also determined and expressed in terms of the initially given physical variables (such as the relative velocity between the reference frames). However, as it has been also noted in Section 2.4 concerning a particular common algebraic property of parametric relations (43-2) and (44-3) which are equivalent to the above expressions, by choosing appropriate integer values for parameters $u_0,u_1,u_2,v_0,v_1,v_2,w$, the parameters $z_0,z_1,z_2$ could take any given integer values. Thus, we may directly determine the relevant expressions for parameters $s_\mu$ in terms of the initially given physical values and variables. Hence, as a particular case, from the isomorphic transformations (63), in addition to these determined expressions for the parameters $s_\mu$ (in terms of the relative velocity between the reference frames in x-direction and the speed of light): $z_0=-\beta/(1+\gamma)$, $z_1=z_2=0$, $\gamma=1/\sqrt{1-\beta^2}$, $\beta = v/c$, we obtain the equivalent form of Lorentz transformations in the standard configuration [38]:

$$\begin{bmatrix} \frac{1+z_0^2}{1-z_0^2} & \frac{2z_0}{1-z_0^2} \\ \frac{2z_0}{1-z_0^2} & \frac{1+z_0^2}{1-z_0^2} \end{bmatrix} \begin{bmatrix} p_0 \\ p_1 \end{bmatrix} = \begin{bmatrix} \gamma & -\beta\gamma \\ -\beta\gamma & \gamma \end{bmatrix} \begin{bmatrix} p_0 \\ p_1 \end{bmatrix} = \begin{bmatrix} p_0' \\ p_1' \end{bmatrix} \tag{63-1}$$

Similar to the derived transformations (63-1), the Lorentz transformations (in standard configuration) are derived by the same approach for higher-dimensional space-times.

### *3.3. Matrix Formalism of the Lorentz Invariant Systems of Linear Equations (59)–(62),…, as Equivalent Forms of the Relativistic Energy-Momentum Quadratic Relation in Various Space-Time Dimensions*

The Lorentz invariant systems of linear Equations (59)–(62), …, (obtained on the basis of the axiom (17) and relevant general results presented in Sections 2.2 and 2.4 for homogeneous quadratic equations) as equivalent forms of the Lorentz invariant energy-momentum quadratic relation (52), could be expressed generally by the following matrix formulation in (1 + $N$)-dimensional space-time:

$$(\alpha^\mu p_\mu - m_0\tilde{\alpha}^\mu k_\mu)S=0, \tag{64}$$

where

$$\alpha^\mu=\beta^\mu+\beta'^\mu,\ \tilde{\alpha}^\mu=\beta^\mu-\beta'^\mu, \tag{65}$$

$m_0$ is the invariant mass of a relativistic particle and $k_\mu=(c/\sqrt{g^{00}},0,...,0)$ is its covariant velocity (that is a time-like covariant vector) in the stationary reference frame, $\alpha^\mu$ and $\tilde{\alpha}^\mu$ are two contravariant $2^N \times 2^N$ square matrices (corresponding with the matrix representations of Clifford algebras Cℓ1,2, Cℓ1,3, Cℓ1,4, …, Cℓ1,N (for $N \geq 2$) and their generalizations [11,24,32], see also Appendix A) that by the isomorphic linear relations (65) are expressed in terms of two corresponding contravariant $2^N \times 2^N$ matrices $\beta$ and $\beta'^{\mu}$, and $S$ is a $2^N \times 1$ parametric column matrix. These matrices in (1 + 1), (1 + 2), (1 + 3), (1 + 4) and (1 + 5) space-time dimensions are given uniquely as follows, respectively:

For (1 + 1)-dimensional case we get:

$$\beta^0 = \begin{bmatrix} 0 & 0 \\ 0 & -1 \end{bmatrix},\ \beta'_0 = \begin{bmatrix} 1 & 0 \\ 0 & 0 \end{bmatrix},\ \beta^1 = \begin{bmatrix} 0 & 1 \\ 0 & 0 \end{bmatrix},\ \beta'_1 = \begin{bmatrix} 0 & 0 \\ 1 & 0 \end{bmatrix},\ S = \begin{bmatrix} (u_0 v_1 - u_1 v_0) w \\ s \end{bmatrix}; \tag{66}$$

where $u_0, u_1,\ v_0, v_1, w;\ s$ are arbitrary parameters.

For (1 + 2)-dimensional case we obtain where $u_0, u_1, u_2,\ v_0, v_1, v_2, w;\ s$ are arbitrary parameters):

$$\beta^0 = \begin{bmatrix} 0 & 0 \\ 0 & -(\sigma^0 + \sigma^1) \end{bmatrix},\ \beta'_0 = \begin{bmatrix} \sigma^0 + \sigma^1 & 0 \\ 0 & 0 \end{bmatrix},\ \beta^1 = \begin{bmatrix} 0 & \sigma^2 \\ -\sigma^2 & 0 \end{bmatrix},\ \beta'_1 = \begin{bmatrix} 0 & \sigma^3 \\ -\sigma^3 & 0 \end{bmatrix},$$
$$\beta^2 = \begin{bmatrix} 0 & -\sigma^1 \\ -\sigma^0 & 0 \end{bmatrix},\ \beta'_2 = \begin{bmatrix} 0 & -\sigma^0 \\ -\sigma^1 & 0 \end{bmatrix},$$
$$S = \begin{bmatrix} (u_0 v_1 - u_1 v_0) w \\ (u_2 v_0 - u_0 v_2) w \\ (u_1 v_2 - u_2 v_1) w \\ s \end{bmatrix}, \tag{67}$$
$$\sigma^0 = \begin{bmatrix} 1 & 0 \\ 0 & 0 \end{bmatrix},\ \sigma^1 = \begin{bmatrix} 0 & 0 \\ 0 & 1 \end{bmatrix},\ \sigma^2 = \begin{bmatrix} 0 & 1 \\ 0 & 0 \end{bmatrix},\ \sigma^3 = \begin{bmatrix} 0 & 0 \\ -1 & 0 \end{bmatrix}$$

For (1 + 3)-dimensional case we obtain:

$$\beta^0 = \begin{bmatrix} 0 & 0 \\ 0 & -(\gamma^0 + \gamma^1) \end{bmatrix},\ \beta'_0 = \begin{bmatrix} (\gamma^0 + \gamma^1) & 0 \\ 0 & 0 \end{bmatrix},\ \beta^1 = \begin{bmatrix} 0 & \gamma^2 \\ -\gamma^3 & 0 \end{bmatrix},\ \beta'_1 = \begin{bmatrix} 0 & \gamma^3 \\ -\gamma^2 & 0 \end{bmatrix},$$
$$\beta^2 = \begin{bmatrix} 0 & \gamma^4 \\ \gamma^5 & 0 \end{bmatrix},\ \beta'_2 = \begin{bmatrix} 0 & -\gamma^5 \\ -\gamma^4 & 0 \end{bmatrix},\ \beta^3 = \begin{bmatrix} 0 & \gamma^6 \\ -\gamma^7 & 0 \end{bmatrix},\ \beta'_3 = \begin{bmatrix} 0 & \gamma^7 \\ -\gamma^6 & 0 \end{bmatrix},$$
$$S = \begin{bmatrix} (u_0 v_1 - u_1 v_0) w \\ (u_0 v_2 - u_2 v_0) w \\ (u_0 v_3 - u_3 v_0) w \\ 0 \\ (u_3 v_2 - u_2 v_3) w \\ (u_1 v_3 - u_3 v_1) w \\ (u_2 v_1 - u_1 v_2) w \\ s \end{bmatrix}, \tag{68}$$

$$\gamma^0 = \begin{bmatrix} 1 & 0 & 0 & 0 \\ 0 & 1 & 0 & 0 \\ 0 & 0 & 0 & 0 \\ 0 & 0 & 0 & 0 \end{bmatrix}, \ \gamma^1 = \begin{bmatrix} 0 & 0 & 0 & 0 \\ 0 & 0 & 0 & 0 \\ 0 & 0 & 1 & 0 \\ 0 & 0 & 0 & 1 \end{bmatrix}, \ \gamma^2 = \begin{bmatrix} 0 & 0 & 0 & 1 \\ 0 & 0 & 0 & 0 \\ 0 & 0 & 0 & 0 \\ -1 & 0 & 0 & 0 \end{bmatrix}, \ \gamma^3 = \begin{bmatrix} 0 & 0 & 0 & 0 \\ 0 & 0 & -1 & 0 \\ 0 & 1 & 0 & 0 \\ 0 & 0 & 0 & 0 \end{bmatrix},$$

$$\gamma^4 = \begin{bmatrix} 0 & 0 & 0 & 0 \\ 0 & 0 & 0 & 1 \\ 0 & 0 & 0 & 0 \\ 0 & -1 & 0 & 0 \end{bmatrix}, \ \gamma^5 = \begin{bmatrix} 0 & 0 & -1 & 0 \\ 0 & 0 & 0 & 0 \\ 1 & 0 & 0 & 0 \\ 0 & 0 & 0 & 0 \end{bmatrix}, \ \gamma^6 = \begin{bmatrix} 0 & 0 & 0 & 0 \\ 0 & 0 & 0 & 0 \\ 0 & 0 & 0 & 1 \\ 0 & 0 & -1 & 0 \end{bmatrix}, \ \gamma^7 = \begin{bmatrix} 0 & -1 & 0 & 0 \\ 1 & 0 & 0 & 0 \\ 0 & 0 & 0 & 0 \\ 0 & 0 & 0 & 0 \end{bmatrix}.$$

where $u_0, u_1, u_2, u_3,\ v_0, v_1, v_2, v_3, w,\ s$ are arbitrary parameters. Moreover, the 4 × 4 matrices $\gamma^i$ (68) generate the Lorentz Lie algebra in (1 + 3) dimensions.

For (1 + 4)-dimensional case we have:

$$\beta^0 = \begin{bmatrix} 0 & 0 \\ 0 & -(\eta^0 + \eta^1) \end{bmatrix}, \ \beta'_0 = \begin{bmatrix} (\eta^0 + \eta^1) & 0 \\ 0 & 0 \end{bmatrix},$$

$$\beta^1 = \begin{bmatrix} 0 & \eta^2 \\ \eta^2 & 0 \end{bmatrix}, \ \beta'_1 = \begin{bmatrix} 0 & \eta^3 \\ \eta^3 & 0 \end{bmatrix}, \beta^2 = \begin{bmatrix} 0 & \eta^4 \\ \eta^5 & 0 \end{bmatrix}, \ \beta'_2 = \begin{bmatrix} 0 & -\eta^5 \\ -\eta^4 & 0 \end{bmatrix},$$

$$\beta^3 = \begin{bmatrix} 0 & \eta^6 \\ -\eta^7 & 0 \end{bmatrix}, \ \beta'_3 = \begin{bmatrix} 0 & \eta^7 \\ -\eta^6 & 0 \end{bmatrix}, \beta^4 = \begin{bmatrix} 0 & \eta^8 \\ \eta^9 & 0 \end{bmatrix}, \ \beta'_4 = \begin{bmatrix} 0 & -\eta^9 \\ -\eta^8 & 0 \end{bmatrix},$$

$$S = \begin{bmatrix} (u_0v_1 - u_1v_0)w \\ (u_2v_0 - u_0v_2)w \\ (u_0v_3 - u_3v_0)w \\ 0 \\ (u_4v_0 - u_0v_4)w \\ 0 \\ 0 \\ 0 \\ 0 \\ (u_3v_4 - u_4v_3)w \\ (u_2v_4 - u_4v_2)w \\ (u_1v_4 - u_4v_1)w \\ (u_2v_3 - u_3v_2)w \\ (u_1v_3 - u_3v_1)w \\ (u_1v_2 - u_2v_1)w \\ s \end{bmatrix}, \quad (69)$$

$$\eta^0=\begin{bmatrix}1&0&0&0&0&0&0&0\\0&1&0&0&0&0&0&0\\0&0&1&0&0&0&0&0\\0&0&0&1&0&0&0&0\\0&0&0&0&0&0&0&0\\0&0&0&0&0&0&0&0\\0&0&0&0&0&0&0&0\\0&0&0&0&0&0&0&0\end{bmatrix},\ \eta^1=\begin{bmatrix}0&0&0&0&0&0&0&0\\0&0&0&0&0&0&0&0\\0&0&0&0&0&0&0&0\\0&0&0&0&0&0&0&0\\0&0&0&0&1&0&0&0\\0&0&0&0&0&1&0&0\\0&0&0&0&0&0&1&0\\0&0&0&0&0&0&0&1\end{bmatrix},$$

$$\eta^2=\begin{bmatrix}0&0&0&0&0&0&0&1\\0&0&0&0&0&0&0&0\\0&0&0&0&0&0&0&0\\0&0&0&0&-1&0&0&0\\0&0&0&0&0&0&0&0\\0&0&1&0&0&0&0&0\\0&-1&0&0&0&0&0&0\\0&0&0&0&0&0&0&0\end{bmatrix},\ \eta^3=\begin{bmatrix}0&0&0&0&0&0&0&0\\0&0&0&0&0&0&-1&0\\0&0&0&0&0&1&0&0\\0&0&0&0&0&0&0&0\\0&0&0&-1&0&0&0&0\\0&0&0&0&0&0&0&0\\0&0&0&0&0&0&0&0\\1&0&0&0&0&0&0&0\end{bmatrix},$$

$$\eta^4=\begin{bmatrix}0&0&0&0&0&0&0&0\\0&0&0&0&0&0&0&-1\\0&0&0&0&0&0&0&0\\0&0&0&0&0&1&0&0\\0&0&0&0&0&0&0&0\\0&0&0&-1&0&0&0&0\\0&0&0&0&0&0&0&0\\0&1&0&0&0&0&0&0\end{bmatrix},\eta^5=\begin{bmatrix}0&0&0&0&0&0&1&0\\0&0&0&0&0&0&0&0\\0&0&0&0&-1&0&0&0\\0&0&0&0&0&0&0&0\\0&0&1&0&0&0&0&0\\0&0&0&0&0&0&0&0\\-1&0&0&0&0&0&0&0\\0&0&0&0&0&0&0&0\end{bmatrix},\eta^6=\begin{bmatrix}0&0&0&0&0&0&0&0\\0&0&0&0&0&0&0&0\\0&0&0&0&0&0&0&1\\0&0&0&0&0&0&-1&0\\0&0&0&0&0&0&0&0\\0&0&0&0&0&0&0&0\\0&0&0&1&0&0&0&0\\0&0&-1&0&0&0&0&0\end{bmatrix},$$

$$\eta^7=\begin{bmatrix}0&0&0&0&0&-1&0&0\\0&0&0&0&1&0&0&0\\0&0&0&0&0&0&0&0\\0&0&0&0&0&0&0&0\\0&-1&0&0&0&0&0&0\\1&0&0&0&0&0&0&0\\0&0&0&0&0&0&0&0\\0&0&0&0&0&0&0&0\end{bmatrix},\eta^8=\begin{bmatrix}0&0&0&0&0&0&0&0\\0&0&0&0&0&0&0&0\\0&0&0&0&0&0&0&0\\0&0&0&0&0&0&0&0\\0&0&0&0&0&0&0&-1\\0&0&0&0&0&0&1&0\\0&0&0&0&0&-1&0&0\\0&0&0&0&1&0&0&0\end{bmatrix},\eta^9=\begin{bmatrix}0&0&0&1&0&0&0&0\\0&0&-1&0&0&0&0&0\\0&1&0&0&0&0&0&0\\-1&0&0&0&0&0&0&0\\0&0&0&0&0&0&0&0\\0&0&0&0&0&0&0&0\\0&0&0&0&0&0&0&0\\0&0&0&0&0&0&0&0\end{bmatrix}.$$

where $u_0,u_1,u_2,u_3,u_4,\ v_0,v_1,v_2,v_3,v_4,w,s$ are arbitrary parameters. Furthermore, similar to the 4 × 4 $\overset{i}{\gamma}$ matrices in (68), the 8 × 8 matrices $\overset{i}{\eta}$ (69) generate the Lorentz Lie algebra in (1 + 4) dimensions.

For (1 + 5)-dimensional case the size of matrices $\beta$ and $\beta'^{\mu}$ is $32\times32$. $S$ is also a $32\times1$ column matrix, given by:

$$S=\begin{bmatrix}S'\\S''\end{bmatrix},\quad S'=\begin{bmatrix}(u_0v_1-u_1v_0)w\\(u_0v_2-u_2v_0)w\\(u_0v_3-u_3v_0)w\\0\\(u_0v_4-u_4v_0)w\\0\\0\\0\\(u_0v_5-u_5v_0)w\\0\\0\\0\\0\\0\\0\\0\end{bmatrix},\quad S''=\begin{bmatrix}0\\0\\0\\(u_5v_4-u_4v_5)w\\0\\(u_3v_5-u_5v_3)w\\(u_5v_2-u_2v_5)w\\(u_1v_5-u_5v_1)w\\0\\(u_4v_3-u_3v_4)w\\(u_2v_4-u_4v_2)w\\(u_4v_1-u_1v_4)w\\(u_2v_3-u_3v_2)w\\(u_3v_1-u_1v_3)w\\(u_1v_2-u_2v_1)w\\s\end{bmatrix}. \tag{70}$$

where $u_0,u_1,u_2,u_3,u_4,u_5,\ v_0,v_1,v_2,v_3,v_4,v_5,w,s$ are arbitrary parameters.

Similar to the formulations (66)–(70), for the higher dimensional cases of invariant quadratic relation (52), the column matrix *S* and square matrices $\beta$ and $\beta'^{\mu}$ (defining the square matrices $\alpha^{\mu}$ and $\alpha'^{\mu}$ that correspond to the matrix representations of Clifford algebras and their generalization, see Section 3.3 and also Appendix A) are obtained using similar algebraic structures, where in (1 + *N*) space-time dimensions the size of square matrices $\beta$ and $\beta'^{\mu}$ is $2^N\times 2^N$ and the size of column matrix *S* is $2^N\times 1$.

Remark 1. General Algebraic Formulation of the Column Matrix S Given bythe Matrix Equation (64)

In addition, as noted in Section 3.3, the matrix Equation (64) represents uniquely the equivalent form of the Lorentz invariant energy-momentum quadratic relation (52) (as the norm of the *D*-momentum), based on the axiomatic relations (17) and (17-1) and relevant general results obtained in Sections 2.2 and 2.4 for homogeneous quadratic equations over the integral domain over $\mathbb{Z}$**.** Therefore (as mentioned in Section 3.3), the general algebraic formulation of the entries of column matrices $S$ obtaining in subsequent higher space-time dimensions, are similar to formulations of the obtained matrices $S$ (66)–(70) corresponding, respectively, to the (1 + 0), (1 + 1), (1 + 2), (1 + 3), (1 + 4) and (1 + 5)-dimensional cases of Lorentz invariant matrix equation (64). Hence, the algebraic formulation of column matrix $S$ in (1 + *N*) space-time dimensions would be generally defined as follows: the last entry of $S$ is represented solely by the arbitrary parameter *s*, $2^{N-1}$ entries are definitely zero (see below) and all the other $2^{N-1}-1$ entries of $S$ could be represented uniformly by the following unique algebraic formulation (expressing in terms of the arbitrary parameters: $u_0,u_1,u_2,u_3,\ldots,u_N,\ v_0,v_1,v_2,v_3,\ldots,v_N,w$) given on the basis of a one-to-one correspondence between these (non-zero) entries of matrix $S$ and the entries $h_{\mu\nu}$ (for $\mu>\nu$) of a $2^{N+1}\times 2^{N+1}$ square matrix $H[h_{\mu\nu}]$ defined in (1 + *N*) dimensions, by:

$$h_{\mu\nu} = (u_\nu v_\mu - u_\mu v_\nu)w \tag{70-1}$$

where $\mu$, $\nu = 0, 1, 2, \ldots, N$, and $h_{\mu\nu} = 0$ for $\mu = \nu$.

Note that the algebraic form (70-1) is equivalent to form (45-2) which, as it has been noted in Section 2.4, generates a symmetric algebra Sym($V$) on the vector space $V$, where $(u_\mu, v_\nu) \in V$ [37].

Hence, as a basic algebraic property of the form (70-1), a natural unique isomorphism is defined between the underlying vector space $V$ of the symmetric algebra Sym($V$) (which is generated by algebraic form (70-1)) and the Weyl algebra W($V$). Moreover, based on this isomorphism, the Weyl algebra W($V$) could be defined as a (first) quantization of the symmetric algebra Sym($V$), where the generators of the Weyl algebra W($V$) would be represented by the corresponding (covariant) differential operators (such as $i\hbar\nabla_\mu$, as per quantum mechanics usage).

In Section 3.4, we use these general and basic algebraic properties of the column matrix $S$, in particular, in the procedure of quantization of the algebraic matrix Equation (64).

In addition to the above algebraic properties of the parametric entries of column matrix $S$, that are represented uniformly by the algebraic formula (70-1), in terms of the arbitrary parameters: $u_0, u_1, u_2, u_3, \ldots, u_N,\ v_0, v_1, v_2, v_3, \ldots, v_N, w$, the following basic properties hold as well:

Displaying the column matrix $S$ by two half-sized $2^{N-1} \times 1$ column matrices $S'$ and $S''$ (containing respectively the upper and lower entries of $S$, similar to the formulas (57-2) and (62-2) representing the (1 + 5)-dimensional case of matrix $S$) such that: $S = \begin{bmatrix} S' \\ S'' \end{bmatrix}$, then we have:

(1) The number of entries of the column matrix $S'$ that are zero, is exactly: $(2^{N-1} - N)$, and the other $N$ entries are represented solely either by the formulation: $h_{\mu 0} = (u_0 v_\mu - u_\mu v_0)w$; or by its negative form, i.e.,: $-h_{\mu 0} = h_{0\mu} = (u_\mu v_0 - u_0 v_\mu)w$, where $\mu = 1, 2, \ldots, N$, and $h_{\mu 0}$ denote the $N$ entries (except the first entry $h_{00}$ that is zero) of the first column of square matrix $H[h_{\mu\nu}]$ (defined by the formula (70-1));

(2) The number of entries of the column matrix $S''$ that are zero, is exactly: $(2^{N-1} - \frac{N(N-1)}{2} - 1)$, and except the last entry (represented by arbitrary parameter $s$), all the other $(\frac{N(N-1)}{2})$ entries are represented solely either by the formulation: $h_{\mu\nu} = (u_\nu v_\mu - u_\mu v_\nu)w$; or by its negative form, i.e.,: $-h_{\mu\nu} = h_{\nu\mu} = (u_\mu v_\nu - u_\nu v_\mu)w$, where $\mu > \nu$, $\mu$, $\nu = 1, 2, \ldots, N$, and $h_{\mu\nu}$ denote the components of square matrix $H[h_{\mu\nu}]$, and the last entry of column matrix $S''$ is also represented by the arbitrary parameter $s$.

(3) If we exchange $S'$ and $S''$ in the column matrix $S = \begin{bmatrix} S' \\ S'' \end{bmatrix}$, that could be shown by,

$$S^{(Ch)} = \begin{bmatrix} 0 & I \\ I & 0 \end{bmatrix} S = \begin{bmatrix} 0 & I \\ I & 0 \end{bmatrix} \begin{bmatrix} S' \\ S'' \end{bmatrix} = \begin{bmatrix} S'' \\ S' \end{bmatrix} \tag{70-2}$$

then based on the general formulation of matrix $S$ (defined uniquely by formulas (66)–(70) for various space-time dimensions) and its algebraic properties (1) and (2) (mentioned above), it is concluded directly that the matrix Equation (64) given with the new column matrix $S^{(Ch)}$ (70-2), i.e., equation: $(\alpha^{\mu}p_{\mu}-m_0\tilde{\alpha}^{\mu}k_{\mu})S^{(Ch)}=0$, is which could be defined solely in (1 + 2) space-time dimensions for $s=0,\ u_1=0,\ v_1=0$, in (1 + 3) space-time dimensions for $s=0$, and in (1 + 4) space-time dimensions for $s=0,\ u_1=0,\ v_1=0$ (which is reduced and be equivalent to the (1 + 3)-dimensional case of matrix Equation (64)). In (1 + 1) and (1 + 5) and higher space-time dimensions, the matrix equation $(\alpha^{\mu}p_{\mu}-m_0\tilde{\alpha}^{\mu}k_{\mu})S^{(Ch)}=0$ would be which are defined if and only if all the entries of column matrix $S^{(Ch)}$ are zero. This means that the matrix Equation (64): $(\alpha^{\mu}p_{\mu}-m_0\tilde{\alpha}^{\mu}k_{\mu})S=0$, is symmetric in the exchange of $S'$ and $S''$ (in the column matrix $S=\begin{bmatrix} S' \\ S'' \end{bmatrix}$), solely in (1 + 2)-dimensional space-time for $s=0, u_1=0, v_1=0$, and in (1 + 3)-dimensional space-time for $s=0$. In Section 3.12, this particular algebraic property of the column matrix $S$ would be used for concluding a new crucial and essential issue in fundamental physics.

In the following Section, the natural isomorphism between the symmetric algebra **Sym**($V$) (generated uniquely by the algebraic form (70-1)) and the Weyl algebra **W**($V$), in addition to the general algebraic properties of column matrix $S$ mentioned above (in Remark 1), would be used and applied directly in the procedure of first quantization of the Lorentz invariant system of linear (64).

### *3.4. A New Unique Mathematical Derivation of the Fundamental (Massive) Field Equations of Physics (Representing the Laws Governing the Fundamental Forces of Nature)*

By first quantization (followed by a basic procedure of minimal coupling to space-time geometry) of the Lorentz invariant system of linear Equation (64) (representing uniquely the equivalent form of energy-momentum quadratic relation (52), see Section 3.3) expressed in terms of the Clifford algebraic matrices (65)–(70),..., two classes of general covariant field equations are derived uniquely as follows (given by (1 + $N$) space-time dimensions):

$$(i\hbar\alpha^{\mu}\nabla_{\mu}-m_0^{(R)}\tilde{\alpha}^{\mu}k_{\mu})\Psi_R=0, \tag{71}$$

$$(i\hbar\alpha^{\mu}D_{\mu}-m_0^{(F)}\tilde{\alpha}^{\mu}k_{\mu})\Psi_F=0 \tag{72}$$

where $i\hbar\nabla_{\mu}$ and $i\hbar D_{\mu}$ are the general relativistic forms of energy-momentum quantum operator (where $\nabla_{\mu}$ is the general covariant derivative, and $D_{\mu}$ is gauge covariant derivative, for detail see the ordinary tensor formalisms of these equations, represented by formulas (78-1)–(79-3), in Section 3.5), $m_0^{(R)}$ and $m_0^{(F)}$ are the fields' invariant masses, $k_{\mu}=(c/\sqrt{g^{00}},0,...,0)$ is the general covariant velocity in stationary reference frame (that is a time-like covariant vector), $\alpha^{\mu}$ and $\tilde{\alpha}^{\mu}$ are two contravariant $2^N\times 2^N$ square matrices (compatible with the matrix representations of certain Clifford algebras, see Section 3.3 and also Appendix A) defined by formulas (65)–(70) in Section 3.3. In the field Equation (72), $\Psi_R$ is a column matrix as a (first) quantized form of the algebraic column matrix $S$ (defined by relations (64)–(70-2)), determined and represented uniquely by formulas (73)–(77), …, in various space-time dimensions. The column matrix $\Psi_R$ contains the components of field strength tensor $R_{\mu\nu\rho\sigma}$ (equivalent to the Riemann curvature tensor), and also the components of covariant quantity $\phi^{(G)}_{\rho\sigma}$ that defines the corresponding source current tensor by relation: $J^{(R)}_{\rho\sigma\nu} = -(\breve{\nabla}_\nu + \frac{im_0^{(R)}}{\hbar}k_\nu)\phi^{(R)}_{\rho\sigma}$ (which appears in the course of the derivation of field Equation (71), see Section 3.6 for details). In a similar manner, in the tensor field Equation (72), $\Psi_F$ is also a column matrix as a (first) quantized form of the algebraic column matrix $S$ (defined by relations (64)–(70-2)), determined and represented uniquely by formulas (73)–(77), …, in to various space-time dimensions. The column matrix $\Psi_F$ contains both the components of tensor field $F_{\mu\nu}$ (defined as the gauge field strength tensor), and also the components of covariant quantity $\phi^{(F)}$ that define the corresponding source current vector by relation: relation: $J^{(F)}_\nu = -(\breve{D}_\nu + \frac{im_0^{(F)}}{\hbar}k_\nu)\phi^{(F)}$ (which appears in the course of the derivation of field Equation (72), see Section 3.6 for details). Moreover, the general covariance formalism of the field Equations (71) and (72), would be also shown in Section 3.5.

In addition, in Section 3.11, based on a basic class of discrete symmetries for the field Equations (71) and (72), along with definite mathematical axiomatic formalism of the derivation of these equations, it is shown that these equations could be defined solely in (1 + 2) and (1 + 3) space-time dimensions. It is shown that (1 + 3) dimensional cases of these equations represent uniquely a new formalism of bispinor fields of spin-2 and spin-1 particles, respectively. It is also shown that the (1 + 2)-dimensional cases of these equations, represent asymptotically new massive forms of bispinor fields of spin-3/2 and spin-1/2 particles, respectively.

Moreover, in Section 3.12, based on the definite mathematical formalism of this axiomatic derivation approach, the basic assumption (3) in Section 3.1, along with the C, P and T symmetries (represented basically by their corresponding quantum matrix operators) of the fundamentally derived general covariant field Equations (71) and (72), it is concluded that the universe could be realized solely by the (1 + 2) and (1 + 3)-dimensional space-times (where this conclusion, in particular, is based on the T-symmetry). In Sections 3.13 and 3.14, it is proved that 'CPT' is the only (unique) combination of C, P, and T symmetries that could be defined as a symmetry for interacting fields. In addition, in Section 3.14, on the basis of these discrete symmetries of the field equations (71) and (72), it is shown that only left-handed particle fields (along with their complementary right-handed fields) could be coupled to the corresponding (any) source currents.

Furthermore, in Section 3.15, it is argued that the metric of background curved space-time is diagonalized for the spin-1/2 fermion field equations (defined by the field Equation (110) as a generalized form of (1 + 2)-dimensional case of Equation (72)), where this property generates a certain set of additional symmetries corresponding uniquely to the $SU(2)_L \otimes U(2)_R$ symmetry group for spin-1/2 fermion fields (represented by two main groups of "1 + 3" generations, corresponding respectively to two subgroups of leptons and two subgroups of quarks), in addition to the $SU(2)_L \otimes U(2)_R$ and SU(3) gauge symmetry groups for spin-1 boson fields coupled to the spin-1/2 fermionic source currents. Moreover, based on these uniquely determined gauge symmetries, four new charge-less spin-1/2 fermions (represented by "$z_e$ , $z_n$ ; $z_u$ , $z_d$", where two right-handed charhe-less quarks $z_u$ and $z_d$ emerge specifically in two subgroups with anti-quarks such that: ( $\overline{s}, \overline{u}, \overline{b}$ , $z_u$) and ( $\overline{c}, \overline{d}, \overline{t}$ , $z_d$), and also three new massive spin-1 bosons (represented by $\tilde{W}^+, \tilde{W}^-, \vec{Z}$, where in particular $\vec{Z}$ is the complementary right-handed particle of ordinary $Z$ boson), are predicted by this new mathematical axiomaticapproach.

As a particular result, in Section 3.6, based on the definite and unique formulation of the derived Maxwell's equations (and also Yang-Mills equations, defined by the (1 + 3)-dimensional case of the field Equation (72), compatible with specific gauge symmetry groups as shown in Section 3-15, formulas (109)–(110-12)), it is also concluded that magnetic monopoles could not exist in nature.

*3.5. Axiomatic Derivation of General Covariant Massive Field Equations (71) and (72)*

First, it should be noted that via initial quantization (followed by a basic procedure of minimal coupling to space-time geometry) of the algebraic systems of linear Equation (64) (as a matrix equation given by the Clifford algebraic matrices (65)–(70), …, in various space-time dimensions), two categories of general covariant field Equations (with a definite matrix formalism compatible with the Clifford algebras and their generalizations, see Section 3.3 and also Appendix A) are derived solely, represented by the tensor Equations (71) and (72) in terms of two tensor fields $R_{\rho\sigma\mu\nu}$ and $F_{\mu\nu}$, respectively. In fact, as it has been mentioned in Remark 1 (in Section 3.3), there is a natural isomorphism between the Weyl algebra and the symmetric algebra generated by the algebraic form (70-1) which represents the general formulation of the entries of algebraic column matrix $S$ in the matrix Equation (64). In addition, the procedure of minimal coupling to space-time geometry would be simply defined as a procedure which, starting from a theory in flat space-time, substitutes all partial derivatives by corresponding covariant derivatives and the flat space-time metric by the curved space-time (pseudo-Riemannian) metric. Moreover, as mentioned in Remark 1 (in Section 3.3), on the basis of this natural isomorphism, the Weyl algebra could be also represented as a quantization of the symmetric algebra generated by the algebraic form (70-1). Hence, using this natural isomorphism, by first quantization (followed by a basic procedure of minimal coupling to space-time geometry) of matrix Equation (64), two definite classes of general covariant massive (tensor) field equations are determined uniquely, expressed in terms of two basic connection forms (denoted by two derivatives $\nabla_\mu$ and $D_\mu$ corresponding respectively to the diffeomorphism (or metric) invariance and gauge invariance), along with their corresponding curvature forms, denoted respectively by $R_{\rho\sigma\mu\nu}$ (as the gravitational field strength tensor, equivalent to Riemann curvature tensor) and $F_{\mu\nu}$ (as the gauge field strength tensor). This natural isomorphism could be represented by the following mappings (corresponding to the curvature forms $R_{\rho\sigma\mu\nu}$ and $F_{\mu\nu}$, respectively):

$$(u_\mu v_\nu - u_\nu v_\mu)w \mapsto (\nabla_\mu \nabla_\nu - \nabla_\nu \nabla_\mu)\omega_R = R(\mu,\nu)\omega_R \mapsto R_{\rho\sigma\mu\nu}, \quad (71\text{-}1)$$

$$(u_\mu v_\nu - u_\nu v_\mu)w \mapsto (D_\mu D_\nu - D_\nu D_\mu)\omega_F \mapsto (ig_F)F_{\mu\nu}. \quad (72\text{-}1)$$

where $R^\rho{}_{\sigma\mu\nu} = (\partial_\nu \Gamma^\rho{}_{\sigma\mu} + \Gamma^\rho{}_{\lambda\nu}\Gamma^\lambda{}_{\sigma\mu}) - (\partial_\mu \Gamma^\rho{}_{\sigma\nu} + \Gamma^\rho{}_{\lambda\mu}\Gamma^\lambda{}_{\sigma\nu})$, $F_{\mu\nu} = D_\nu A_\mu - D_\mu A_\nu$, and $g_F$, $A_\mu$ are respectively the corresponding coupling constant and gauge field (that is defined generally as a Lie algebra-valued 1-form represented by a unique vector field [39]). Based on this natural unique isomorphism represented by the mappings (71-1) and (72-1), the column matrices $\Psi_R$ and $\Psi_F$ (in the expressions of field Equations (71) and (72), respectively) would be determined uniquely various dimensional space-times, represented by formulas (73)–(77), … .

In addition, as detailed in Section 3.4, the last entry of algebraic column matrix $S$ in matrix Equation (64) (as it has been shown in the relations (64)–(70)), is represented by the arbitrary algebraic parameter $s$. In the course of the derivation of field Equations (71) and (72) (via the first quantization procedure mentioned above, and the mappings (71-1) and (72-1)), the arbitrary parameter $s$ could be substituted solely by two covariant quantities $\phi^{(R)}_{\rho\sigma}$ and $\phi^{(F)}$ that define the corresponding covariant source currents $\phi^{(R)}_{\rho\sigma}$ and $J^{(F)}_\nu$ (given bythe field Equations (71) and (72), respectively) by the conditional relations: $J_{\rho\sigma\nu}^{(R)} = -(\nabla_\nu + \frac{im_0^{(R)}}{\hbar} k_\nu)\phi_{\rho\sigma}^{(R)}$ and $J_\nu^{(F)} = -(D_\nu + \frac{im_0^{(F)}}{\hbar} k_\nu)\phi^{(F)}$.

As another basic issue concerning the general covariance formulation of tensor field Equations (71) and (72), we should note that each of these Equations (as a system of equations) includes also an equation corresponding to the 2nd Bianchi identity, as follows, respectively:

$$(\nabla_\lambda + \frac{im_0^{(R)}}{\hbar} k_\lambda)R_{\rho\sigma\mu\nu} + (\nabla_\mu + \frac{im_0^{(R)}}{\hbar} k_\mu)R_{\rho\sigma\nu\lambda} + (\nabla_\nu + \frac{im_0^{(R)}}{\hbar} k_\nu)R_{\rho\sigma\lambda\mu} = 0, \tag{71-2}$$

$$(D_\lambda + \frac{im_0^{(F)}}{\hbar} k_\lambda)F_{\mu\nu} + (D_\mu + \frac{im_0^{(F)}}{\hbar} k_\mu)F_{\nu\lambda} + (D_\nu + \frac{im_0^{(F)}}{\hbar} k_\nu)F_{\lambda\mu} = 0 \tag{72-2}$$

However, the tensor field $R_{\rho\sigma\mu\nu}$ as the Riemann curvature tensor, obeys the relation (71-2) tensor, if and only if a torsion tensor is defined in as: $T_{\tau\mu\nu} = (im_0^{(R)}/2\hbar)(g_{\tau\mu}k_\nu - g_{\tau\nu}k_\mu)$, and subsequently the relation (71-2) be equivalent to the 2nd Bianchi identity of the Riemann tensor. Consequently, the covariant derivative $\nabla_\nu$ should be also defined with this torsion, that we may show it by $\breve{\nabla}_\nu$. Moreover, as presented in Section 3.6, concerning the relation (72-2), we may also define a torsion field as: $Z_{\tau\mu\nu} = (im_0^{(F)}/2\hbar)(g_{\tau\mu}k_\nu - g_{\tau\nu}k_\mu)$, and write the relations (71-2) and (72-2) (representing the 2nd Bianchi identities) as follows:

$$\breve{\nabla}_\lambda R_{\rho\sigma\mu\nu} + \breve{\nabla}_\mu R_{\rho\sigma\nu\lambda} + \breve{\nabla}_\nu R_{\rho\sigma\lambda\mu} = T^\tau{}_{\lambda\mu} R_{\rho\sigma\tau\nu} + T^\tau{}_{\mu\nu} R_{\rho\sigma\tau\lambda} + T^\tau{}_{\nu\lambda} R_{\rho\sigma\tau\mu}, \tag{71-2-a}$$

$$\breve{D}_\lambda F_{\mu\nu} + \breve{D}_\mu F_{\nu\lambda} + \breve{D}_\nu F_{\lambda\mu} = 0 \tag{72-2-a}$$

where the general relativistic form of gauge derivative $\breve{D}_\mu$ has been defined with torsion field $Z_{\tau\mu\nu}$. We use the derivatives $\breve{\nabla}_\mu$ and $\breve{D}_\mu$ in the ordinary tensor representations (i.e., the formulas (78-1)–(79-3)) of the field Equations (71) and (72) in Section 3.6. In addition, based on the formulations of torsions $T_{\tau\mu\nu}$ and $Z_{\tau\mu\nu}$ (that have appeared naturally in the course of derivation of the field Equations (71) and (72)) and the general properties of torsion tensors (in particular, the property that allows a torsion tensor to always be treated as an independent tensor field, or equivalently, as part of the space-time geometry [40–42]), it could be concluded directly that torsion field $T_{\tau\mu\nu}$ generates the invariant mass of corresponding gravitational field, and torsion field $Z_{\tau\mu\nu}$ generates the invariant mass of corresponding gauge field, respectively. Hence, based on our axiomatic derivation approach including the mappings (71-1) and (72-1) (mentioned above), the (1 + 1), (1 + 2), (1 + 3), (1 + 4), (1 + 5), …, dimensional cases of column matrices $\Psi_R$ and $\Psi_F$ in the specific expressions of general covariant massive (tensor) field Equations (71) and (72), are determined uniquely as follows, respectively; For (1 + 1)-dimensional space-time we have:

$$\Psi_R = \begin{bmatrix} R_{\rho\sigma 10} \\ \phi_{\rho\sigma}^{(R)} \end{bmatrix}, \quad \Psi_F = \begin{bmatrix} F_{10} \\ \phi^{(F)} \end{bmatrix}, \quad \begin{array}{l} J_{\rho\sigma\nu}^{(R)} = -(\breve{\nabla}_\nu + \frac{im_0^{(R)}}{\hbar} k_\nu)\phi_{\rho\sigma}^{(R)}, \\ J_\nu^{(F)} = -(\breve{D}_\nu + \frac{im_0^{(F)}}{\hbar} k_\nu)\phi^{(F)}; \end{array} \quad (73)$$

For (1 + 2)-dimensional space-time we obtain:

$$\Psi_R = \begin{bmatrix} R_{\rho\sigma 10} \\ R_{\rho\sigma 02} \\ R_{\rho\sigma 21} \\ \phi_{\rho\sigma}^{(R)} \end{bmatrix}, \quad \Psi_F = \begin{bmatrix} F_{10} \\ F_{02} \\ F_{21} \\ \phi^{(F)} \end{bmatrix}, \quad \begin{array}{l} J_{\rho\sigma\nu}^{(R)} = -(\breve{\nabla}_\nu + \frac{im_0^{(R)}}{\hbar} k_\nu)\phi_{\rho\sigma}^{(R)}, \\ J_\nu^{(F)} = -(\breve{D}_\nu + \frac{im_0^{(F)}}{\hbar} k_\nu)\phi^{(F)}; \end{array} \quad (74)$$

For (1 + 3)-dimensional space-time we have:

$$\Psi_R = \begin{bmatrix} R_{\rho\sigma 10} \\ R_{\rho\sigma 20} \\ R_{\rho\sigma 30} \\ 0 \\ R_{\rho\sigma 23} \\ R_{\rho\sigma 31} \\ R_{\rho\sigma 12} \\ \phi_{\rho\sigma}^{(R)} \end{bmatrix}, \quad \Psi_F = \begin{bmatrix} F_{10} \\ F_{20} \\ F_{30} \\ 0 \\ F_{23} \\ F_{31} \\ F_{12} \\ \phi^{(F)} \end{bmatrix}, \quad \begin{array}{l} J_{\rho\sigma\nu}^{(R)} = -(\breve{\nabla}_\nu + \frac{im_0^{(R)}}{\hbar} k_\nu)\phi_{\rho\sigma}^{(R)}, \\ J_\nu^{(F)} = -(\breve{D}_\nu + \frac{im_0^{(F)}}{\hbar} k_\nu)\phi^{(F)}; \end{array} \quad (75)$$

For (1 + 4)-dimensional space-time we get:

$$\Psi_R = \begin{bmatrix} R_{\rho\sigma 10} \\ R_{\rho\sigma 02} \\ R_{\rho\sigma 30} \\ 0 \\ R_{\rho\sigma 04} \\ 0 \\ 0 \\ 0 \\ 0 \\ R_{\rho\sigma 43} \\ R_{\rho\sigma 42} \\ R_{\rho\sigma 41} \\ R_{\rho\sigma 32} \\ R_{\rho\sigma 31} \\ R_{\rho\sigma 21} \\ \phi_{\rho\sigma}^{(R)} \end{bmatrix}, \quad \Psi_F = \begin{bmatrix} F_{10} \\ F_{02} \\ F_{30} \\ 0 \\ F_{04} \\ 0 \\ 0 \\ 0 \\ 0 \\ F_{43} \\ F_{42} \\ F_{41} \\ F_{32} \\ F_{31} \\ F_{21} \\ \phi^{(F)} \end{bmatrix}, \quad \begin{array}{l} J_{\rho\sigma\nu}^{(R)} = -(\breve{\nabla}_\nu + \frac{im_0^{(R)}}{\hbar} k_\nu)\phi_{\rho\sigma}^{(R)}, \\ J_\nu^{(F)} = -(\breve{D}_\nu + \frac{im_0^{(F)}}{\hbar} k_\nu)\phi^{(F)}; \end{array} \quad (76)$$

For (1 + 5)-dimensional space-time we obtain:

$$\Psi_R=\begin{bmatrix} R_{\rho\sigma10}\\ R_{\rho\sigma02}\\ R_{\rho\sigma30}\\ 0\\ R_{\rho\sigma04}\\ 0\\ 0\\ 0\\ 0\\ R_{\rho\sigma43}\\ R_{\rho\sigma42}\\ R_{\rho\sigma41}\\ R_{\rho\sigma32}\\ R_{\rho\sigma31}\\ R_{\rho\sigma21}\\ \phi^{(R)}_{\rho\sigma} \end{bmatrix},\quad \Psi_F=\begin{bmatrix} F_{10}\\ F_{02}\\ F_{30}\\ 0\\ F_{04}\\ 0\\ 0\\ 0\\ 0\\ F_{43}\\ F_{42}\\ F_{41}\\ F_{32}\\ F_{31}\\ F_{21}\\ \phi^{(F)} \end{bmatrix},\quad \begin{matrix} J^{(R)}_{\rho\sigma\nu}=-(\breve{\nabla}_\nu+\frac{im_0^{(R)}}{\hbar}k_\nu)\phi^{(R)}_{\rho\sigma},\\ J^{(F)}_\nu=-(\breve{D}_\nu+\frac{im_0^{(F)}}{\hbar}k_\nu)\phi^{(F)}; \end{matrix}\quad \Psi_R=\begin{bmatrix} R_{\rho\sigma10}\\ R_{\rho\sigma20}\\ R_{\rho\sigma30}\\ 0\\ R_{\rho\sigma40}\\ 0\\ 0\\ 0\\ R_{\rho\sigma50}\\ 0\\ 0\\ 0\\ 0\\ 0\\ 0\\ 0\\ 0\\ 0\\ 0\\ R_{\rho\sigma45}\\ 0\\ R_{\rho\sigma53}\\ R_{\rho\sigma25}\\ R_{\rho\sigma51}\\ 0\\ R_{\rho\sigma34}\\ R_{\rho\sigma42}\\ R_{\rho\sigma14}\\ R_{\rho\sigma32}\\ R_{\rho\sigma13}\\ R_{\rho\sigma21}\\ \phi^{(R)}_{\rho\sigma} \end{bmatrix},\quad \Psi_F=\begin{bmatrix} F_{10}\\ F_{20}\\ F_{30}\\ 0\\ F_{40}\\ 0\\ 0\\ 0\\ F_{50}\\ 0\\ 0\\ 0\\ 0\\ 0\\ 0\\ 0\\ 0\\ 0\\ 0\\ F_{45}\\ 0\\ F_{53}\\ F_{25}\\ F_{51}\\ 0\\ F_{34}\\ F_{42}\\ F_{14}\\ F_{32}\\ F_{13}\\ F_{21}\\ \phi^{(F)} \end{bmatrix},\quad \begin{matrix} J^{(R)}_{\rho\sigma\nu}=-(\breve{\nabla}_\nu+\frac{im_0^{(R)}}{\hbar}k_\nu)\phi^{(R)}_{\rho\sigma},\\ J^{(F)}_\nu=-(\breve{D}_\nu+\frac{im_0^{(F)}}{\hbar}k_\nu)\phi^{(F)} \end{matrix} \tag{77}$$

where in the relations (73)–(77), $J^{(R)}_{\nu\rho\sigma}$ and $J^{(F)}_\nu$ are the source currents expressed, necessarily, in terms of the covariant quantities $\phi^{(G)}_{\rho\sigma}$ and $\phi^{(F)}$ (as the initially given quantities), respectively. For higher-dimensional space-times, the column matrices $\Psi_R$ and $\Psi_F$ (with similar formulations) are determined uniquely as well.

### *3.6. General Covariant (Tensor) Representations of the Field Equations (71) and (72)*

From the Field Equations (71) and (72) (derived uniquely with certain matrix formalisms compatible with the Clifford and Weyl algebras), the following general covariant field equations, with ordinary tensor formalisms, are obtained (but not vice versa), respectively:

$$\begin{cases} \breve{\nabla}_\lambda R_{\rho\sigma\mu\nu}+\breve{\nabla}_\mu R_{\rho\sigma\nu\lambda}+\breve{\nabla}_\nu R_{\rho\sigma\lambda\mu}=T^{\tau}{}_{\lambda\mu}R_{\rho\sigma\tau\nu}+T^{\tau}{}_{\mu\nu}R_{\rho\sigma\tau\lambda}+T^{\tau}{}_{\nu\lambda}R_{\rho\sigma\tau\mu}\ , & \text{(78-1)}\\ \breve{\nabla}_\mu R_{\rho\sigma}{}^{\mu\nu}-(im_0^{(R)}/\hbar)k_\mu R_{\rho\sigma}{}^{\mu\nu}=-J^{(R)\nu}_{\rho\sigma} & \text{(78-2)} \end{cases}$$

$$R^{\rho}{}_{\sigma\mu\nu}=(\partial_\nu\Gamma^{\rho}{}_{\sigma\mu}+\Gamma^{\rho}{}_{\lambda\nu}\Gamma^{\lambda}{}_{\sigma\mu})-(\partial_\mu\Gamma^{\rho}{}_{\sigma\nu}+\Gamma^{\rho}{}_{\lambda\mu}\Gamma^{\lambda}{}_{\sigma\nu}),$$
$$J^{(R)}_{\rho\sigma\nu}=-(\breve{\nabla}_\nu+\frac{im_0^{(R)}}{\hbar}k_\nu)\phi^{(R)}_{\rho\sigma}\ ,\ T_{\tau\mu\nu}=\frac{im_0^{(R)}}{2\hbar}(g_{\tau\mu}k_\nu-g_{\tau\nu}k_\mu). \tag{78-3}$$

and

$$\begin{cases} \breve{D}_\lambda F_{\mu\nu} + \breve{D}_\mu F_{\nu\lambda} + \breve{D}_\nu F_{\lambda\mu} = 0, & (79\text{-}1) \\ \breve{D}_\mu F^{\mu\nu} = -J^{\nu(F)} & (79\text{-}2) \end{cases}$$

$$F_{\mu\nu} = \breve{D}_\nu A_\mu - \breve{D}_\mu A_\nu$$

$$J_\nu^{(F)} = -(\breve{D}_\nu + \frac{im_0^{(F)}}{\hbar} k_\nu)\phi^{(F)}, \; Z_{\tau\mu\nu} = \frac{im_0^{(F)}}{2\hbar}(g_{\tau\mu}k_\nu - g_{\tau\nu}k_\mu). \quad (79\text{-}3)$$

where $\Gamma^\rho{}_{\sigma\mu}$ is the affine connection: $\Gamma^\rho{}_{\sigma\mu} = \bar{\Gamma}^\rho{}_{\sigma\mu} - K^\rho{}_{\sigma\mu}$, $\bar{\Gamma}^\rho{}_{\sigma\mu}$ is the Christoffel symbol (or the torsion-free connection), $K^\rho{}_{\sigma\mu}$ is the contorsion tensor defined by: $K_{\rho\sigma\mu} = (im_0^{(R)}/2\hbar) g_{\rho\mu} k_\sigma$ (that is anti-symmetric in the first and last indices), $T_{\rho\sigma\mu}$ is the torsion given by: $T_{\rho\sigma\mu} = K_{\rho\mu\sigma} - K_{\rho\sigma\mu}$ (that generates the invariant mass of the gravitational field), $k_\mu = (c/\sqrt{g^{00}}, 0, ..., 0)$ (where we supposed $c = 1$) is the covariant velocity of particle (or the static observer) in the stationary reference frame, and $A_\mu$ is the gauge potential vector field. Moreover, in general covariant field Equations (79-1)–(79-3), the covariant derivative $\breve{D}_\mu$ has been defined specifically with the torsion field $Z_{\tau\mu\nu}$ (generating the invariant mass of gauge field strength tensor $F_{\mu\nu}$ ).

It should be emphasized again that the tensor field Equations (78-1)–(78-3) and (79-1)–(79-3) (which are obtained respectively from the original Equations (71) and (72), *but not* vice versa) show merely the general covariance formalism (including torsions fields $T^\tau{}_{\mu\nu}$ and $Z^\tau{}_{\mu\nu}$ ) of the axiomatically derived field Equations (71) and (72). The crucial issue here is that the original field Equations (71) and (72) could not be obtained from the tensor Equations (78-1)–(78-3) and (79-1)–(79-3). In fact, the tensor Equations (78-1)–(78-3) and (79-1)–(79-3) don't represent completely the definite matrix formalism (compatible with certain Clifford and Weyl algebras) of the axiomatic field Equations (71) and (72). Hence, based on this mathematical axiomatic formalism and derivation approach of Equations (71) and (72) (presented in Sections 3.3–3.5), it is concluded that the fundamental force fields of physics cannot be described completely via the ordinary tensor representations of these fields (in the current standard classic and quantum relativistic field theoretic formalism of physics), such as the representations (78-1)–(78-3) and (79-1)–(79-3); and as shown in Sections 3.3–3.5, on the basis of this new mathematical formalism, all the fundamental force fields of physics could be represented (and described) solely by the axiomatically determined and formulated field Equations (71) and (72) with their definite covariant matrix formalisms (given and specified by formulas (65)–(70) for various space-time dimensions, compatible with certain Clifford and Weyl algebras).

### *3.7. Derivation of the Einstein field Equations*

Along with the massive gravitational field Equations (78-1)–(78-3) (obtained uniquely from the originally derived field Equation (71)) that are expressed solely in terms of $R_{\mu\nu\rho\sigma}$ as the field strength tensor and also torsion's depended terms, we also assume the following relation as basic definition for the Ricci tensor (where the Riemann curvature tensor and Ricci tensor obey the interchange symmetries: $R_{\mu\nu\rho\sigma} \neq R_{\rho\sigma\mu\nu}$, $R_{\mu\nu} \neq R_{\nu\mu}$, because of the torsion [43]):

$$(\breve{\nabla}_\sigma + \frac{im_0^{(R)}}{\hbar} k_\sigma) R_{\mu\nu\rho}{}^{\sigma} = (\breve{\nabla}_\nu + \frac{im_0^{(R)}}{\hbar} k_\nu) R_{\mu\rho} - (\breve{\nabla}_\mu + \frac{im_0^{(R)}}{\hbar} k_\mu) R_{\nu\rho} \tag{78-4}$$

where the relation (78-4) particularly remains unchanged by the transformation:

$$R_{\mu\nu} \to R_{\mu\nu} + \Lambda g_{\mu\nu} \tag{78-5}$$

(where as would be shown, $\Lambda$ is equivalent to the cosmological constant). It should be noted that by taking $\Lambda = 0$, from the 2nd Bianchi identity of the Riemann curvature tensor and relation (78-4) it could be shown that the Ricci tensor is also the contraction of the Riemann tensor, i.e., $R_{\mu\nu} = R_{\mu\ \nu\sigma}^{\ \sigma}$ (which is equivalent to the ordinary definition of the Ricci tensor). However, this ordinary definition for the Ricci tensor, necessarily, doesn't imply the above transformation. In fact, in the following, we show that this basic transformation is necessary for having the cosmological constant in the gravitational field Equations (including the Einstein field equations which could be derived from the above equations and relations) expressed in terms of the Ricci and stress-energy tensors. As a direct result, a unique equivalent expression of gravitational field equations, in terms of the Ricci tensor $R_{\mu\nu}$ and stress-energy tensor $T_{\mu\nu}$, could be also determined from the basic definition (78-4) (for Ricci curvature tensor, based on this axiomatic formalism), and field Equations (78-1)–(78-3), along with the following expression for current $J_{\rho\sigma\nu}^{(R)}$ (defined in terms of the stress-energy tensor $T_{\mu\nu}$, $T(=T^{\mu}{}_{\mu})$, and metric $g_{\mu\nu}$, in $D$-dimensional space-time):

$$J_{\rho\sigma\nu}^{(R)} = -8\pi[(\nabla_\sigma + \frac{im_0^{(R)}}{\hbar} k_\sigma) T_{\rho\nu} - (\nabla_\rho + \frac{im_0^{(R)}}{\hbar} k_\rho) T_{\sigma\nu}] + 8\pi B[(\nabla_\sigma + \frac{im_0^{(R)}}{\hbar} k_\sigma) T g_{\rho\nu} - (\nabla_\rho + \frac{im_0^{(R)}}{\hbar} k_\rho) T g_{\sigma\nu}]' \tag{78-6}$$

where $T_{\mu\nu} \neq T_{\nu\mu}$ for $m_0^{(R)} \neq 0$, $B = 0$ for $D = 1, 2$, and $B = 1/(D-2)$ for $D \geq 3$, the Einstein field Equations (as the massless case) are determined directly as follows:

$$R_{\mu\nu} = -8\pi(T_{\mu\nu} - BTg_{\mu\nu}) - \Lambda g_{\mu\nu} \tag{78-7}$$

### *3.8. Showing that Magnetic Monopoles Cannot Exist in Nature*

As a direct consequence of the uniquely derived general covariant field Equation (72) that are specified by the matrices (73)–(77) and (65)–(70) (or the general covariant field Equations (79-1)–(79-3) obtained from the original Equation (72)), which, in fact, represent the electromagnetic fields equivalent to a generalized massive form of the Maxwell's Equations (as well as a generalized massive form of the Yang-Mills fields corresponding to certain gauge symmetry groups, see Section 3.15), *it is concluded straightforwardly that magnetic monopoles could not exist in nature.*

### *3.9. Gauge Invariance of (Massive) Tensor Field Equations(79-1)–(79-3) Corresponding to the Maxwell's (and Yang-Mills) Equations in (1 + 3) Dimensions, and Dirac Equation in (1 + 2) Dimensions*

The Lagrangian density specified for the tensor field $F_{\mu\nu}$ in the field Equations (79-1)–(79-3) is (supposing $J_\nu^{(F)} = 0$) [39]:

$$L^{(F)} = -(1/4\sqrt{-g}) F^{*\mu\nu} F_{\mu\nu} \tag{80}$$

where $g$ is the metric's determinant. Moreover, the trace part of torsion field $Z_{\tau\mu\nu}$ in (79-3) is obtained as:

$$Z^{\mu}{}_{\mu\nu} = Z_{\nu} = N(im_0^{(F)}/2\hbar)k_{\nu} = N\alpha k_{\nu} \quad (81)$$

where (1 + $N$) is the number of space-time dimensions and $\alpha = \frac{im_0^{(F)}}{2\hbar}$. Now based on the definition of covariant vector $k_{\mu}$ (as a time-like covariant vector), we simply get: $\exists\phi : k_{\nu} = \partial_{\nu}\phi$. This basic property, along with and formula (81), imply the general covariant massive field Equations (79-1)–(79-3) (formulated originally with the torsion field (79-3) generating the invariant mass $m_0^{(F)}$ of field $F_{\mu\nu}$ ), and the corresponding Lagrangian density (80), be invariant under the U(1) Abelian gauge group [17,39,44–48]. However, in Section 3.15, we show that assuming the spin-1/2 fermion fields (describing generally by the field Equation (110-9) compatible with specific gauge symmetry group (110-12), as shown in Section 3.15, formulas (110-9)–(110-12)) and their compositions as the source currents of the (1 + 3)-dimensional cases of general covariant massive field Equation (72) (describing the spin-1 boson field), then this field equation would be invariant under two types of gauge symmetry groups, including: $SU(2)_L \otimes U(2)_R$ and SU(3), corresponding with a group of seven bosons and a groups of eight bosons (as shown in Section 3.15, formulas (114-4)–(114-9)).

### *3.10. Identifying a New Particular Massive Gauge Boson*

According to Refs. [44–48], in agreement with the recent astronomical data, we can directly establish a lower bound for a constant quantity which is equivalent to the constant $\alpha = \frac{im_0^{(F)}}{2\hbar}$ (defined by the relation (80)) as: $|\alpha| \geq 21$. Hence, a new massive particle (corresponding to the U(1) symmetry group) would be identified with the invariant mass:

$$m_{\gamma} \approx 4.90571 \times 10^{-50}\,\text{kg} \quad (82)$$

that is generated by a coupling torsion field of the type (79-3) of the background curved space-time. In addition, it should be noted that, in general, based on the covariant massive field Equations (71) and (72) derived by our axiomatic approach (or field Equations (78-1)–(78-3) and (79-1)–(79-3) obtained from (71) and (72)), the invariant masses of the elementary particles are generated by torsion fields of the types (78-3) (for spin-3/2 and spin-2 particles) and (79-3) (for spin-1/2 and spin-1 particles, see Section 3.15). Hence, this approach could be also applied for massive neutrinos concluding that their masses are generated by the coupling torsion fields (of the type (79-3)). Such massive particle fields coupled with the torsions (of the type (79-3)) of the background space-time geometry could be completely responsible for the mysteries of dark energy and dark matter [49,50].

### *3.11. Quantum Representations of C, P and T Symmetries of the Axiomatically Derived General Covariant Massive (tensor) Field Equations (71) and (72)*

As shown in Sections 3.3–3.5, the general covariant massive (tensor) field Equations (71) and (72) as the unique axiomatically determined Equations (representing the fundamental field of physics, as assumed in Section 3.1), are represented originally with definite matrix formalisms constructed from the combination of two specific matrix classes including the column matrices (73)–(77),… compatible with the Weyl algebras (based on the isomorphism (71-1)–(72-1)), and the square matrices (65)–(70), … that are compatible with the Clifford algebras and their generalizations; see Sections 3.3–3.5 and also Appendix A for detail).

In agreement with the principles of relativistic quantum theory [6], and also as another primary assumption in addition to the basic assumption (3) defined in Section 3.1, we basically represent the *C, P* and *T symmetries* of the source-free cases of by the following quantum matrix operators (with the same forms in both flat and curved space-time), respectively:

(**Note**: In Section 3.13, we show that only a certain simultaneous combination of the C, P and T transformations could be defined for the field Equations (71) and (72) with non-zero source currents.)

**(1)- Parity Symmetry (P-Symmetry)**:

$$\hat{P} = \gamma^P = \begin{bmatrix} -I & 0 \\ 0 & I \end{bmatrix} \tag{83}$$

where $I$ is the identity matrix, and the size of matrix $\gamma^P$ in (1 + *N*)-dimensional space-time is $2^N \times 2^N$. The operator $\hat{P}$ obeys the relations:

$$\det(\hat{P}) = -1,\ \hat{P}^2 = 1,\ \hat{P} = \hat{P}^{-1} = \hat{P}^* = \hat{P}^T \tag{83-1}$$

**(2)- Time-Reversal Symmetry (T-Symmetry)**:

$$\hat{T} = \hat{T}_0 \hat{K} = i\gamma^P \gamma^{Ch} \hat{K} \tag{84}$$

where the operator $\hat{K}$ denotes complex conjugation, the operator $\gamma^P$ defined by formula (83) and the operator $\gamma^{Ch}$ in (1 + 1) and (1 + 2) space-time dimensions, is given by:

$$\gamma^{Ch} = \begin{bmatrix} 0 & I \\ I & 0 \end{bmatrix}, \tag{84-1}$$

and in (1 + 3) and higher space-time dimensions, $\gamma^{Ch}$ is denoted by:

$$\gamma^{Ch} = \begin{bmatrix} 0 & iI \\ -iI & 0 \end{bmatrix} \tag{84-2}$$

where the size of matrix $\gamma^{Ch}$ in (1 + *N*)-dimensional space-time is: $2^N \times 2^N$. Moreover, in (1 + 1) and (1 + 2) space-time dimensions, the time reversal operator $\hat{T}$ (84) and the Hermitian operator $\hat{T}_0 = i\gamma^P \gamma^{Ch}$ (specified in the formula (84)) obey the relations:

$$\hat{T}^2 = -1,\ \hat{T}_0 = \hat{T}_0^{-1} = \hat{T}_0^* = -\hat{T}_0^T, \tag{84-3}$$

and in (1 + 3) and higher space-time dimensions, $\hat{T}$ and $\hat{T}_0$ obey the relations:

$$\hat{T}^2 = 1,\ \hat{T}_0 = \hat{T}_0^{-1} = \hat{T}_0^* = \hat{T}_0^T \tag{84-4}$$

Concerning the time reversal symmetry, it should be noted that relations (84-3) are solely compatible with the fermionic fields and relations (84-4) are solely compatible with the bosonic fields. In addition, it should be noted that these basic quantum mechanical properties (i.e., the relations (84-3) and (84-4)) of the time reversal symmetry (84), are fully compatible with corresponding properties of the field tensors $F_{\mu\nu}$ and $R_{\rho\sigma\mu\nu}$ presented in Section 3.15, where the tensor field $F_{\mu\nu}$ (describing by general covariant field Equation (72)) represents (asymptotically) solely a massive bispinor field of spin-1/2 particles (as a new general covariant massive formulation of the Dirac equation) in (1 + 2) space-time dimensions, and also represents a massive bispinor field of spin-1—as new massive general covariant (matrix) formulations of both Maxwell and Yang-Mills field equations are compatible with specified gauge symmetry groups—in (1 + 3) space-time dimensions; and tensor field $R_{\rho\sigma\mu\nu}$ (describing by general covariant field Equation (71)) represents (asymptotically) solely a bispinor field of spin-3/2 particles (as a new massive general covariant form of the Rarita–Schwinger equation) in (1 + 2) space-time dimensions , and also represents a massive bispinor field spin-2 particles (equivalent to a generalized massive form of the Einstein equations) in (1 + 3) space-time dimensions.

**(3)- Charge Conjugation Symmetry (C-Symmetry)**:

$$(\Psi_R)_C = \tilde{C}\Psi_R = iI\hat{K}\Psi_R,\ (\Psi_F)_C = \tilde{C}\Psi_F = iI\hat{K}\Psi_F \tag{85}$$

where $\tilde{C} = iI\hat{K}$ , $I$ is the identity matrix, the operator $\hat{K}$ denotes complex conjugation, and the charge conjugation operator $\hat{C}$ defined by: $(\Psi_R)_C = \hat{C}(\bar{\Psi}_R)^{\mathsf{T}}$, $(\Psi_F)_C = \hat{C}(\bar{\Psi}_F)^{\mathsf{T}}$. The charge conjugation operator $\hat{C}$ obeys the following relations:

$$\hat{C}\hat{C}^{*} = 1,\ \hat{C} = -\hat{C}^{-1} = -\hat{C}^{*} = \hat{C}^{\mathsf{T}} \tag{85-1}$$

As a basic additional issue, it is worth noting that the time-reversal operator (84) could be also expressed basically in terms of the parity matrix operator $\gamma^{P}$ (83), matrix operator $\tilde{C} = iI\hat{K}$ given bythe definition of charge-conjugated transformation (85), and matrix $\gamma^{Ch}$ defined by formulas (84-1) and (84-2), as follows:

$$\gamma^{P}\gamma^{Ch}\tilde{C} = \hat{T} \tag{85-2}$$

where we have: $\gamma^{P}\gamma^{Ch} = -\gamma^{Ch}\gamma^{P}$.

Remark 2. The Main Properties of Matrix Operator $\gamma^{Ch}$ (Defined by Unitary Matrices (84-1) and (84-2))

In this Section, have been presented. Each of thegeneral covariant (tensor) field Equations (71) and (72) (including their source-free and non-source-free cases), as a system of differential equations, is symmetric and has the same spectrum as a result of multiplying by matrix $\gamma^{Ch}$ . The multiplied column matrices $\Psi_R^{(Ch)} = \gamma^{Ch}\Psi_R$ and $\Psi_F^{(Ch)} = \gamma^{Ch}\Psi_F$ then obey the Equations (71) and (72), respectively, but with opposite sign in mass term such that: $(i\hbar\alpha^{\mu}\nabla_{\mu} + m_0^{(R)}\tilde{\alpha}^{\mu}k_{\mu})\Psi_R^{(Ch)} = 0$, $(i\hbar\alpha^{\mu}\nabla_{\mu} + m_0^{(F)}\tilde{\alpha}^{\mu}k_{\mu})\Psi_F^{(Ch)} = 0$.

As a general additional issue concerning the column matrices $\Psi_R^{(Ch)} = \gamma^{Ch}\Psi_R$ and $\Psi_F^{(Ch)} = \gamma^{Ch}\Psi_F$, should be also added that the sign change of the mass terms introduced in the field Equations (71) and (72) is immaterial (the same property also holds for the ordinary formulation of Dirac equation, and so on [51]). In other words, the field Equation (71) of the form $(i\hbar\alpha^{\mu}\nabla_{\mu} \pm m_0^{(R)}\tilde{\alpha}^{\mu}k_{\mu})\Psi_R = 0$ are equivalent, and similarly the field Equation (72) of the form $(i\hbar\alpha^{\mu}\nabla_{\mu}\pm m_0^{(F)}\tilde{\alpha}^{\mu}k_{\mu})\Psi_F=0$ would be equivalent as well. However, since the algebraic column matrix $S$ in the matrix Equation (64) (derived and represented uniquely in terms of the matrices (66)–(70),… corresponding to various space-time dimensions), is not symmetric by multiplying by matrix $\gamma^{Ch}$ (84-1) and (84-2) (except for (1 + 2) and (1 + 3)-dimensional cases of column matrix $S$, based on the definite algebraic properties of matrix $S$ presented in Remark 1, (in Section 3.3), it is concluded that except the (1 + 2) and (1 + 3)-dimensional cases of the fundamental field Equations (71) and (72), these field equations could not be defined with the column matrices of the types $\Psi_R^{(Ch)}(=\gamma^{Ch}\Psi_R)$ and $\Psi_F^{(Ch)}(=\gamma^{Ch}\Psi_F)$ (if assuming that the column matrices $\Psi_R$ and $\Psi_F$ are defined with field Equations (71) and (72), i.e., they have the formulations similar to the formulations of originally derived column matrices (73)–(77),… corresponding to various space-time dimensions). This conclusion follows from this fact that the filed Equations (71) and (72) have been derived (and defined) uniquely from the matrix Equation (64) via the axiomatic derivation approach (including the first quantization procedure) presented in Sections 3.4 and 3.5. In Section 3.12, using this property (i.e., multiplication of column matrices $\Psi_R$ and $\Psi_F$ defined in the fundamental field Equations (71) and (72), by matrix $\gamma^{Ch}$ (84-1) and (84-2) from the left), this crucial and essential issue would be concluded directly that by assuming the time-reversal invariance of the general covariant filed Equations (71) and (72) (represented by the transformations $\hat{T}\Psi_R$ and $\hat{T}\Psi_F$, where the quantum operator $\hat{T}$ is given uniquely by formula (84), i.e.,: $\hat{T}=\hat{T}_0\hat{K}=i\gamma^P\gamma^{Ch}\hat{K}$), these fundamental field equations could be defined solely in (1 + 2) and (1 + 3) space-time dimensions (with the column matrices of the forms (96-1) and (98-2), respectively). Subsequently, in Section 3.13, it would be also shown that only a definite simultaneous combination of all the transformations $\hat{C}$, $\hat{P}$, $\hat{T}$ and also matrix $\gamma^{Ch}$ (given by quantum operators (83)–(87)) could be defined for the field Equations (71) and (72) with non-zero source currents. In addition, the matrix operator $\gamma^{Ch}$ in (1 + 1) and (1 + 2) space-time dimensions obeys the relations:

$$(\gamma^{Ch})^2=1,\ \gamma^{Ch}=(\gamma^{Ch})^{-1}=(\gamma^{Ch})^*=(\gamma^{Ch})^T, \tag{86}$$

and in the (1 + 3) and higher dimensions obeys the following relations as well:

$$(\gamma^{Ch})^2=1,\ \gamma^{Ch}=(\gamma^{Ch})^{-1}=(\gamma^{Ch})^*=-(\gamma^{Ch})^T \tag{87}$$

Furthermore, in Section 3.14, the matrix $\gamma^{Ch}$ would be also used basically for defining and representing the left-handed and right-handed components of the column field matrices $\Psi_R$ and $\Psi_F$ defined originally in the field Equations (71) and (72).

### *3.12. Showing That the Universe Could Be Realized Solely by the (1 + 2) and (1 + 3)-Dimensional Space-Times*

The proof of this essential property of nature within the new mathematical axiomatic formalism presented in this article is mainly based on the T-symmetry (represented by quantum matrix operators (84)) of the fundamentally derived general covariant field Equations (71) and (72). As shown in Section 3.11, the source-free cases (as basic cases) of field Equations (71) and (72) are invariant under the time-reversal transformation defined by matrix operator (84). Moreover, in Section 3.13, it would be also shown that these field equations with non-zero source currents are solely invariant under the simultaneous transformations of all the $\hat{C}$, $\hat{P}$, and $\hat{T}$ (83)–(85), multiplied by matrix $\gamma^{Ch}$ (given by formulas (84-1) and (84-2)). Now, following the definite mathematical formalism of the axiomatic derivation approach of fundamental field Equations (71) and (72), assuming that any column matrix $X_R$, or $Y_F$, expressible in the tensor formulation of general covariant field Equation (71) or (72), is basically definable, if and only if, it could be also derived originally as a column matrix via the axiomatic derivation approach presented in Sections 3.4 and 3.5.

On this basis, it could be shown that the time-reversal transformed forms of the column matrices $\Psi_R$ and $\Psi_F$ given bythe expressions of source-free cases of field Equations (71) and (72), are definable solely in (1 + 2) and (1 + 3)-dimensional space-times. Based on this result, along with the basic assumption (3) in Section 3.1, it is concluded directly that the universe could be realized solely with the (1 + 2) and (1 + 3)-dimensional space-times. We show this in the following in detail.

As noted, in fact, the above conclusion follows directly from the formulations of uniquely determined time-reversal transformed forms of column matrices $\Psi_R$ and $\Psi_F$ given by the expressions of source-free cases of field Equations (71) and (72). Denoting these column matrices by $\Upsilon_R = \hat{T}\Psi_R$ and $\Upsilon_F = \hat{T}\Psi_F$, where the time-reversal operator (84) is defined by: $\hat{T} = \hat{T}_0\hat{K} = i\gamma^P\gamma^{Ch}\hat{K}$, they would be determined as follows in various dimensions:

For (1 + 1)-dimensional space-time we have:

$$\Upsilon_R(\boldsymbol{x},t) = \hat{T}\Psi_R(\boldsymbol{x},t) = \hat{T}_0\Psi_R^*(\boldsymbol{x},t) = \begin{bmatrix} 0 \\ iR^*_{\rho\sigma 10}(\boldsymbol{x},t) \end{bmatrix},\ \Upsilon_F(\boldsymbol{x},t) = \hat{T}\Psi_F(\boldsymbol{x},t) = \hat{T}_0\Psi_F^*(\boldsymbol{x},t) = \begin{bmatrix} 0 \\ iF^*_{10}(\boldsymbol{x},t) \end{bmatrix}; \quad (88)$$

For (1 + 2)-dimensional space-time it is obtained:

$$\Upsilon_R(\boldsymbol{x},t) = \hat{T}\Psi_R(\boldsymbol{x},t) = \hat{T}_0\Psi_R^*(\boldsymbol{x},t) = \begin{bmatrix} -iR^*_{\rho\sigma 21}(\boldsymbol{x},t) \\ 0 \\ iR^*_{\rho\sigma 10}(\boldsymbol{x},t) \\ iR^*_{\rho\sigma 02}(\boldsymbol{x},t) \end{bmatrix},\ \Upsilon_F(\boldsymbol{x},t) = \hat{T}\Psi_F(\boldsymbol{x},t) = \hat{T}_0\Psi_F^*(\boldsymbol{x},t) = \begin{bmatrix} -iF^*_{21}(\boldsymbol{x},t) \\ 0 \\ iF^*_{10}(\boldsymbol{x},t) \\ iF^*_{02}(\boldsymbol{x},t) \end{bmatrix}; \quad (89)$$

For (1 + 3)-dimensional space-time we get:

$$\Upsilon_R(\boldsymbol{x},t) = \hat{T}\Psi_R(\boldsymbol{x},t) = \hat{T}_0\Psi_R^*(\boldsymbol{x},t) = \begin{bmatrix} -iR^*_{\rho\sigma 23}(\boldsymbol{x},t) \\ -iR^*_{\rho\sigma 31}(\boldsymbol{x},t) \\ -iR^*_{\rho\sigma 12}(\boldsymbol{x},t) \\ 0 \\ iR^*_{\rho\sigma 10}(\boldsymbol{x},t) \\ iR^*_{\rho\sigma 20}(\boldsymbol{x},t) \\ iR^*_{\rho\sigma 30}(\boldsymbol{x},t) \\ 0 \end{bmatrix},\ \Upsilon_F(\boldsymbol{x},t) = \hat{T}\Psi_F(\boldsymbol{x},t) = \hat{T}_0\Psi_F^*(\boldsymbol{x},t) = \begin{bmatrix} -iF^*_{23}(\boldsymbol{x},t) \\ -iF^*_{31}(\boldsymbol{x},t) \\ -iF^*_{12}(\boldsymbol{x},t) \\ 0 \\ iF^*_{10}(\boldsymbol{x},t) \\ iF^*_{20}(\boldsymbol{x},t) \\ iF^*_{30}(\boldsymbol{x},t) \\ 0 \end{bmatrix}; \quad (90)$$

For (1 + 4)-dimensional space-time we have:

$$\Upsilon_R(\boldsymbol{x},t)=\hat{T}\Psi_R(\boldsymbol{x},t)=\hat{T}_0\Psi_R^*(\boldsymbol{x},t)=\begin{bmatrix}0\\-iR^*_{\rho\sigma43}(\boldsymbol{x},t)\\-iR^*_{\rho\sigma42}(\boldsymbol{x},t)\\-iR^*_{\rho\sigma41}(\boldsymbol{x},t)\\-iR^*_{\rho\sigma32}(\boldsymbol{x},t)\\-iR^*_{\rho\sigma31}(\boldsymbol{x},t)\\-iR^*_{\rho\sigma21}(\boldsymbol{x},t)\\0\\iR^*_{\rho\sigma10}(\boldsymbol{x},t)\\iR^*_{\rho\sigma02}(\boldsymbol{x},t)\\iR^*_{\rho\sigma30}(\boldsymbol{x},t)\\0\\iR^*_{\rho\sigma04}(\boldsymbol{x},t)\\0\\0\\0\end{bmatrix},\quad \Upsilon_F(\boldsymbol{x},t)=\hat{T}\Psi_F(\boldsymbol{x},t)=\hat{T}_0\Psi_F^*(\boldsymbol{x},t)=\begin{bmatrix}0\\-iF^*_{43}(\boldsymbol{x},t)\\-iF^*_{42}(\boldsymbol{x},t)\\-iF^*_{41}(\boldsymbol{x},t)\\-iF^*_{32}(\boldsymbol{x},t)\\-iF^*_{31}(\boldsymbol{x},t)\\-iF^*_{21}(\boldsymbol{x},t)\\0\\iF^*_{10}(\boldsymbol{x},t)\\iF^*_{02}(\boldsymbol{x},t)\\iF^*_{30}(\boldsymbol{x},t)\\0\\iF^*_{04}(\boldsymbol{x},t)\\0\\0\\0\end{bmatrix}; \qquad (91)$$

For (1 + 5)-dimensional space-time we obtain:

$$\Upsilon_R(\boldsymbol{x},t)=\hat{T}\Psi_R(\boldsymbol{x},t)=\hat{T}_0\Psi_R^*(\boldsymbol{x},t)=\begin{bmatrix}0\\0\\0\\-iR^*_{\rho\sigma45}(\boldsymbol{x},t)\\0\\-iR^*_{\rho\sigma53}(\boldsymbol{x},t)\\-iR^*_{\rho\sigma25}(\boldsymbol{x},t)\\-iR^*_{\rho\sigma51}(\boldsymbol{x},t)\\0\\-iR^*_{\rho\sigma34}(\boldsymbol{x},t)\\-iR^*_{\rho\sigma42}(\boldsymbol{x},t)\\-iR^*_{\rho\sigma14}(\boldsymbol{x},t)\\-iR^*_{\rho\sigma32}(\boldsymbol{x},t)\\-iR^*_{\rho\sigma13}(\boldsymbol{x},t)\\-iR^*_{\rho\sigma21}(\boldsymbol{x},t)\\0\\iR^*_{\rho\sigma10}(\boldsymbol{x},t)\\iR^*_{\rho\sigma20}(\boldsymbol{x},t)\\iR^*_{\rho\sigma30}(\boldsymbol{x},t)\\0\\iR^*_{\rho\sigma40}(\boldsymbol{x},t)\\0\\0\\0\\iR^*_{\rho\sigma50}(\boldsymbol{x},t)\\0\\0\\0\\0\\0\\0\\0\end{bmatrix},\quad \Upsilon_F(\boldsymbol{x},t)=\hat{T}\Psi_F(\boldsymbol{x},t)=\hat{T}_0\Psi_F^*(\boldsymbol{x},t)=\begin{bmatrix}0\\0\\0\\-iF^*_{45}(\boldsymbol{x},t)\\0\\-iF^*_{53}(\boldsymbol{x},t)\\-iF^*_{25}(\boldsymbol{x},t)\\-iF^*_{51}(\boldsymbol{x},t)\\0\\-iF^*_{34}(\boldsymbol{x},t)\\-iF^*_{42}(\boldsymbol{x},t)\\-iF^*_{14}(\boldsymbol{x},t)\\-iF^*_{32}(\boldsymbol{x},t)\\-iF^*_{13}(\boldsymbol{x},t)\\-iF^*_{21}(\boldsymbol{x},t)\\0\\iF^*_{10}(\boldsymbol{x},t)\\iF^*_{20}(\boldsymbol{x},t)\\iF^*_{30}(\boldsymbol{x},t)\\0\\iF^*_{40}(\boldsymbol{x},t)\\0\\0\\0\\iF^*_{50}(\boldsymbol{x},t)\\0\\0\\0\\0\\0\\0\\0\end{bmatrix}; \qquad (92)$$

Now based on the formulations of the derived time-reversal transformed column matrices $\Upsilon_R$ and $\Upsilon_F$ (88)–(92), although they could be expressed merely by the tensor formulations of field Equations (71) and (72), however, except the (1 + 2) and (1 + 3)-dimensional cases of these transformed column matrices, all the other cases cannot be derived originally as a column matrix via the axiomatic derivation approach presented in Sections 3.4 and 3.5 (following the formulation of originally derived column matrices (73)–(77)). Below this conclusion (and subsequent remarkable results) is discussed in more detail.

In addition, it is also worth noting that on the basis of our derivation approach, since there are not the corresponding isomorphism (that could be represented by the unique mappings (71-1) and (72-1), in Section 3.5) between the entries of column matrices $\Upsilon_R$ and $\Upsilon_F$ (88), (91), (92),… and the entries (with the exactly same indices) of column matrix $S$ (in the algebraic matrix Equation (64), where its last entry, i.e., arbitrary parameter "$s$" is zero compatible with the source-free cases of the field Equations (71) and (72)) that are given uniquely as follows in (1 + 1) and (1 + 4), (1 + 5),... and higher space-time dimensions, respectively, using the definitions (66)–(70),… (in Section 3.3), and also the algebraic properties of column matrix $S$ (presented in Remark 1 in Section 3.3) representing in terms of two half-sized $2^{N-1}\times 1$ column matrices $S'$ and $S''$ such that: $S=\begin{bmatrix} S' \\ S'' \end{bmatrix}$ (where $u_0,u_1,u_2,u_3,\ldots,u_N,\ v_0,v_1,v_2,v_3,\ldots,v_N,w$ are arbitrary parameters):

For (1 + 1)-dimensional space-time we have:

$$S=\begin{bmatrix} S' \\ S'' \end{bmatrix}=\begin{bmatrix} (u_0v_1-u_1v_0)w \\ 0 \end{bmatrix}; \tag{93}$$

For (1 + 4), (1 + 5)-dimensional space-times we get, respectively:

$$S=\begin{bmatrix} S' \\ S'' \end{bmatrix},\ S'=\begin{bmatrix} (u_0v_1-u_1v_0)w \\ (u_2v_0-u_0v_2)w \\ (u_0v_3-u_3v_0)w \\ 0 \\ (u_4v_0-u_0v_4)w \\ 0 \\ 0 \\ 0 \end{bmatrix},\ S''=\begin{bmatrix} 0 \\ (u_3v_4-u_4v_3)w \\ (u_2v_4-u_4v_2)w \\ (u_1v_4-u_4v_1)w \\ (u_2v_3-u_3v_2)w \\ (u_1v_3-u_3v_1)w \\ (u_1v_2-u_2v_1)w \\ 0 \end{bmatrix};\ S=\begin{bmatrix} S' \\ S'' \end{bmatrix},S'=\begin{bmatrix} (u_0v_1-u_1v_0)w \\ (u_0v_2-u_2v_0)w \\ (u_0v_3-u_3v_0)w \\ 0 \\ (u_0v_4-u_4v_0)w \\ 0 \\ 0 \\ 0 \\ (u_0v_5-u_5v_0)w \\ 0 \\ 0 \\ 0 \\ 0 \\ 0 \\ 0 \\ 0 \end{bmatrix},S''=\begin{bmatrix} 0 \\ 0 \\ 0 \\ (u_5v_4-u_4v_5)w \\ 0 \\ (u_3v_5-u_5v_3)w \\ (u_5v_2-u_2v_5)w \\ (u_1v_5-u_5v_1)w \\ 0 \\ (u_4v_3-u_3v_4)w \\ (u_2v_4-u_4v_2)w \\ (u_4v_1-u_1v_4)w \\ (u_2v_3-u_3v_2)w \\ (u_3v_1-u_1v_3)w \\ (u_1v_2-u_2v_1)w \\ 0 \end{bmatrix};\ \ldots\ , \tag{94}$$

it would be directly concluded that in (1 + 1) and (1 + 4), (1 + 5),... and higher space-time dimensions, the column matrices $\Upsilon_R$ and $\Upsilon_F$ could not be defined as the column matrices in unique formulations of the axiomatically derived general covariant field Equations (71) and (72). In other words, for the (1 + 2)-dimensional cases of the transformed column matrices $\Upsilon_R$ and $\Upsilon_F$ (89), the corresponding isomorphism (represented uniquely by the mappings (71-1) and (72-1)) could be defined between the components of these matrices and the entries of column matrix $S$ (67), for $s = 0$ (compatible with $\phi_{\rho\sigma}^{(R)} = 0$, $\phi^{(F)} = 0$), if and only if: $iR_{\rho\sigma 02}^{*}(x,t) = 0$ and $iF_{02}^{*}(x,t) = 0$. This could be shown as follows:

$$\left( S = \begin{bmatrix} (u_0 v_1 - u_1 v_0)w \\ (u_2 v_0 - u_0 v_2)w \\ (u_1 v_2 - u_2 v_1)w \\ 0 \end{bmatrix} \overset{\substack{\text{Derivation Procedure} \\ \text{(First Quantization)}}}{\mapsto} \Upsilon_R = \begin{bmatrix} -iR_{\rho\sigma 21}^{*}(x,t) \\ 0 \\ iR_{\rho\sigma 10}^{*}(x,t) \\ iR_{\rho\sigma 02}^{*}(x,t) \end{bmatrix}, \Upsilon_F = \begin{bmatrix} -iF_{21}^{*}(x,t) \\ 0 \\ iF_{10}^{*}(x,t) \\ iF_{02}^{*}(x,t) \end{bmatrix} \right) \Rightarrow \begin{matrix} u_2 v_0 - u_0 v_2 = 0, \\ iR_{\rho\sigma 02}^{*}(x,t) = 0, \; iF_{02}^{*}(x,t) = 0. \end{matrix} \tag{95}$$

where for appeared parametric condition: $u_2 v_0 - u_0 v_2 = 0$, as it would be shown in Remark 3 (in Section 3.12), it could be supposed solely: $u_2 = v_2$, $v_0 = u_0$, implying conditions: $R_{\rho\sigma 02} = 0$ and $F_{02} = 0$, which could be assumed for the field strength tensors $R_{\rho\sigma\mu\nu}$ and $F_{\mu\nu}$ in (1 + 2)-dimensional space-time (without vanishing these tensor fields), based on their basic definitions given by formulas (71-1) and (72-1).

Hence, definite mathematical framework of our axiomatic derivation approach (presented in Section 3.4), in addition to the time-reversal invariance (represented by the quantum operator (84)) of source-free cases of general covariant field Equations (71) and (72), imply the (1 + 2)-dimensional case of column matrices $\Psi_R$ and $\Psi_F$ given by relations (74) (where we assumed $\phi_{\rho\sigma}^{(R)} = 0$, $\phi^{(F)} = 0$), could be given solely as follows, to be compatible with the above assumed conditions (i.e., being compatible with the mathematical framework of axiomatic derivation of field Equations (71) and (72), and also the time-reversal invariance defined by quantum operator (84)), and consequently, as the column matrices could be defined in the formulations of the fundamental tensor field Equations (71) and (72), respectively:

$$\Psi_R = \begin{bmatrix} R_{\rho\sigma 10} \\ 0 \\ R_{\rho\sigma 21} \\ 0 \end{bmatrix}, \quad \Psi_F = \begin{bmatrix} F_{10} \\ 0 \\ F_{21} \\ 0 \end{bmatrix} \tag{96}$$

The formulations (96) that are represented the column matrices $\Psi_R$ and $\Psi_F$ in the field Equations (71) and (72) compatible with the above basic conditions, are also represented these matrices in the field Equations (71) and (72) with non-zero source currents compatible with two basic conditions (similar to above conditions) including a unique combination of the C, P and T symmetries (that have been represented by quantum operators (83)–(85)) for these cases of field Equations (71) and (72), and also the mathematical framework of axiomatic derivation of Equations (71) and (72). In fact, as it has been shown in Remark 2 (in Section 3.11), the field Equations (71) and (72) with non-zero source currents could have solely a certain combination (given by formulas (86) and (87)) of the C, P and T symmetries (that are represented by the operators (83), (84) and (85)). This unique combined symmetry in addition to the unique formulations (96) of source-free cases of column matrices $\Psi_R$ and $\Psi_F$ in (1 + 2)-dimensional space-time, implies these matrices could take solely the following forms to be defined in the formulations of the fundamental tensor field Equations (71) and (72) (with non-zero source currents):

$$\Psi_R = \begin{bmatrix} R_{\rho\sigma 10} \\ 0 \\ R_{\rho\sigma 21} \\ \phi_{\rho\sigma}^{(R)} \end{bmatrix}, \quad \Psi_F = \begin{bmatrix} F_{10} \\ 0 \\ F_{21} \\ \phi^{(F)} \end{bmatrix}, \quad \begin{matrix} J_{\rho\sigma\nu}^{(R)} = -(\nabla_\nu + \dfrac{i m_0^{(R)}}{\hbar} k_\nu) \phi_{\rho\sigma}^{(R)}, \\ J_\nu^{(F)} = -(D_\nu + \dfrac{i m_0^{(F)}}{\hbar} k_\nu) \phi^{(F)} \end{matrix} \tag{96-1}$$

In the same manner, concerning the (1 + 4)-dimensional cases of column matrices $\Upsilon_R$ and $\Upsilon_F$ (91), there would be a mapping between the entries of these matrices and entries (with the same indices) of algebraic column matrix *S* (69), where *s* = 0 (compatible with $\phi_{\rho\sigma}^{(R)}=0,\ \phi^{(F)}=0$), if and only if: $iR_{\rho\sigma10}^{*}(\boldsymbol{x},t)=0,\ iR_{\rho\sigma41}^{*}(\boldsymbol{x},t)=0,\ iR_{\rho\sigma31}^{*}(\boldsymbol{x},t)=0,\ iR_{\rho\sigma21}^{*}(\boldsymbol{x},t)=0,\ iF_{10}^{*}(\boldsymbol{x},t)=0,$ $iF_{41}^{*}(\boldsymbol{x},t)=0,\ iF_{31}^{*}(\boldsymbol{x},t)=0,\ iF_{21}^{*}(\boldsymbol{x},t)=0,$ i.e.,:

$$\left(S=\begin{bmatrix}(u_0v_1-u_1v_0)w\\(u_2v_0-u_0v_2)w\\(u_0v_3-u_3v_0)w\\0\\(u_4v_0-u_0v_4)w\\0\\0\\0\\0\\(u_3v_4-u_4v_3)w\\(u_2v_4-u_4v_2)w\\(u_1v_4-u_4v_1)w\\(u_2v_3-u_3v_2)w\\(u_1v_3-u_3v_1)w\\(u_1v_2-u_2v_1)w\\0\end{bmatrix}\overset{\substack{\text{Derivation Procedure}\\\text{(First Quantization)}}}{\mapsto}\Upsilon_R=\begin{bmatrix}0\\-iR_{\rho\sigma43}^{*}(\boldsymbol{x},t)\\-iR_{\rho\sigma42}^{*}(\boldsymbol{x},t)\\-iR_{\rho\sigma41}^{*}(\boldsymbol{x},t)\\-iR_{\rho\sigma32}^{*}(\boldsymbol{x},t)\\-iR_{\rho\sigma31}^{*}(\boldsymbol{x},t)\\-iR_{\rho\sigma21}^{*}(\boldsymbol{x},t)\\0\\iR_{\rho\sigma10}^{*}(\boldsymbol{x},t)\\iR_{\rho\sigma02}^{*}(\boldsymbol{x},t)\\iR_{\rho\sigma30}^{*}(\boldsymbol{x},t)\\0\\iR_{\rho\sigma04}^{*}(\boldsymbol{x},t)\\0\\0\\0\end{bmatrix},\ \Upsilon_F=\begin{bmatrix}0\\-iF_{43}^{*}(\boldsymbol{x},t)\\-iF_{42}^{*}(\boldsymbol{x},t)\\-iF_{41}^{*}(\boldsymbol{x},t)\\-iF_{32}^{*}(\boldsymbol{x},t)\\-iF_{31}^{*}(\boldsymbol{x},t)\\-iF_{21}^{*}(\boldsymbol{x},t)\\0\\iF_{10}^{*}(\boldsymbol{x},t)\\iF_{02}^{*}(\boldsymbol{x},t)\\iF_{30}^{*}(\boldsymbol{x},t)\\0\\iF_{04}^{*}(\boldsymbol{x},t)\\0\\0\\0\end{bmatrix}\right)\Rightarrow\begin{matrix}u_1=0,\ v_1=0,\\iR_{\rho\sigma10}^{*}(\boldsymbol{x},t)=0,\ iF_{10}^{*}(\boldsymbol{x},t)=0,\\iR_{\rho\sigma21}^{*}(\boldsymbol{x},t)=0,\ iF_{21}^{*}(\boldsymbol{x},t)=0,\\iR_{\rho\sigma31}^{*}(\boldsymbol{x},t)=0,\ iF_{31}^{*}(\boldsymbol{x},t)=0,\\iR_{\rho\sigma41}^{*}(\boldsymbol{x},t)=0,\ iF_{41}^{*}(\boldsymbol{x},t)=0.\end{matrix}\tag{97}$$

This means that in (1 + 4) space-time dimensions, the mathematical framework of our axiomatic derivation approach (described in Section 3.4) in addition to the time reversal invariance (defined by the quantum operator (84)) of the source-free case of the derived general covariant fundamental field Equations (71) and (72) imply the column matrices $\Psi_R$ and $\Psi_F$ (76) (for $\phi_{\rho\sigma}^{(R)}=0,\ \phi^{(F)}=0$) could take solely the following forms (in general) to be defined in the formulations of the field Equations (71) and (72):

$$\Psi_R=\begin{bmatrix}0\\R_{\rho\sigma02}\\R_{\rho\sigma30}\\0\\R_{\rho\sigma04}\\0\\0\\0\\0\\R_{\rho\sigma43}\\R_{\rho\sigma42}\\0\\R_{\rho\sigma32}\\0\\0\\0\end{bmatrix},\quad\Psi_F=\begin{bmatrix}0\\F_{02}\\F_{30}\\0\\F_{04}\\0\\0\\0\\0\\F_{43}\\F_{42}\\0\\F_{32}\\0\\0\\0\end{bmatrix}\tag{98}$$

which are equivalent to the (1 + 3)-dimensional source-free cases of column matrices $\Psi_R$ and $\Psi_F$ (represented uniquely by formulas (75)). In addition, similar to the formulations (96-1), as it has been shown in Remark 2 (in Section 3.11), the field Equations (71) and (72) with non-zero source currents have a certain (and unique) combination of the C, P and T symmetries (that have been defined by the operators (83), (84) and (85)). This combined symmetry in addition to the forms (98), imply also the (1 + 4)-dimensional cases of column matrices $\Psi_R$ and $\Psi_F$ represented by formula (76) could take solely the following forms (in general) to be defined in the formulations of fundamental field Equations (71) and (72):

$$\Psi_R = \begin{bmatrix} 0 \\ R_{\rho\sigma 02} \\ R_{\rho\sigma 30} \\ 0 \\ R_{\rho\sigma 04} \\ 0 \\ 0 \\ 0 \\ 0 \\ R_{\rho\sigma 43} \\ R_{\rho\sigma 42} \\ 0 \\ R_{\rho\sigma 32} \\ 0 \\ 0 \\ 0 \\ \phi^{(R)}_{\rho\sigma} \end{bmatrix}, \quad \Psi_F = \begin{bmatrix} 0 \\ F_{02} \\ F_{30} \\ 0 \\ F_{04} \\ 0 \\ 0 \\ 0 \\ 0 \\ F_{43} \\ F_{42} \\ 0 \\ F_{32} \\ 0 \\ 0 \\ 0 \\ \phi^{(F)} \end{bmatrix}, \quad \begin{aligned} J^{(R)}_{\rho\sigma\nu} &= -(\nabla_\nu + \frac{im^{(R)}_0}{\hbar} k_\nu)\phi^{(R)}_{\rho\sigma}, \\ J^{(F)}_\nu &= -(D_\nu + \frac{im^{(F)}_0}{\hbar} k_\nu)\phi^{(F)} \end{aligned} \tag{98-1}$$

Consequently, the (1 + 4)-dimensional cases of column matrices $\Psi_R$ and $\Psi_F$ that are originally given by formulations (76), are reduced to formulas (98-1) which are equivalent to the (1 + 3)-dimensional cases of these matrices (given originally by column matrices of the forms (75)), i.e.,:

$$\Psi_R = \begin{bmatrix} R_{\rho\sigma 10} \\ R_{\rho\sigma 20} \\ R_{\rho\sigma 30} \\ 0 \\ R_{\rho\sigma 23} \\ R_{\rho\sigma 31} \\ R_{\rho\sigma 12} \\ \phi^{(R)}_{\rho\sigma} \end{bmatrix}, \quad \Psi_F = \begin{bmatrix} F_{10} \\ F_{20} \\ F_{30} \\ 0 \\ F_{23} \\ F_{31} \\ F_{12} \\ \phi^{(F)} \end{bmatrix}, \quad \begin{aligned} J^{(R)}_{\rho\sigma\nu} &= -(\nabla_\nu + \frac{im^{(R)}_0}{\hbar} k_\nu)\phi^{(R)}_{\rho\sigma}, \\ J^{(F)}_\nu &= -(D_\nu + \frac{im^{(F)}_0}{\hbar} k_\nu)\phi^{(F)}. \end{aligned} \tag{98-2}$$

Moreover, as it would be also noted in Section 3.15, it is noteworthy to add that the tensor field $R_{\rho\sigma\mu\nu}$ in column matrix $\Psi_R$ (98-2) (expressing the general representation of column matrices definable in the formulation of (1 + 3)-dimensional case of general covariant field Equation (71)), in fact, equivalently represents a massive bispinor field of spin-2 particles in (1 + 3) space-time dimensions (which could be identified as a definite generalized massive matrix formulation of the Einstein gravitational field, as it has been also shown in Section 3.6), and the tensor field $F_{\mu\nu}$ in the column matrix $\Psi_F$ (98-2) (expressing the general representation of column matrices definable in the formulation of (1 + 3)-dimensional case of general covariant field Equation (72)), in fact, equivalently represents a massive bispinor field of spin-1 particles in (1 + 3) space-time dimensions (which could be identified as definite generalized massive formulation of the Maxwell electromagnetic field, as it has been also shown in Sections 3.6 and 3.9; and also Yang-Mills fields compatible with specific gauge groups, as it would be shown in Section 3.15).

Summing up, in this Section (Section 3.12) we showed that the axiomatic approach of derivation of the field Equations (71) and (72) (described in Sections 3.1, 3.3 and 3.4) in addition to their time reversal invariance (represented basically by the quantum operator (84)), imply these fundamentally derived equations could be solely defined in (1 + 2) and (1 + 3) space-time dimensions. "*Hence, based on the later conclusion and also the basic assumption* (3) (*defined in Section 3.1*), *we may conclude directly that the universe could be realized solely with the* (1 + 2) *and* (1 + 3)-*dimensional space-times, and cannot have more than four space-time dimensions.*"

Based on the axiomatic arguments and relevant results presented and obtained in this Section, in the following Sections we consider solely the (1 + 2) and (1 + 3)-dimensional cases of general covariant field Equations (71) and (72) that are defined solely with the column matrices of the forms (96-1) and (98-2).

Equivalent (Asymptotically) Representations of the Bispinor Fields of Spin-3/2 and Spin-1/2 Particles, Respectively, by General Covariant Field Equations (71) and (72) (Formulated Solely with Column Matrices of the Types (96-1)) in (1 + 2) Space-Time Dimensions

It is noteworthy that according to the Ref. [52] and also based on the basic properties of the Riemann curvature tensor $R_{\rho\sigma\mu\nu}$ in (1 + 2) space-time dimensions [53] (in particular the identity: $R_{\rho\sigma\mu\nu} = \varepsilon_{\rho\sigma}{}^{\alpha}\varepsilon_{\mu\nu\beta}G^{\beta}{}_{\alpha}$, where $G^{\beta}{}_{\alpha}$ is the Einstein tensor and), it would be concluded that $R_{\rho\sigma\mu\nu}$ which is defined by (1 + 2)-dimensional case of the general covariant massive field Equation (71) (which could be defined solely with a column matrix of the type $\Psi_R$ (96-1)), represents asymptotically a general covariant bispinor field of spin-3/2 particles (that would be asymptotically equivalent to the Rarita–Schwinger equation). In a similar manner, according to the Ref. [52], and also following the basic properties of field strength tensor $F_{\mu\nu}$ in (1 + 2) space-time dimensions (that as a rank two anti-symmetric with three independent components holding, in particular, the identities: $F_{\mu\nu} = \varepsilon_{\mu\nu\alpha}T^{\alpha}$ , $T^{\alpha} = (1/2)\varepsilon^{\alpha\mu\nu}F_{\mu\nu}$ , showing that $F_{\mu\nu}$ could be represented equivalently by a vector $T^{\alpha}$ with three independent components as well) it would be concluded that $F_{\mu\nu}$ which is defined by (1 + 2)-dimensional case of the general covariant massive (tensor) field Equation (72) (which could be defined solely with a column matrix of the type $\Psi_F$ (96-1)), represents asymptotically a general covariant bispinor field of spin-1/2 particles (that would be asymptotically equivalent to the Dirac equation [52]). Furthermore, as shown in Section 3.15, the general covariant field Equations (72) (representing asymptotically the spin-1/2 fermion fields) is also compatible with the $SU(2)_L \otimes U(2)_R$ symmetry group (representing "1 + 3" generations for both lepton and quark fields including a new charge-less fermion).

Remark 3. Equivalent Representations of the Bispinor Fields of Spin-2 and Spin-1 Particles, Respectively, by General Covariant Field Equations (71) and (72) (Defined Solely with Column Matrices of the Types (98-2)) in (1 + 3) Space-Time Dimensions

It should be also noted that according to the Refs. [5–7,51,54–56], the basic properties of the Riemann curvature tensor including the relevant results presented in Section 3.6, it would be concluded that the field strength tensor $R_{\rho\sigma\mu\nu}$ (i.e., the Riemann tensor) the in (1 + 3) space-time dimensions by general covariant massive (tensor) field Equation (71) (formulated solely with a column matrix of the type $\Psi_R$ (98-2)), represents a general covariant bispinor field of spin-2 particles (as a generalized massive formulation of the Einstein gravitational field equation). In a similar manner, according to the Refs. [5–7,51,54–56], the field strength tensor $F_{\mu\nu}$ which is defined in (1 +

3) space-time dimensions by the general covariant massive (tensor) field Equation (72) (formulated solely by a column matrix of the type $\Psi_F$ (98-2)), and represents a general covariant bispinor field of spin-1 particles (representing new generalized massive formulations of the Maxwell's equations, and also Yang-Mills field equations). Furthermore, as it would be shown in Section 3.15, the general covariant field Equation (72) (representing the spin-1 boson fields coupling to the spin-1/2 fermionic currents) is also compatible with the $SU(2)_L \otimes U(2)_R$ and SU(3) symmetry groups.

Moreover, based on these determined gauge symmetries for the derived fermion and boson field equations, four new charge-less spin-1/2 fermions (represented by "$z_e$ , $z_n$ ; $z_u$ , $z_d$", where two right-handed charhe-less quarks $z_u$ and $z_d$ emerge specifically in two subgroups with anti-quarks such that: ( $\overline{s},\overline{u},\overline{b}$, $z_u$) and ( $\overline{c},\overline{d},\overline{t}$ , $z_d$)), and also three new massive spin-1 bosons (represented by $\tilde{W}^+,\tilde{W}^-,\vec{Z}$, where in particular $\vec{Z}$ is the complementary right-handed particle of ordinary $Z$ boson), are predicted uniquely by this new mathematical axiomatic approach.

### *3.13. Showing that Only a Definite Simultaneous Combination of the Quantum Mechanical Transformations $\hat{C}$ , $\hat{P}$ , $\hat{T}$ and $\gamma^{Ch}$ . (Given Uniquely by the Matrix Operators (83)–(87)) Could Be Defined for the General Covariant Massive (Tensor) Field Equations (71) and (72) with Non-Zero Source Currents*

As it has been shown in Remark 1 (in Section 3.11) and Section 3.12, since the algebraic column matrix $S$ in the matrix Equation (64) (derived and represented uniquely in terms of the matrices (66)–(70), … corresponding to various space-time dimensions), is not symmetric by multiplying by matrix $\gamma^{Ch}$ (84-1) and (84-2) (except for (1 + 2) and (1 + 3)-dimensional cases of column matrix $S$ , based on the definite algebraic properties of matrix $S$ presented in Remark 1 in Section 3.3), it is concluded that except the (1 + 2) and (1 + 3)-dimensional cases of the fundamental field Equations (71) and (72), these field equations could not be defined with column matrices of the type $\Psi_R^{(Ch)}(=\gamma^{Ch}\Psi_R)$ and $\Psi_F^{(Ch)}(=\gamma^{Ch}\Psi_F)$ (if assuming that the column matrices $\Psi_R$ and $\Psi_F$ are defined with field Equations (71) and (72), i.e., they have the formulations similar to the formulations of originally derived column matrices (73)–(77), … corresponding to various space-time dimensions). This conclusion follows from this fact that the filed Equations (71) and (72) have been derived (and defined) uniquely from the matrix Equation (64) via the axiomatic derivation approach (including the first quantization procedure) presented in Sections 3.4 and 3.5. As it has been shown in Section 3.12, using this property (i.e., multiplication of column matrices $\Psi_R$ and $\Psi_F$ , defined in the unique expressions of fundamental field Equations (71) and (72), by matrix $\gamma^{Ch}$ from the left), this crucial and essential issue is concluded directly that by assuming the time-reversal invariance of the general covariant filed Equations (71) and (72) (represented by the transformations $\hat{T}\Psi_R$ and $\hat{T}\Psi_F$ , where the quantum operator $\hat{T}$ is given uniquely by formula (84), i.e.,: $\hat{T}=\hat{T}_0\hat{K}=i\gamma^P\gamma^{Ch}\hat{K}$), these fundamental field equations could be defined solely in (1 + 2) and (1 + 3) space-time dimensions (with the column matrices of the forms (96-1) and (98-2), respectively).

Hence, the definite mathematical formalism of the axiomatic approach of derivation of fundamental field Equations (71) and (72), along with the C, P and T symmetries (represented by the quantum matrix operators (83)–(87), in Section 3.11) of source-free cases (as basic cases) of these equations, in fact, imply these equations with non-zero source currents, would be invariant solely under the simultaneous combination of all the transformations $\hat{C}$, $\hat{P}$ , and $\hat{T}$ (83)–(85), multiplied by matrix $\gamma^{Ch}$ (defined by formulas (84-1) and (84-2)). This unique combined transformation could be expressed uniquely as follows, respectively, for the particle fields (represented by column matrices $\Psi_R(-\vec{r},t)$, $\Psi_F(-\vec{r},t)$) and their corresponding antiparticle fields (represented by column matrices $\Psi'_R(\vec{r},-t)$, $\Psi'_F(\vec{r},-t)$ given solely with reversed signs of the temporal and spatial coordinates):

$$\begin{aligned}\hat{Z}_{COMB}\Psi_R(-\vec{r},t) &= -\gamma^{Ch}\hat{T}\hat{P}\hat{C}\ \Psi_R(-\vec{r},t),\\ \hat{Z}_{COMB}\Psi_F(-\vec{r},t) &= -\gamma^{Ch}\hat{T}\hat{P}\hat{C}\ \Psi_F(-\vec{r},t);\end{aligned} \tag{99}$$

$$\begin{aligned}\hat{\hat{Z}}_{COMB}\Psi'_R(\vec{r},-t) &= \gamma^{Ch}\hat{T}\hat{P}\hat{C}\ \Psi'_R(\vec{r},-t),\\ \hat{\hat{Z}}_{COMB}\Psi'_F(\vec{r},-t) &= \gamma^{Ch}\hat{T}\hat{P}\hat{C}\ \Psi'_F(\vec{r},-t).\end{aligned} \tag{100}$$

The unique combined form of transformation $\hat{Z}_{COMB}$ (99) (and also $\hat{\hat{Z}}_{COMB}$ (100), where $\hat{\hat{Z}}_{COMB} = -\hat{Z}_{COMB}$) is based on the following two basic issues:

Firstly, it follows from the definite formulations of uniquely determined column matrices (73)–(77), … (corresponding to various space-time dimensions, however, as noted above, based on the arguments presented in Section 3.12, the only definable column matrices in the formulations of field Equations (71) and (72), are of the types $\Psi_R$ and $\Psi_F$ represented by formulas (96-1) and (98-2), in (1 + 2) and (1 + 3) space-time dimensions, respectively), where the source currents $J^{(R)}_{\rho\sigma\nu}$ and $J^{(F)}_{\nu}$ should be expressible by these conditional relations (in terms of the arbitrary covariant quantities $\phi^{(R)}_{\rho\sigma}$ and $\phi^{(F)}$), respectively: $J^{(R)}_{\rho\sigma\nu} = -(\breve{\nabla}_\nu + \frac{im_0^{(R)}}{\hbar}k_\nu)\phi^{(R)}_{\rho\sigma}$, $J^{(F)}_{\nu} = -(\breve{\nabla}_\nu + \frac{im_0^{(F)}}{\hbar}k_\nu)\phi^{(F)}$. In other words, the unique formulation of derived combined symmetries $Z_{COMB}$ and $\tilde{Z}_{COMB}$ represented by the quantum operators $\hat{Z}_{COMB}$ (99) and $\hat{\hat{Z}}_{COMB}$ (100), in particular, is a direct consequent of the above conditional expressions for source currents $J^{(R)}_{\rho\sigma\nu}$ and $J^{(F)}_{\nu}$. As noted in Section 3.5, these relations appear as necessary conditions in the course of the axiomatic derivation of general covariant field Equations (71) and (72). In fact, in the field Equations (71) and (72) the uniquely derived column matrices $\Psi_R$ and $\Psi_F$ (73)–(77),…, not only contain all the components of tensor fields $R_{\rho\sigma\mu\nu}$ and $F_{\mu\nu}$, but also contain the components of arbitrary covariant quantities $\phi^{(R)}_{\rho\sigma}$ and $\phi^{(F)}$ (as the initially given quantities) which define the source currents $J^{(R)}_{\rho\sigma\nu}$ and $J^{(F)}_{\nu}$ by the above expressions, respectively, i.e.,: $J^{(R)}_{\rho\sigma\nu} = -(\breve{\nabla}_\nu + \frac{im_0^{(R)}}{\hbar}k_\nu)\phi^{(R)}_{\rho\sigma}$, $J^{(F)}_{\nu} = -(\breve{\nabla}_\nu + \frac{im_0^{(F)}}{\hbar}k_\nu)\phi^{(F)}$. Now based on these conditional expressions in addition to this natural and basic circumstance that the source currents $J^{(R)}_{\rho\sigma\nu}$ and $J^{(F)}_{\nu}$ should be also transferred respectively as a rank three tensor and a vector, under the parity, time-reversal and charge conjugation transformations (defined by formulas (83)–(85)) of the field Equations (71) and (72), it would be concluded directly that the transformations (99) and (100) are the only simultaneous combinations of transformations $\hat{C}$, $\hat{P}$, $\hat{T}$ (also including the matrix $\gamma^{Ch}$, necessarily, as it would be shown in the following paragraph), which could be defined for the field Equations (71) and (72) with "non-zero" source currents.

Secondly, appearing the matrix operator $\gamma^{Ch}$ in simultaneous combinations $-\gamma^{Ch}\hat{T}\hat{P}\hat{C}$ and $\gamma^{Ch}\hat{T}\hat{P}\hat{C}$ in the combined transformations (99) and (100), follows simply from the basic arguments presented in Section 3.12. In fact, in these uniquely determined combinations, the simultaneous multiplication by matrix $\gamma^{Ch}$ (from the left) is a necessary condition for that the transformed column matrices: $\hat{Z}_{COMB}\Psi_R(-\vec{r},t)$, $\hat{\hat{Z}}_{COMB}\Psi'_R(\vec{r},-t)$, $\hat{Z}_{COMB}\Psi_F(-\vec{r},t)$, $\hat{\hat{Z}}_{COMB}\Psi'_F(\vec{r},-t)$ (given bythe transformations (99) and (100)) could be also defined in the field Equations (71) and (72), based on the formulations of column matrices (96-1) and (98-2), as mentioned in Section 3.12 (however, it is worth notingthat this argument is not merely limited to the definability of column matrices of the types (96-1) and (98-2), and it could be also represented on the basis of unique formulations of all the originally derived column matrices (73)–(77), … corresponding to various space-time dimensions).

In the next Section, we show how the 'CPT' theorem in addition to the unique formulations of the combined transformations (99) and (100) (representing the only definable transformation forms, including C, P and T quantum mechanical transformations, for the field Equations (71) and (72) with "non-zero" source currents), imply only the left-handed particle fields (along with their complementary right-handed fields) could be coupled to the corresponding (any) source currents.

### *3.14. Showing that Only the Left-Handed Particle (along with Their Complementary Right-Handed Antiparticle) Fields Could Be Coupled to the Corresponding Source Currents*

On the basis of the 'CPT' theorem [6,7], it would be concluded directly that the unique combined forms of transformations (99) and (100) (representing the only combination of $\hat{C}$, $\hat{P}$, and $\hat{T}$ transformations multiplied by matrix $\gamma^{Ch}$, that could be defined as a symmetry for general covariant field Equations (71) and (72) with non-zero source currents), should be equivalent only to simultaneous combination of $\hat{C}$, $\hat{P}$, and $\hat{T}$ transformations (that have been defined uniquely by formulas (83)–(85)). Moreover, based on the 'CPT' theorem, the simultaneous combination of transformations $\hat{C}$, $\hat{P}$, and $\hat{T}$ should: "interchange the particle field and its corresponding antiparticle field; inverts the spatial coordinates $\vec{r} \to -\vec{r}$; reverse the spin of all particle fields; leave the direction of the momentum invariant; and, therefore, should interchange the left-handed and right-handed components of both particle field and its corresponding antiparticle field". Hence, we should have:

$$\begin{aligned}\hat{Z}_{COMB}[\Psi_R(-\vec{r})]_{(Left)} &= -\gamma^{Ch}\hat{T}\hat{P}\hat{C}\ [\Psi_R(-\vec{r})]_{(Left)} = -\gamma^{Ch}[\Psi'_R(\vec{r})]_{(Right)} = [\Psi'_R(\vec{r})]_{(Right)},\\ \hat{Z}_{COMB}[\Psi_F(-\vec{r})]_{(Left)} &= -\gamma^{Ch}\hat{T}\hat{P}\hat{C}\ [\Psi_F(-\vec{r})]_{(Left)} = -\gamma^{Ch}[\Psi'_F(\vec{r})]_{(Right)} = [\Psi'_F(\vec{r})]_{(Right)};\end{aligned} \quad (101)$$

$$\begin{aligned}\hat{Z}_{COMB}[\Psi_R(-\vec{r})]_{(Right)} &= -\gamma^{Ch}\hat{T}\hat{P}\hat{C}\ [\Psi_R(-\vec{r})]_{(Right)} = -\gamma^{Ch}[\Psi'_R(\vec{r})]_{(Left)} = [\Psi'_R(\vec{r})]_{(Left)},\\ \hat{Z}_{COMB}[\Psi_F(-\vec{r})]_{(Right)} &= -\gamma^{Ch}\hat{T}\hat{P}\hat{C}\ [\Psi_F(-\vec{r})]_{(Right)} = -\gamma^{Ch}[\Psi'_F(\vec{r})]_{(Left)} = [\Psi'_F(\vec{r})]_{(Left)};\end{aligned} \quad (102)$$

$$\begin{aligned}\hat{\hat{Z}}_{COMB}[\Psi'_R(\vec{r})]_{(Left)} &= \gamma^{Ch}\hat{T}\hat{P}\hat{C}\ [\Psi'_R(\vec{r})]_{(Left)} = \gamma^{Ch}[\Psi_R(-\vec{r})]_{(Right)} = [\Psi_R(-\vec{r})]_{(Right)},\\ \hat{\hat{Z}}_{COMB}[\Psi'_F(\vec{r})]_{(Left)} &= \gamma^{Ch}\hat{T}\hat{P}\hat{C}\ [\Psi'_F(\vec{r})]_{(Left)} = \gamma^{Ch}[\Psi_F(-\vec{r})]_{(Right)} = [\Psi_F(-\vec{r})]_{(Right)};\end{aligned} \quad (103)$$

$$\begin{aligned}\hat{\hat{Z}}_{COMB}[\Psi'_R(\vec{r})]_{(Right)} &= \gamma^{Ch}\hat{T}\hat{P}\hat{C}\ [\Psi'_R(\vec{r})]_{(Right)} = \gamma^{Ch}[\Psi_R(-\vec{r})]_{(Left)} = [\Psi_R(-\vec{r})]_{(Left)},\\ \hat{\hat{Z}}_{COMB}[\Psi'_F(\vec{r})]_{(Right)} &= \gamma^{Ch}\hat{T}\hat{P}\hat{C}\ [\Psi'_F(\vec{r})]_{(Right)} = \gamma^{Ch}[\Psi_F(-\vec{r})]_{(Left)} = [\Psi_F(-\vec{r})]_{(Left)}.\end{aligned} \quad (104)$$

where the column matrices $\Psi_R(-\vec{r})$ and $\Psi_F(-\vec{r})$ represent the particle field and $\Psi'_R(\vec{r})$ and $\Psi'_F(\vec{r})$ denote the transformed forms of column matrices of $\Psi_R(-\vec{r})$ and $\Psi_F(-\vec{r})$, respectively, under the simultaneous combination of transformations $\hat{C}$, $\hat{P}$, and $\hat{T}$ (83)–(85). Furthermore, in agreement and based on the definitions and properties of quantum operators $\hat{C}$, $\hat{P}$, $\hat{T}$ and matrix $\gamma^{Ch}$ given by formulas (83)–(87), the left-handed and right-handed components of column matrices of the types (96-1) and (98-2) (representing the unique formulations of column matrices that could be defined in the field Equations (71) and (72), as mentioned in Section 3.12) are defined solely as follows for the column matrices $\Psi_R(-\vec{r})$, $\Psi_F(-\vec{r})$ and also $\Psi'_R(\vec{r})$, $\Psi'_F(\vec{r})$ (as the transformed forms of column matrices $\Psi_R(-\vec{r})$ and $\Psi_F(-\vec{r})$ under the $\hat{C}\hat{P}\hat{T}$ transformation, respectively):

$$\begin{aligned}&[\Psi_R(-\vec{r})]_{(Left)}=\frac{1}{2}[\Psi_R(-\vec{r})+\gamma^{Ch}\Psi_R(-\vec{r})],\ [\Psi_R(-\vec{r})]_{(Right)}=\frac{1}{2}[\Psi_R(-\vec{r})-\gamma^{Ch}\Psi_R(-\vec{r})],\\&[\Psi_F(-\vec{r})]_{(Left)}=\frac{1}{2}[\Psi_F(-\vec{r})+\gamma^{Ch}\Psi_F(-\vec{r})],\ [\Psi_F(-\vec{r})]_{(Right)}=\frac{1}{2}[\Psi_F(-\vec{r})-\gamma^{Ch}\Psi_F(-\vec{r})];\end{aligned}\qquad(105)$$

$$\begin{aligned}&[\Psi'_R(\vec{r})]_{(Left)}=\frac{1}{2}[\Psi'_R(\vec{r})+\gamma^{Ch}\Psi'_R(\vec{r})],\ [\Psi'_R(\vec{r})]_{(Right)}=\frac{1}{2}[\Psi'_R(\vec{r})-\gamma^{Ch}\Psi'_R(\vec{r})],\\&[\Psi'_F(\vec{r})]_{(Left)}=\frac{1}{2}[\Psi'_F(\vec{r})+\gamma^{Ch}\Psi'_F(\vec{r})],\ [\Psi'_F(\vec{r})]_{(Right)}=\frac{1}{2}[\Psi'_F(\vec{r})-\gamma^{Ch}\Psi'_F(\vec{r})].\end{aligned}\qquad(106)$$

where we have:

$$\begin{aligned}&\Psi_R(-\vec{r})=[\Psi_R(-\vec{r})]_{(Left)}+[\Psi_R(-\vec{r})]_{(Right)},\quad \Psi'_R(-\vec{r})=[\Psi'_R(-\vec{r})]_{(Left)}+[\Psi'_R(-\vec{r})]_{(Right)},\\&\Psi_F(-\vec{r})=[\Psi_F(-\vec{r})]_{(Left)}+[\Psi_F(-\vec{r})]_{(Right)},\quad \Psi'_F(-\vec{r})=[\Psi'_F(-\vec{r})]_{(Left)}+[\Psi'_F(-\vec{r})]_{(Right)}.\end{aligned}\qquad(107)$$

Now using the definitions (105) and (106) in the formulas (101)–(104), we obtain:

$$\begin{aligned}&\gamma^{Ch}[\Psi_R(-\vec{r})]_{(Left)}=\frac{1}{2}\gamma^{Ch}[\Psi_R(-\vec{r})+\gamma^{Ch}\Psi_R(-\vec{r})]=\frac{1}{2}[\gamma^{Ch}\Psi_R(-\vec{r})+\Psi_R(-\vec{r})]=[\Psi_R(-\vec{r})]_{(Left)},\\&\gamma^{Ch}[\Psi_F(-\vec{r})]_{(Left)}=\frac{1}{2}\gamma^{Ch}[\Psi_F(-\vec{r})+\gamma^{Ch}\Psi_F(-\vec{r})]=\frac{1}{2}[\gamma^{Ch}\Psi_F(-\vec{r})+\Psi_F(-\vec{r})]=[\Psi_F(-\vec{r})]_{(Left)};\end{aligned}\qquad(101\text{-}1)$$

$$\begin{aligned}&\gamma^{Ch}[\Psi_R(-\vec{r})]_{(Right)}=\frac{1}{2}\gamma^{Ch}[\Psi_R(-\vec{r})-\gamma^{Ch}\Psi_R(-\vec{r})]=\frac{1}{2}[\gamma^{Ch}\Psi_R(-\vec{r})-\Psi_R(-\vec{r})]\neq[\Psi_R(-\vec{r})]_{(Right)},\\&\gamma^{Ch}[\Psi_F(-\vec{r})]_{(Right)}=\frac{1}{2}\gamma^{Ch}[\Psi_F(-\vec{r})-\gamma^{Ch}\Psi_F(-\vec{r})]=\frac{1}{2}[\gamma^{Ch}\Psi_F(-\vec{r})-\Psi_F(-\vec{r})]\neq[\Psi_F(-\vec{r})]_{(Right)};\end{aligned}\qquad(102\text{-}1)$$

$$\begin{aligned}&-\gamma^{Ch}[\Psi'_R(\vec{r})]_{(Left)}=-\frac{1}{2}\gamma^{Ch}[\Psi'_R(\vec{r})+\gamma^{Ch}\Psi'_R(\vec{r})]=\frac{1}{2}[-\gamma^{Ch}\Psi'_R(\vec{r})-\Psi'_R(\vec{r})]\neq[\Psi'_R(\vec{r})]_{(Left)},\\&-\gamma^{Ch}[\Psi'_F(\vec{r})]_{(Left)}=-\frac{1}{2}\gamma^{Ch}[\Psi'_F(\vec{r})+\gamma^{Ch}\Psi'_F(\vec{r})]=\frac{1}{2}[-\gamma^{Ch}\Psi'_F(\vec{r})-\Psi'_F(\vec{r})]\neq[\Psi'_F(\vec{r})]_{(Left)};\end{aligned}\qquad(103\text{-}1)$$

$$\begin{aligned}&-\gamma^{Ch}[\Psi'_R(\vec{r})]_{(Right)}=-\frac{1}{2}\gamma^{Ch}[\Psi'_R(\vec{r})-\gamma^{Ch}\Psi'_R(\vec{r})]=\frac{1}{2}[-\gamma^{Ch}\Psi'_R(\vec{r})+\Psi'_R(\vec{r})]=[\Psi'_R(\vec{r})]_{(Right)},\\&-\gamma^{Ch}[\Psi'_F(\vec{r})]_{(Right)}=-\frac{1}{2}\gamma^{Ch}[\Psi'_F(\vec{r})-\gamma^{Ch}\Psi'_F(\vec{r})]=\frac{1}{2}[-\gamma^{Ch}\Psi'_F(\vec{r})+\Psi'_F(\vec{r})]=[\Psi'_F(\vec{r})]_{(Right)}.\end{aligned}\qquad(104\text{-}1)$$

Based on the relations (101-1)–(104-1), it would be concluded directly that only the left-handed components of particle fields represented by $[\Psi_R(-\vec{r})]_{(Left)}$, $[\Psi_F(-\vec{r})]_{(Left)}$, and the right-handed components of their corresponding antiparticle fields represented by $[\Psi'_R(\vec{r})]_{(Right)}$, $[\Psi'_F(\vec{r})]_{(Right)}$ (as the transformed forms of column matrices $[\Psi_R(-\vec{r})]_{(Left)}$ and $[\Psi_F(-\vec{r})]_{(Left)}$ under the $\hat{C}\hat{P}\hat{T}$ transformation, respectively), obey the transformations (101-1) and (104-1) (as the necessary conditions given respectively by relations (101) and (104)). On the other hand, the right-handed components of particle fields represented by $[\Psi_R(-\vec{r})]_{(Right)}$, $[\Psi_F(-\vec{r})]_{(Right)}$, and the left-handed components of their corresponding antiparticle fields represented by $[\Psi'_R(\vec{r})]_{(Left)}$, $[\Psi'_F(\vec{r})]_{(Left)}$ (as the transformed forms of column matrices $[\Psi_R(-\vec{r})]_{(Right)}$ and $[\Psi_F(-\vec{r})]_{(Right)}$ under the $\hat{C}\hat{P}\hat{T}$ transformation, respectively), obey the transformations (102-1) and (103-1) (as the necessary conditions given respectively by relations (102) and (103)). Hence (and also following the basic assumption (3) defined in Section 3.1), it is concluded directly that only the left-handed particle fields (along with their complementary right-handed fields) could be coupled to the corresponding (any) source currents. This means that only left-handed bosonic fields (along with their complementary right-handed fields) could be coupled to the corresponding fermionic source currents; which also means that only left-handed fermions (along with their complementary right-handed fermions) can participate in any interaction with the bosons (which consequently would be only left-handed bosons or their complementary right-handed bosons).

### *3.15. Showing the gauge invariance of axiomatically derived general covariant (tensor) field Equation (72) in (1 + 2)-dimensional space-time (definable with column matrices of the type $\Psi_F$ (96-1), representing the spin-1/2 fermion fields) under the $SU(2)_L \otimes U(2)_R$ symmetry group, and also invariance of this equation in (1 + 3)-dimensional space-time (definable column matrices of the type $\Psi_F$ (98-2), representing the spin-1 boson fields coupled to the fermionic source currents) under the $SU(2)_L \otimes U(2)_R$ and SU(3) symmetry group*

One of the natural and basic properties of the (1 + 2)-dimensional space-time geometry is that the metric tensor can be "diagonalized" [57]. Using this basic property, the invariant energy-momentum quadratic relation (52) (in Section 3.1) would be expressed as follows:

$$g^{00}(p_0)^2 - g^{00}(p_0^{st})^2 + g^{11}(p_1)^2 + g^{22}(p_2)^2 = 0 \tag{108}$$

that is equivalent to: $g^{00}(p_0)^2 + g^{11}(p_1)^2 + g^{22}(p_2)^2 = (m_0 c)^2$, where (as defined in Section 3.1) $m_0$ and $p_\mu$ are the particle's rest mass and momentum (3-momentum), $p_\mu^{st} = m_0 k_\mu$, and $k_\mu = (k_0, 0, 0) = (c/\sqrt{g^{00}}, 0, 0)$ denotes the covariant form of the 3-velocity of particle in stationary reference frame. As it would be shown in the following, a crucial and essential property of the quadratic relation (108) is its invariance under a certain set of sign inversions of the components of particle's momentum: $(p_0, p_1, p_2)$, along with similar inversions for the components: $(p_0^{st}, p_1^{st}, p_2^{st})$, where $p_0^{st} = m_0 k_0$, $p_1^{st} = p_2^{st} = 0$. This set includes seven different types of the sign inversions (in total), which could be represented simply by the following symmetric group of transformations (based on the formalism of the corresponding Lorentz symmetry group of invariant relation (108)), respectively:

$$(p_0, p_0^{st}, p_1, p_2) \mapsto (p_0, p_0^{st}, -p_1, p_2) = (p_0^{(1)}, p_0^{st(1)}, p_1^{(1)}, p_2^{(1)}) \tag{108-1}$$

$$(p_0, p_0^{st}, p_1, p_2) \mapsto (p_0, p_0^{st}, p_1, -p_2) = (p_0^{(2)}, p_0^{st(2)}, p_1^{(2)}, p_2^{(2)}) \quad (108\text{-}2)$$

$$(p_0, p_0^{st}, p_1, p_2) \mapsto (p_0, p_0^{st}, -p_1, -p_2) = (p_0^{(3)}, p_0^{st(3)}, p_1^{(3)}, p_2^{(3)}) \quad (108\text{-}3)$$

$$(p_0, p_0^{st}, p_1, p_2) \mapsto (-p_0, -p_0^{st}, -p_1, -p_2) = (p_0^{(4)}, p_0^{st(4)}, p_1^{(4)}, p_2^{(4)}) \quad (108\text{-}4)$$

$$(p_0, p_0^{st}, p_1, p_2) \mapsto (-p_0, -p_0^{st}, p_1, -p_2) = (p_0^{(5)}, p_0^{st(5)}, p_1^{(5)}, p_2^{(5)}) \quad (108\text{-}5)$$

$$(p_0, p_0^{st}, p_1, p_2) \mapsto (-p_0, -p_0^{st}, -p_1, p_2) = (p_0^{(6)}, p_0^{st(6)}, p_1^{(6)}, p_2^{(6)}) \quad (108\text{-}6)$$

$$(p_0, p_0^{st}, p_1, p_2) \mapsto (-p_0, -p_0^{st}, p_1, p_2) = (p_0^{(7)}, p_0^{st(7)}, p_1^{(7)}, p_2^{(7)}) \quad (108\text{-}7)$$

Moreover, although, following noncomplex- algebraic values of momentum's components $p_\mu$ $(p_\mu^* = p_\mu)$, the corresponding complex representations of transformations (108-1)–(108-7) is not a necessary issue in general, however, if the invariant relation (108) is represented formally by equivalent complex form:

$$g^{00}(p_0 p_0^*) - g^{00}(p_0^{st} p_0^{st*}) + g^{11}(p_1 p_1^*) + g^{22}(p_2 p_2^*) = 0 \quad (108\text{-}8)$$

then, along with the set seven real-valued transformations (108-1)–(108-7), this relation would be also invariant under these corresponding sets of complex transformations (for $a = 1, 2, 3, \ldots, 7$ ):

$$\begin{aligned} &(p_0, p_0^{st}, p_1, p_2) \to (\pm i p_0^{(a)*}, \pm i p_0^{st(a)*}, \pm i p_1^{(a)*}, \pm i p_2^{(a)*}), \\ &(p_0^*, p_0^{st*}, p_1^*, p_2^*) \to (\mp i p_0^{(a)}, \mp i p_0^{st(a)}, \mp i p_1^{(a)}, \mp i p_2^{(a)}) \end{aligned} \quad (108\text{-}9)$$

As the next step, in the following, using the transformations (108-1)–(108-7) (along with their corresponding complex forms (108-9)), a certain set of seven simultaneous (different) general covariant field Equations (corresponding to a group of seven bispinor fields of spin-1/2 particles) would be determined as particular cases of the (1 + 2)-dimensional form of general covariant field Equation (72) (defined with a column matrix of the type (96-1)).

Based on the definite formulation of (1 + 2)-dimensional case of system of linear Equation (64) (formulated in terms of the matrices (67)), for the energy-momentum relation (108) (along with the transformations (108-1)–(108-7)), the following set of seven systems of linear Equations (with different parametric formalisms) is determined uniquely. The general parametric solution of each of these systems of linear equations, obeys also the quadratic relation (108) (representing a set of seven forms, with different parametric formulations, of the general parametric solutions of quadratic relation (108)). This set of the seven systems of linear equations could be represented uniformly by a matrix equation as follows:

$$(\alpha^\mu p_\mu^{(a)} - m_0^{(a)} \tilde{\alpha}^\mu k_\mu) S^{(a)} = 0 \quad (109)$$

where $a = 1, 2, 3, \ldots, 7$, $p_\mu^{st(a)} = m_0^{(a)} k_\mu$, $\alpha^\mu$ and $\tilde{\alpha}^\mu$ are two contravariant $4 \times 4$ real matrices (compatible with matrix representations of the Clifford algebra $C\ell_{1,2}$) defined solely by formulas (65) and (67), and parametric column matrix $S^{(a)}$ is also given uniquely as follows (formulated on the basis of definite parametric formulation of column matrix $S$ (67) in (1 + 2) space-time dimensions):

$$S^{(a)} = \begin{bmatrix} (u_0^{(a)}v_1^{(a)} - u_1^{(a)}v_0^{(a)})w \\ (u_2^{(a)}v_0^{(a)} - u_0^{(a)}v_2^{(a)})w \\ (u_1^{(a)}v_2^{(a)} - u_2^{(a)}v_1^{(a)})w \\ s \end{bmatrix} \quad (109\text{-}1)$$

which includes seven cases with specific parametric formulations expressed respectively in terms of seven groups of independent arbitrary parameters: $u_0^{(a)}, u_1^{(a)}, u_2^{(a)}, v_0^{(a)}, v_1^{(a)}, v_2^{(a)}$, and two common arbitrary parameters $s$ and $w$ (i.e., having the same forms in all of the seven cases of column matrix $S^{(a)}$). In addition, concerning the specific parametric expression (109-1) of column matrix $S^{(a)}$ in the formulation of matrix Equation (109), it is necessary to add that this parametric expression has been determined specifically by assuming (as a basic assumption in addition to the systematic natural approach of formulating the matrix Equation (109), based on the definite formulation of axiomatically determined matrix Equation (64)) the minimum value for total number of the arbitrary parameters in all of seven cases of column matrix $S^{(a)}$, which implies equivalently the minimum value for total number of the arbitrary parameters in all of seven simultaneous (different) cases of matrix Equation (109) (necessarily with seven independent parametric solutions representing a certain set of seven different equivalent forms of the general parametric solution of quadratic relation (108), based on the general conditions of basic definition of the systems of linear equations corresponding to homogeneous quadratic and higher degree equations, presented in Section 2, and Sections 2.2–2.4 and 3.1 concerning the homogenous quadratic equations).

In the following, in the derivation of the corresponding field Equations (from matrix Equation (109)), we will also use the above particular algebraic property of parameters $s$ which has been expressed commonly in the expressions of all of seven simultaneous cases of matrix Equation (109) (and as shown below, concerning the (1 + 3)-dimensional corresponding form of matrix Equation (109), which holds the similar property).

In addition, along with the transformations (108-1)–(108-7) and algebraic matrix Equation (109), using the corresponding complex transformations (108-9), we may also formally have the following equivalent matrix Equation (with the complex expression):

$$(i\alpha^{\mu} p_{\mu}^{(a)*} - i m_0^{(a)} \tilde{\alpha}^{\mu} k_{\mu}) S^{(a)} = 0 \quad (109\text{-}2)$$

where $a = 1, 2, 3, \ldots, 7$, and $p_{\mu}^{st(a)} = m_0^{(a)} k_{\mu}$. Although, based on the real value of momentum $p_{\mu}^{(\alpha)}$ $(p_{\mu}^{(a)} = p_{\mu}^{*(a)})$, the complex expression of each of the seven cases of algebraic matrix Equation (109), definitely, is not a necessary issue at the present stage. However, since the corresponding momentum operator $\hat{p}_{\mu}^{(a)}$ has a complex value (where $\hat{p}_{\mu}^{(a)} \neq \hat{p}_{\mu}^{*(a)}$), in the following, using this basic property of the momentum operator, we derive a certain set of seven different simultaneous general covariant field equations from the matrix Equations (109) and (109-2) (based on the general axiomatic approach of derivation of general covariant massive field Equation (72), presented in Sections 3.4–3.10, in addition to certain forms of quantum representations of the C, P and T symmetries of this field equation, presented in Sections 3.11–3.14). Furthermore, in the following, it would be also shown that the uniform representation of this determined set of seven simultaneous field equations, describe a certain group of seven simultaneous bispinor fields of spin-1/2 particles (corresponding, respectively, to a new right-handed charge-less fermion in addition to three right-handed anti-fermions, along with their three complementary left-handed fermions).

Furthermore, concerning the gravitational field Equation (71), it should be noted that following from the fact that the general covariant field Equation (71) should describe, uniquely and uniformly, the background space-time geometry via a certain form of the Riemann curvature tensor (which should be determined from the tensor field Equation (71)), the matrix Equation (109) could not be used for the derivation of a set of simultaneous different spin-3/2 fermion fields in (1 + 2) dimensions (there would be the same condition for the field Equation (71) in higher-dimensional space-times).

Hence, based on the axiomatic approach of derivation of (1 + 2)-dimensional case of field Equation (72) (defined solely by a column matrix of the form (96-1) in (1 + 2) space-time dimensions, as shown in Section 3.12), from the matrix Equation (109) and (109-2) (defined solely by column matrix (109-1)), and also taking into account the momentum operator's property: $\hat{p}_\mu \neq \hat{p}_\mu^*$, the following group of seven simultaneous (different) general covariant field equations could be determined:

$$(i\hbar\alpha^\mu D_\mu^{(f)} - m_0^{(f)}\tilde{\alpha}^\mu k_\mu)\Psi_F^{(f)} = 0 \tag{110}$$

specifying by the following group of transformations (based on the corresponding group of transformations (108-1)–(108-7) and (108-9)), for $f = 1, 2, 3, \ldots, 7$, respectively

$$(D_0^{(1)}, m_0^{(1)}, D_1^{(1)}, D_2^{(1)}) = (D_0, m_0, -D_1, D_2) \tag{110-1}$$

$$(D_0^{(2)}, m_0^{(2)}, D_1^{(2)}, D_2^{(2)}) = (D_0, m_0, D_1, -D_2) \tag{110-2}$$

$$(D_0^{(3)}, m_0^{(3)}, D_1^{(3)}, D_2^{(3)}) = (D_0, m_0, -D_1, -D_2) \tag{110-3}$$

$$(D_0^{(4)}, m_0^{(4)}, D_1^{(4)}, D_2^{(4)}) = (-iD_0^*, -im_0, -iD_1^*, -iD_2^*) \tag{110-4}$$

$$(D_0^{(5)}, m_0^{(5)}, D_1^{(5)}, D_2^{(5)}) = (-iD_0^*, -im_0, iD_1^*, -iD_2^*) \tag{110-5}$$

$$(D_0^{(6)}, m_0^{(6)}, D_1^{(6)}, D_2^{(6)}) = (-iD_0^*, -im_0, -iD_1^*, iD_2^*) \tag{110-6}$$

$$(D_0^{(7)}, m_0^{(7)}, D_1^{(7)}, D_2^{(7)}) = (-iD_0^*, -im_0, iD_1^*, iD_2^*) \tag{110-7}$$

where the column matrix $\Psi_F^{(f)}$ would be also given as follows (based on the definite formulation of column matrix $\Psi_F$ (96-1) in Section 3.12, expressing the general representation of column matrices definable in the formulation of (1 + 2)-dimensional case of general covariant field Equation (72)):

$$\Psi_F^{(f)} = \begin{bmatrix} F_{10}^{(f)} \\ 0 \\ F_{21}^{(f)} \\ \phi_F \end{bmatrix}, \quad J_\nu^{(f)} = -(D_\nu^{(f)} + \frac{im_0^{(f)}}{\hbar} k_\nu)\phi_F \tag{110-8}$$

where in all of the seven simultaneous cases of field Equation (110) defined respectively by the column matrices $\Psi_F^{(f)}$ (110-8) (for $f = 1, 2, 3, \ldots, 7$), the scalar quantity $\phi_F$ (that as a given initial quantity, defines the source currents $J_\nu^{(f)}$ (110-8)), necessarily, has the same value, based on the definite parametric formulation of the algebraic column matrix (109-1) (in particular, the common form of the corresponding arbitrary parameter $s$ in the expressions of all of the seven simultaneous cases of matrix Equation (109)).

Following the definite formulations of set of seven general covariant (massive) field Equations (110) (specified, respectively, by the group of seven transformations (110-1)–(110-7)), the set of these could be represented uniformly by the following general covariant field equation as well (defined solely in (1 + 2) space-time dimensions):

$$(i\hbar\alpha^{\mu}D_{\mu}-m_0\tilde{\alpha}^{\mu}k_{\mu})\Psi_F=0 \tag{110-9}$$

where the column matrix $\Psi_F$ given by:

$$\Psi_F=\begin{bmatrix}F_{10}\\0\\F_{21}\\\breve{\phi}_F\end{bmatrix},\ J_{\mu}=-(D_{\mu}+\frac{im_0}{\hbar}k_{\mu})\breve{\phi}_F \tag{110-10}$$

and the field strength tensor $F_{\mu\nu}$, scalar $\breve{\phi}_F$, along with the source current $J_{\mu}$ are defined as follows:

$$\begin{gathered}F_{\mu\nu}=\sum_{f=1}^{7}F_{\mu\nu}^{(f)}\tau_f\,,\\-I_2(D_{\mu}+\frac{im_0}{\hbar}k_{\mu})\breve{\phi}_F=\sum_{f=1}^{7}-(D_{\mu}^{(f)}+\frac{im_0^{(f)}}{\hbar}k_{\mu})\phi_F=\sum_{f=1}^{7}J_{\mu}^{(f)}\tau_f=J_{\mu}.\end{gathered} \tag{110-11}$$

where $I$ is the $2\times2$ identity matrix, and $\tau_f=\frac{l_f}{2}$ (for $f=1,2,3,\ldots,7$) are a set of seven $2\times2$ complex matrices given by,

$$\tau_f=\frac{l_f}{2},\ \begin{Bmatrix}l_1, & l_2, & l_3,\\ \circ & \circ & l_4,\\ l_5, & l_6, & l_7\end{Bmatrix}=\begin{Bmatrix}\begin{bmatrix}0&-i\\-i&0\end{bmatrix}, & \begin{bmatrix}0&1\\-1&0\end{bmatrix}, & \begin{bmatrix}-i&0\\0&i\end{bmatrix},\\ \circ & \circ & \begin{bmatrix}-1&0\\0&-1\end{bmatrix},\\ \begin{bmatrix}0&1\\1&0\end{bmatrix}, & \begin{bmatrix}0&-i\\i&0\end{bmatrix}, & \begin{bmatrix}1&0\\0&-1\end{bmatrix}\end{Bmatrix} \tag{110-12}$$

which as would be shown in the following, represents uniformly a combined gauge symmetry group of the form: $SU(2)_L\otimes U(2)_R$, where the sub-set of three matrices "$\tau_1,\tau_2,\tau_3$" corresponds to $SU(2)_L$ group, and subset of four matrices "$\tau_4,\tau_5,\tau_6,\tau_7$" corresponds to $U(2)_R$ group.

Now based on the matrix formulation of field strength tensor $F_{\mu\nu}$ (defined by the general covariant field Equation (110-9)), and on the basis of C, P and T symmetries of this field Equation (as a particular form of the (1 + 2)-dimensional case of field Equation (72)) that have been represented basically by their corresponding quantum operators (in Sections 3.5–3.5.4), it would be concluded that the general covariant field Equation (110-9) describes uniformly a group of seven spin-1/2 fermion fields corresponding to, respectively: "three left-handed fermions (for $f=1,2,3$), in addition to their three complementary right-handed anti-fermions (for $f=5,6,7$), and also a new single charge-less right-handed spin-1/2 fermion (for $f=4$)." Hence, following the basic algebraic properties of seven matrices $\tau_f$ (110-12), and the gauge symmetry group of the type: $SU(2)_L \otimes U(2)_R$ generated by these matrices, the three matrices $\tau_1, \tau_2, \tau_3$ (corresponding with $SU(2)_L$) represent respectively "three left-handed fermions", and four matrices $\tau_4, \tau_5, \tau_6, \tau_7$ (corresponding with $U(2)_R$) represent respectively: "a new single right-handed charge-less spin-1/2 fermion, and three right-handed spin-1/2 fermions as the complementary particles of the three left-handed spin-1/2 fermions represented by matrices $\tau_1, \tau_2, \tau_3$".

Furthermore, as noted previously and shown below, as a natural assumption, by assuming the seven types of spin-1/2 fermion fields that are described by general covariant field Equation (110-9), as the source currents of spin-1 boson fields (that will be represented by two determined unique groups describing respectively by general covariant field Equations (114-4) and (114-5)), it would be concluded that there should be, in total, four specific groups of seven spin-1/2 fermion fields (each) with certain properties, corresponding to "1 + 3" generations of four fermions, including two groups of four leptons each, and two groups of four quarks each. Moreover, based on this basic circumstances, two groups of leptons would be represented uniquely by: "[($\nu_\mu$, $e^-$, $\nu_\tau$), ($\bar{\nu}_\mu$, $e^+$, $\bar{\nu}_\tau$ $z_e$)] and [($\mu^-$, $\nu_e$, $\tau^-$), ($\mu^+$, $\bar{\nu}_e$, $\tau^+$, $z_n$)], respectively, where each group includes a new single right-handed charge-less lepton, represented by: $z_e$ and $z_n$"; and two groups of quarks would be also represented uniquely by: "[(s ,u ,b ), ($\bar{s}, \bar{u}, \bar{b}$, $z_u$)] and [(c ,d ,t ), ($\bar{c}, \bar{d}, \bar{t}$ , $z_d$)], respectively, where similar to leptons, each group includes a new single right-handed charge-less quark, represented by: $z_u$ and $z_d$". In addition, emerging two right-handed charhe-less quarks $z_u$ and $z_d$ specifically in two subgroups with anti-quarks ( $\bar{s}, \bar{u}, \bar{b}$, $z_u$) and ( $\bar{c}, \bar{d}, \bar{t}$ , $z_d$), could explain the baryon asymmetry, and subsequently, the asymmetry between matter and antimatter in the universe.

Assuming the spin-1/2 fermion fields describing by general covariant massive field Equations (110-9) (defined by column matrix (110-10) in (1 + 2) space-time dimensions with a digonalized metric), as the coupling source currents of spin-1 boson fields (describing generally by (1 + 3)-dimensional case of general covariant field Equation (72) formulated with a column matrix of the type $\Psi_F$ (98-2)), it is concluded that the (1 + 3)-dimensional metric could be also diagonalized for corresponding spin-1 boson fields. This conclusion follows directly from the above assumption that the (1 + 3)-dimensional metric of spin-1 boson fields (coupled to the corresponding fermionic source currents) would be also partially diagonalized such that for *μ*, *ν* = 0, 1, 2 and *μ* ≠ *ν*: $g_{\underline{\mu\nu}}=0$, which subsequently impliy $g_{03}=g_{13}=g_{23}=0$. Hence, the invariant energy-momentum relation (52) will be expressed as follows in (1 + 3)-dimensional space-time with diagonalized metric:

$$g^{00}(p_0)^2 - g^{00}(p_0^{st})^2 + g^{11}(p_1)^2 + g^{22}(p_2)^2 + g^{33}(p_3)^2 = 0 \quad (111)$$

that is equivalent to: $g^{00}(p_0)^2 + g^{11}(p_1)^2 + g^{22}(p_2)^2 + g^{33}(p_3)^2 = (m_0 c)^2$, where (similar to the (1 + 2)-dimensional case in Section 3.1) $m_0$ and $p_\mu$ are the particle's rest mass and momentum (4-momentum), $p_\mu^{st} = m_0 k_\mu$, and $k_\mu = (k_0, 0, 0, 0) = (c/\sqrt{g^{00}}, 0, 0, 0)$ denotes the covariant form of the 4-velocity of particle in stationary reference frame. Now similar to the transformations (108-1)–(108-7), as it would be shown in the following, a crucial property of the quadratic relation (111) would be also its invariance under two certain sets of sign inversions of the components of particle's momentum: $(p_0, p_1, p_2, p_3)$, along with similar inversions for the components:

$(p_0^{st}, p_1^{st}, p_2^{st}, p_3^{st})$ (as particular cases), where $p_0^{st} = m_0 k_0$, $p_1^{st} = p_2^{st} = p_3^{st} = 0$. The first set of these includes seven different odd types of the sign inversions (i.e., with odd inversions), and the second set includes eight different even types of the sign inversions (i.e., with even inversions), which could be represented simply by the following two symmetric groups of transformations (based on the formalism of the Lorentz symmetry group of invariant relation (111)), respectively:

The first group includes,

$$(p_0, p_0^{st}, p_1, p_2, p_3) \mapsto (p_0, p_0^{st}, -p_1, -p_2, p_3) = (p_0^{(1)}, p_0^{st(1)}, p_1^{(1)}, p_2^{(1)}, p_3^{(1)}) \quad (111\text{-}1)$$

$$(p_0, p_0^{st}, p_1, p_2, p_3) \mapsto (p_0, p_0^{st}, p_1, -p_2, -p_3) = (p_0^{(2)}, p_0^{st(2)}, p_1^{(2)}, p_2^{(2)}, p_3^{(2)}) \quad (111\text{-}2)$$

$$(p_0, p_0^{st}, p_1, p_2, p_3) \mapsto (p_0, p_0^{st}, -p_1, p_2, -p_3) = (p_0^{(3)}, p_0^{st(3)}, p_1^{(3)}, p_2^{(3)}, p_3^{(3)}) \quad (111\text{-}3)$$

$$(p_0, p_0^{st}, p_1, p_2, p_3) \mapsto (-p_0, -p_0^{st}, -p_1, -p_2, -p_3) = (p_0^{(4)}, p_0^{st(4)}, p_1^{(4)}, p_2^{(4)}, p_3^{(4)}) \quad (111\text{-}4)$$

$$(p_0, p_0^{st}, p_1, p_2, p_3) \mapsto (-p_0, -p_0^{st}, p_1, p_2, -p_3) = (p_0^{(5)}, p_0^{st(5)}, p_1^{(5)}, p_2^{(5)}, p_3^{(5)}) \quad (111\text{-}5)$$

$$(p_0, p_0^{st}, p_1, p_2, p_3) \mapsto (-p_0, -p_0^{st}, -p_1, p_2, p_3) = (p_0^{(6)}, p_0^{st(6)}, p_1^{(6)}, p_2^{(6)}, p_3^{(6)}) \quad (111\text{-}6)$$

$$(p_0, p_0^{st}, p_1, p_2, p_3) \mapsto (-p_0, -p_0^{st}, p_1, -p_2, p_3) = (p_0^{(7)}, p_0^{st(7)}, p_1^{(7)}, p_2^{(7)}, p_3^{(7)}) \quad (111\text{-}7)$$

And the second group is given by, respectively:

$$(p_0, p_0^{st}, p_1, p_2, p_3) \mapsto (p_0, p_0^{st}, -p_1, p_2, p_3) = (p_0^{(8)}, p_0^{st(8)}, p_1^{(8)}, p_2^{(8)}, p_3^{(8)}) \quad (111\text{-}8)$$

$$(p_0, p_0^{st}, p_1, p_2, p_3) \mapsto (p_0, p_0^{st}, p_1, -p_2, p_3) = (p_0^{(9)}, p_0^{st(9)}, p_1^{(9)}, p_2^{(9)}, p_3^{(9)}) \quad (111\text{-}9)$$

$$(p_0, p_0^{st}, p_1, p_2, p_3) \mapsto (p_0, p_0^{st}, p_1, p_2, -p_3) = (p_0^{(10)}, p_0^{st(10)}, p_1^{(10)}, p_2^{(10)}, p_3^{(10)}) \quad (111\text{-}10)$$

$$(p_0, p_0^{st}, p_1, p_2, p_3) \mapsto (p_0, p_0^{st}, -p_1, -p_2, -p_3) = (p_0^{(11)}, p_0^{st(11)}, p_1^{(11)}, p_2^{(11)}, p_3^{(11)}) \quad (111\text{-}11)$$

$$(p_0, p_0^{st}, p_1, p_2, p_3) \mapsto (-p_0, -p_0^{st}, p_1, p_2, p_3) = (p_0^{(12)}, p_0^{st(12)}, p_1^{(12)}, p_2^{(12)}, p_3^{(12)}) \quad (111\text{-}12)$$

$$(p_0, p_0^{st}, p_1, p_2, p_3) \mapsto (-p_0, -p_0^{st}, p_1, -p_2, -p_3) = (p_0^{(13)}, p_0^{st(13)}, p_1^{(13)}, p_2^{(13)}, p_3^{(13)}) \quad (111\text{-}13)$$

$$(p_0, p_0^{st}, p_1, p_2, p_3) \mapsto (-p_0, -p_0^{st}, -p_1, p_2, -p_3) = (p_0^{(14)}, p_0^{st(14)}, p_1^{(14)}, p_2^{(14)}, p_3^{(14)}) \quad (111\text{-}14)$$

$$(p_0, p_0^{st}, p_1, p_2, p_3) \mapsto (-p_0, -p_0^{st}, -p_1, -p_2, p_3) = (p_0^{(15)}, p_0^{st(15)}, p_1^{(15)}, p_2^{(15)}, p_3^{(15)}) \quad (111\text{-}15)$$

where, similar to the transformations (108-9) (as equivalent complex representations of the determined group of transformations (108-1)–(108-7), in (1 + 2)-dimensional space-time), following noncomplex-algebraic values of momentum's components $p_\mu$ ( $\overset{*}{p}_\mu = p_\mu$ ), the corresponding complex representations of transformations (111-1)–(111-15) is not a necessary issue in general, however, if the invariant relation (111) is represented formally by equivalent complex form:

$$g^{00}(p_0 p_0^*) - g^{00}(p_0^{st} p_0^{st*}) + g^{11}(p_1 p_1^*) + g^{22}(p_2 p_2^*) + g^{33}(p_3 p_3^*) = 0 \quad (111\text{-}16)$$

then, along with the set fifteen real-valued transformations (111-1)–(111-15), this relation would be also invariant under these corresponding sets of complex transformations (for $b = 1, 2, 3, \ldots, 15$ ):

$$\begin{gathered} (p_0, p_0^{st}, p_1, p_2, p_3) \to (\pm i p_0^{(b)*}, \pm i p_0^{st(b)*}, \pm i p_1^{(b)*}, \pm i p_2^{(b)*}, \pm i p_3^{(b)*}), \\ (p_0^*, p_0^{st*}, p_1^*, p_2^*, p_3^*) \to (\mp i p_0^{(b)}, \mp i p_0^{st(b)}, \mp i p_1^{(b)}, \mp i p_2^{(b)}, \mp i p_3^{(b)}) \end{gathered} \quad (111\text{-}17)$$

In addition, in the following, we show that using the transformations (111-1)–(111-15) (along with their corresponding complex forms (111-17)), a set of fifteen different general covariant field equations would be determined, including two certain groups of simultaneous field Equations (corresponding, respectively, to a group of seven bispinor fields and a group of eight bispinor fields of spin-1 particles) as the particular cases of the (1 + 3)-dimensional form of field Equation (72) (defined with a column matrix of the type (98-2)).

Similar to the set of seven algebraic matrix Equations (109) (determined uniquely as the algebraic equivalent matrix representation of the energy-momentum relation (108)), based on the definite formulation of the system of linear Equation (64) in (1 + 3) space-time dimensions (formulated in terms of the matrices (68)), for the energy-momentum relation (111) (along with the transformations (111-1)–(111-15)) the following two sets of systems of linear equations are also determined uniquely, including respectively a set of seven and a set of eight systems of Equations (with different parametric formalisms). The general parametric solution of each of these systems of linear equations, obeys also the quadratic relation (111) (representing a set of fifteen forms, with different parametric formulations, of the general parametric solutions of quadratic relation (111)). Each of these sets of the systems of linear equations could be represented uniformly by a matrix equation as follows, respectively:

$$(\alpha^\mu p_\mu^{(b_1)} - m_0^{(b_1)} \tilde{\alpha}^\mu k_\mu) S^{(b_1)} = 0, \quad (112\text{-}1)$$

$$(\alpha^\mu p_\mu^{(b_2)} - m_0^{(b_2)} \tilde{\alpha}^\mu k_\mu) S^{(b_2)} = 0 \quad (112\text{-}2)$$

where $b_1 = 1,2,3,\ldots,7$ , $b_2 = 8,9,\ldots,15$ , $p_\mu^{st(b_1)} = m_0^{(b_1)} k_\mu$ , $p_\mu^{st(b_2)} = m_0^{(b_2)} k_\mu$ , $\alpha^\mu$ and $\tilde{\alpha}^\mu$ are two contravariant $8 \times 8$ real matrices (compatible with matrix representations of the Clifford algebra $C\ell_{1,3}$) defined solely by formulas (65) and (68), and parametric column matrices $S^{(b_1)}$ and $S^{(b_2)}$ are also given uniquely as follows by two distinct expressions (formulated on the basis of definite parametric formulation of column matrix $S$ (68) in (1 + 3) space-time dimensions):

$$S^{(b_1)} = \begin{bmatrix} (u_0^{(b_1)} v_1^{(b_1)} - u_1^{(b_1)} v_0^{(b_1)}) w \\ (u_0^{(b_1)} v_2^{(b_1)} - u_2^{(b_1)} v_0^{(b_1)}) w \\ (u_0^{(b_1)} v_3^{(b_1)} - u_3^{(b_1)} v_0^{(b_1)}) w \\ 0 \\ (u_3^{(b_1)} v_2^{(b_1)} - u_2^{(b_1)} v_3^{(b_1)}) w \\ (u_1^{(b_1)} v_3^{(b_1)} - u_3^{(b_1)} v_1^{(b_1)}) w \\ (u_2^{(b_1)} v_1^{(b_1)} - u_1^{(b_1)} v_2^{(b_1)}) w \\ s \end{bmatrix}, \quad S^{(b_2)} = \begin{bmatrix} (u_0^{(b_2)} v_1^{(b_2)} - u_1^{(b_2)} v_0^{(b_2)}) w \\ (u_0^{(b_2)} v_2^{(b_2)} - u_2^{(b_2)} v_0^{(b_2)}) w \\ (u_0^{(b_2)} v_3^{(b_2)} - u_3^{(b_2)} v_0^{(b_2)}) w \\ 0 \\ (u_3^{(b_2)} v_2^{(b_2)} - u_2^{(b_2)} v_3^{(b_2)}) w \\ (u_1^{(b_2)} v_3^{(b_2)} - u_3^{(b_2)} v_1^{(b_2)}) w \\ (u_2^{(b_2)} v_1^{(b_2)} - u_1^{(b_2)} v_2^{(b_2)}) w \\ s' \end{bmatrix} \quad (113)$$

which column matrix $S^{(b_1)}$ includes seven cases with specific parametric formulations expressed respectively in terms of seven groups of independent arbitrary parameters: $u_0^{(b_1)}, u_1^{(b_1)}, u_2^{(b_1)}, v_0^{(b_1)}, v_1^{(b_1)}, v_2^{(b_1)}$, and two common arbitrary parameters $s$ and $w$ (i.e., having the same forms in all of the seven cases of column matrix $S^{(b_1)}$), and column matrix $S^{(b_2)}$ also includes eight cases with specific parametric formulations expressed respectively in terms of eight groups of independent arbitrary parameters: $u_0^{(b_2)}, u_1^{(b_2)}, u_2^{(b_2)}, v_0^{(b_2)}, v_1^{(b_2)}, v_2^{(b_2)}$, and two common arbitrary parameters $s'$ and $w$ (with the same forms in all of seven cases of the column matrix $S^{(b_2)}$). In addition, similar to the column matrix $S^{(a)}$ represented soley by formula (109-1), the specific parametric expressions (113) of column matrices $S^{(b_1)}$ and $S^{(b_2)}$ in the formulation of matrix Equations (112-1) and (112-2), have been determined specifically by assuming (as a basic assumption in addition to the systematic natural approach of formulating the matrix Equations (112-1) and (112-2), based on the definite formulation of axiomatically determined matrix Equation (64)) the minimum value for total number of arbitrary parameters in both column matrices $S^{(b_1)}$ and $S^{(b_2)}$, which implies equivalently the minimum value for total number of arbitrary parameters in all of the fifteen simultaneous (different) cases of matrix Equations (112-1) and (112-2) (necessarily with fifteen independent parametric solutions representing totally a certain set of fifteen different equivalent forms of the general parametric solution of quadratic relation (111), based on the general conditions of basic definition of the systems of linear equations corresponding to homogeneous quadratic and higher degree equations, presented in Section 2, and Sections 2.2–2.4, 3.1 concerning the homogenous quadratic equations). In the following, similar to the fundamental general covariant field Equation (109-2), in the derivation of the corresponding field Equations (from matrix Equations (112-1) and (112-2), respectively), we will also use the above particular algebraic properties of parameters $s$ and $s'$ which, respectively, have been expressed commonly in the expressions of all of seven simultaneous cases of matrix Equation (112-1), and in the expressions of all of eight cases of matrix Equation (112-1).

Moreover, similar to the invariant relation (108) and derived matrix Equation (109), along with the transformations (111-1)–(111-15) and algebraic matrix Equations (112-1) and (112-2), using the corresponding complex transformations (111-17), we may also formally have the following equivalent matrix Equations (with the complex expression), respectively (for $b_1 = 1,2,3,\ldots,7$, $b_2 = 8,9,\ldots,15$):

$$(i\alpha^{\mu} p_{\mu}^{(b_1)*} - i m_0^{(b_1)} \tilde{\alpha}^{\mu} k_{\mu}) S^{(b_1)} = 0, \quad (112\text{-}3)$$

$$(i\alpha^{\mu} p_{\mu}^{(b_2)*} - i m_0^{(b_2)} \tilde{\alpha}^{\mu} k_{\mu}) S^{(b_2)} = 0 \quad (112\text{-}4)$$

Where $p_{\mu}^{st(b_1)} = m_0^{(b_1)} k_{\mu}$, $p_{\mu}^{st(b_2)} = m_0^{(b_2)} k_{\mu}$. Similar to the matrix Equation (109-2), though based on the real value of momentum $p_{\mu} (p_{\mu}^{*} = p_{\mu})$, the complex expression of each of the seven cases of algebraic matrix Equation (112-1), and also each of the eight cases of algebraic matrix Equation (112-2), definitely, is not a necessary issue at the present stage. However, since the corresponding momentum operator $\hat{p}_{\mu}$ has a complex value (where $\hat{p}_{\mu} \neq \hat{p}_{\mu}^{*}$), in the following, using this basic property of the operator $\hat{p}_{\mu}$, we derive, distinctly, two certain groups of the general covariant field equations, including a group of seven different simultaneous field equations from the matrix

Equations (112-1) and (112-3), and a group of eight different simultaneous field equations from the matrix Equations (112-2) and (112-4) (based on the general axiomatic approach of derivation of general covariant massive field Equations (72) presented in Sections 3.4–3.10, and the quantum representations of C, P and T symmetries of this equation, presented in Sections 3.10–3.14). Furthermore, in the following, it is also shown that each of these determined two sets of seven and eight simultaneous field equations describe, respectively, a uniform group of seven spin-1boson fields (corresponding to two left-handed massive charged bosons, along with their two complementary right-handed bosons; a left-handed massive charge-less boson, along with its complementary right-handed boson; and a single right-handed massless and charge-less boson), and a uniform group of eight spin-1 boson field (corresponding to eight massless charged bosons).

Hence, similar to the (1 + 2)-dimensional general covariant field Equation (114), based on the axiomatic approach of derivation of the (1 + 3)-dimensional case of field Equation (72) (defined solely by a column matrix of the form (98-2) in (1 + 3) space-time dimensions, as shown in Section 3.12), from the matrix Equations (112-1), (112-3) and (112-2), (112-4) (defined solely by column matrices (113)) and taking into account this basic momentum operator's property: $\hat{p}_\mu^* \neq \hat{p}_\mu$, the following two unique groups of seven and eight simultaneous general covariant field equations are determined solely, respectively:

$$(i\hbar\alpha^\mu D_\mu^{(b_1)} - m_0^{(b_1)}\tilde{\alpha}^\mu k_\mu)\Phi_Z^{(b_1)} = 0 \qquad (114\text{-}1)$$

$$(i\hbar\alpha^\mu D_\mu'^{(b_2)} - m_0'^{(b_2)}\tilde{\alpha}^\mu k_\mu)\Phi_G^{(b_2)} = 0 \qquad (114\text{-}2)$$

specifying by the following two groups of transformations (based on their two corresponding groups of (sign) transformations (111-1)–(111-7), (111-8)–(111-15) and (111-17)), for $b_1 = 1,2,3,...,7$ and $b_2 = 1,2,3,...,8$, respectively:

The first group includes,

$$(D_0^{(1)}, m_0^{(1)}, D_1^{(1)}, D_2^{(1)}, D_3^{(1)}) = (D_0, m_0, -D_1, -D_2, D_3) \qquad (114\text{-}1\text{-}1)$$

$$(D_0^{(2)}, m_0^{(2)}, D_1^{(2)}, D_2^{(2)}, D_3^{(2)}) = (D_0, m_0, D_1, -D_2, -D_3) \qquad (114\text{-}1\text{-}2)$$

$$(D_0^{(3)}, m_0^{(3)}, D_1^{(3)}, D_2^{(3)}, D_3^{(3)}) = (D_0, m_0, -D_1, D_2, -D_3) \qquad (114\text{-}1\text{-}3)$$

$$(D_0^{(4)}, m_0^{(4)}, D_1^{(4)}, D_2^{(4)}, D_3^{(4)}) = (-iD_0^*, -im_0, -iD_1^*, -iD_2^*, -iD_3^*) \qquad (114\text{-}1\text{-}4)$$

$$(D_0^{(5)}, m_0^{(5)}, D_1^{(5)}, D_2^{(5)}, D_3^{(5)}) = (-iD_0^*, -im_0, iD_1^*, iD_2^*, -iD_3^*) \qquad (114\text{-}1\text{-}5)$$

$$(D_0^{(6)}, m_0^{(6)}, D_1^{(6)}, D_2^{(6)}, D_3^{(6)}) = (-iD_0^*, -im_0, -iD_1^*, iD_2^*, iD_3^*) \qquad (114\text{-}1\text{-}6)$$

$$(D_0^{(7)}, m_0^{(7)}, D_1^{(7)}, D_2^{(7)}, D_3^{(7)}) = (-iD_0^*, -im_0, iD_1^*, -iD_2^*, iD_3^*) \qquad (114\text{-}1\text{-}7)$$

and the second group is given as follows, respectively:

$$(D_0'^{(1)}, m_0'^{(1)}, D_1'^{(1)}, D_2'^{(1)}, D_3'^{(1)}) = (D_0, m_0, -D_1, D_2, D_3) \qquad (114\text{-}2\text{-}1)$$

$$(D_0'^{(2)}, m_0'^{(2)}, D_1'^{(2)}, D_2'^{(2)}, D_3'^{(2)}) = (D_0, m_0, D_1, -D_2, D_3) \qquad (114\text{-}2\text{-}2)$$

$$(D_0'^{(3)}, m_0'^{(3)}, D_1'^{(3)}, D_2'^{(3)}, D_3'^{(3)}) = (D_0, m_0, D_1, D_2, -D_3) \quad (114\text{-}2\text{-}3)$$

$$(D_0'^{(4)}, m_0'^{(4)}, D_1'^{(4)}, D_2'^{(4)}, D_3'^{(4)}) = (D_0, m_0, -D_1, -D_2, -D_3) \quad (114\text{-}2\text{-}4)$$

$$(D_0'^{(5)}, m_0'^{(5)}, D_1'^{(5)}, D_2'^{(5)}, D_3'^{(5)}) = (-iD_0^*, -im_0, iD_1^*, iD_2^*, iD_3^*) \quad (114\text{-}2\text{-}5)$$

$$(D_0'^{(6)}, m_0'^{(6)}, D_1'^{(6)}, D_2'^{(6)}, D_3'^{(6)}) = (-iD_0^*, -im_0, iD_1^*, -iD_2^*, -iD_3^*) \quad (114\text{-}2\text{-}6)$$

$$(D_0'^{(7)}, m_0'^{(7)}, D_1'^{(7)}, D_2'^{(7)}, D_3'^{(7)}) = (-iD_0^*, -im_0, -iD_1^*, iD_2^*, -iD_3^*) \quad (114\text{-}2\text{-}7)$$

$$(D_0'^{(8)}, m_0'^{(8)}, D_1'^{(8)}, D_2'^{(8)}, D_3'^{(8)}) = (-iD_0^*, -im_0, -iD_1^*, -iD_2^*, iD_3^*) \quad (114\text{-}2\text{-}8)$$

where the column matrices $\Phi_Z^{(b_1)}$ and $\Phi_G^{(b_1)}$ are also given as follows, written on the basis of definite formulations of algebraic column matrices (113) and in addition to the unique formulation of column matrix (98-2) (expressing the general representation of column matrices definable in the formulation of (1 + 3)-dimensional case of general covariant field Equation (72)):

$$\Phi_Z^{(b_1)} = \begin{bmatrix} Z_{10}^{(b_1)} \\ Z_{20}^{(b_1)} \\ Z_{30}^{(b_1)} \\ 0 \\ Z_{23}^{(b_1)} \\ Z_{31}^{(b_1)} \\ Z_{31}^{(b_1)} \\ \phi_Z \end{bmatrix}, \ J_\mu^{(b_1)} = -(D_\mu^{(b_1)} + \frac{im_0^{(b_1)}}{\hbar} k_\mu)\phi_Z, \ \Phi_G^{(b_2)} = \begin{bmatrix} G_{10}^{(b_2)} \\ G_{20}^{(b_2)} \\ G_{30}^{(b_2)} \\ 0 \\ G_{23}^{(b_2)} \\ G_{31}^{(b_2)} \\ G_{31}^{(b_2)} \\ \phi_G \end{bmatrix}, \ J_\mu'^{(b_2)} = -(D_\mu'^{(b_2)} + \frac{im_0'^{(b_2)}}{\hbar} k_\mu)\phi_G \quad (114\text{-}3)$$

where in all of the seven simultaneous (different) field Equations (112-1) formulated with column matrix $\Phi_Z^{(b_1)}$ (for $b_1 = 1,2,3,...,7$), and also in all of the eight simultaneous (different) field Equations (112-2) formulated with column matrix $\Phi_G^{(b_2)}$ (for $b_2 = 1,2,3,...,8$), the scalar quantity $\phi_Z$ (as initially given quantity) defines a commonly set of seven source currents $J_\mu^{(b_1)}$, and scalar quantity $\phi_G$ also defines commonly set of eight source currents $J_\mu'^{(b_2)}$.

**F**ollowing the definite formulations of a set of seven field Equation (114-1), and set of eight field Equation (114-2) specified, respectively, by the transformations (114-1-1)–(114-1-7) and (114-2-1)–(114-2-8), these two sets of the field equations could be represented uniformly by the following general covariant field equations as well (defined solely in (1 + 3) space-time dimensions), respectively:

$$(i\hbar\alpha^\mu D_\mu - m_0\tilde{\alpha}^\mu k_\mu)\Phi_Z = 0 \quad (114\text{-}4)$$

$$(i\hbar\alpha^\mu D_\mu - m_0\tilde{\alpha}^\mu k_\mu)\Phi_G = 0 \quad (114\text{-}5)$$

where the column matrices $\Phi_Z$ and $\Phi_G$ are given by:

$$\Phi_Z = \begin{bmatrix} Z_{10} \\ Z_{20} \\ Z_{30} \\ 0 \\ Z_{23} \\ Z_{31} \\ Z_{31} \\ \breve{\phi}_Z \end{bmatrix}, \ J_\mu = -(D_\mu + \frac{im_0}{\hbar} k_\mu)\breve{\phi}_Z, \ \Phi_G = \begin{bmatrix} G_{10} \\ G_{20} \\ G_{30} \\ 0 \\ G_{23} \\ G_{31} \\ G_{31} \\ \breve{\phi}_G \end{bmatrix}, \ J'_\mu = -(D_\mu + \frac{im_0}{\hbar} k_\mu)\breve{\phi}_G \quad (114\text{-}6)$$

and the field strength tensors $Z_{\mu\nu}$, $G_{\mu\nu}$ and scalars $\breve{\phi}_Z$ and $\breve{\phi}_G$, along with the source currents $J_\mu$ and $J'_\mu$ are defined as follows:

$$Z_{\mu\nu} = \sum_{b_1=1}^{7} Z_{\mu\nu}^{(b_1)} \tau_{b_1}, \ G_{\mu\nu} = \sum_{b_2=1}^{8} G_{\mu\nu}^{(b_2)} \breve{\lambda}_{b_2},$$

$$-I_2(D_\mu + \frac{im_0}{\hbar} k_\mu)\breve{\phi}_Z = \sum_{b_1=1}^{7} -(D_\mu^{(b_1)} + \frac{im_0^{(b_1)}}{\hbar} k_\mu)\phi_Z = \sum_{b_1=1}^{7} J_\mu^{(b_1)} \tau_{b_1} = J_\mu, \quad (114\text{-}7)$$

$$-I_3(D_\mu + \frac{im_0}{\hbar} k_\mu)\breve{\phi}_G = \sum_{b_2=1}^{8} -(D_\mu^{(b_2)} + \frac{im_0^{(b_2)}}{\hbar} k_\mu)\phi_G = \sum_{b_2=1}^{8} J'^{(b_2)}_\mu \breve{\lambda}_{b_2} = J'_\mu .$$

where $I_2$, $I_3$ are $2\times 2$ and $3\times 3$ identity matrices, and $\tau_{b_1} = \frac{l_{b_1}}{2}$ (for $b_1 = 1,2,3,...,7$) are the following set of seven $2\times 2$ complex matrices:

$$\tau_{b_1} = \frac{l_{b_1}}{2}, \begin{Bmatrix} l_1, & l_2, & l_3, \\ \circ & \circ & l_4, \\ l_5, & l_6, & l_7 \end{Bmatrix} = \begin{Bmatrix} \begin{bmatrix} 0 & -i \\ -i & 0 \end{bmatrix}, & \begin{bmatrix} 0 & 1 \\ -1 & 0 \end{bmatrix}, & \begin{bmatrix} -i & 0 \\ 0 & i \end{bmatrix}, \\ \circ & \circ & \begin{bmatrix} -1 & 0 \\ 0 & -1 \end{bmatrix}, \\ \begin{bmatrix} 0 & 1 \\ 1 & 0 \end{bmatrix}, & \begin{bmatrix} 0 & -i \\ i & 0 \end{bmatrix}, & \begin{bmatrix} 1 & 0 \\ 0 & -1 \end{bmatrix} \end{Bmatrix} \quad (114\text{-}8)$$

which are similar to the set of matrices (110-12), and uniformly represent a combined gauge symmetry group of the form: SU(2)L⊗U(2)R, where the subset of three matrices "$\tau_1, \tau_2, \tau_3$" corresponds to SU(2)L group, and subset of four matrices "$\tau_4, \tau_5, \tau_6, \tau_7$" corresponds to U(2)R group.

The matrices $\breve{\lambda}_{b_2} = (1/2)\lambda_{b_2}$ (for $b_2 = 1,2,3,...,8$) are also the following set of eight $3\times 3$ complex matrices equivalent to the Gell-Mann matrices (representing the SU(3) gauge symmetry group):

$$\breve{\lambda}_{b_2} = \frac{1}{2}\lambda_{b_2}, \begin{Bmatrix} \lambda_1, & \lambda_2, & \lambda_3, \\ \lambda_4, & \lambda_5, & \circ \\ \lambda_6, & \lambda_7, & \lambda_8 \end{Bmatrix} = \begin{Bmatrix} \begin{bmatrix} 0 & 1 & 0 \\ 1 & 0 & 0 \\ 0 & 0 & 0 \end{bmatrix}, & \begin{bmatrix} 0 & -i & 0 \\ i & 0 & 0 \\ 0 & 0 & 0 \end{bmatrix}, & \begin{bmatrix} 1 & 0 & 0 \\ 0 & -1 & 0 \\ 0 & 0 & 0 \end{bmatrix}, \\ \begin{bmatrix} 0 & 0 & 1 \\ 0 & 0 & 0 \\ 1 & 0 & 0 \end{bmatrix}, & \begin{bmatrix} 0 & 0 & -i \\ 0 & 0 & 0 \\ i & 0 & 0 \end{bmatrix}, & \circ \\ \begin{bmatrix} 0 & 0 & 0 \\ 0 & 0 & 1 \\ 0 & 1 & 0 \end{bmatrix}, & \begin{bmatrix} 0 & 0 & 0 \\ 0 & 0 & -i \\ 0 & i & 0 \end{bmatrix}, & \frac{1}{\sqrt{3}}\begin{bmatrix} 1 & 0 & 0 \\ 0 & 1 & 0 \\ 0 & 0 & -2 \end{bmatrix} \end{Bmatrix} \quad (114\text{-}9)$$

Now based on the definite matrix formalisms of the field strength tensors $Z_{\mu\nu}$ and $G_{\mu\nu}$ (114-7) (described respectively by general covariant massive field Equations (114-4) and (114-5)), and on the basis of C, P and T symmetries of these field Equations (as two particular forms of the (1 + 3)-dimensional case of tensor field Equation (72)), represented by their corresponding quantum operators (defined in Sections 3.11–3.14), it could be concluded that the field Equation (114-4) describes uniformly a definite group of seven simultaneous bispinor fields of spin-1 particles (corresponding to seven matrices $\tau_1, \tau_2, \tau_3, \tau_4, \tau_5, \tau_6, \tau_7$ (114-8)), including, respectively: "three left-handed massive bosons that could be denoted by $W^-, \tilde{W}^+, \overleftarrow{Z}$ (represented respectively by three matrices $\tau_1, \tau_2, \tau_3$, corresponding with SU(2)L), a right-handed charge-less spin-1 boson and also three right-handed spin-1 (massive) bosons denoted by $\psi, W^+, \tilde{W}^-, \vec{Z}$ (represented respectively by four matrices $\tau_4, \tau_5, \tau_6, \tau_7$, corresponding with U(2)R), as the complementary particles of three left-handed bosons represented by matrices $\tau_1, \tau_2, \tau_3$".

In addition, following the definite representations of these determined seven bosons, it could be concluded that four bosons $(\psi, W^-, W^+, \overleftarrow{Z})$ correspond to the known bosons including photon (determined as a right-handed charge-less boson, compatible with the positive-frequency corresponding to the right-handed circular polarization state of photon), and $W^-, W^+, Z$ bosons. Hence, particles $\tilde{W}^+, \tilde{W}^-, \vec{Z}$ represent three new massive spin-1 bosons (where, in particular, $\vec{Z}$ is the complementary right-handed particle of ordinary $Z$ boson), predicted uniquely by this new mathematical axiomatic approach. Furthermore, the field Equation (114-5) would also describe uniformly a definite group of eight spin-1 boson fields (corresponding respectively to eight matrices $\breve{\lambda}_1, \breve{\lambda}_2, \breve{\lambda}_3, \breve{\lambda}_4, \breve{\lambda}_5, \breve{\lambda}_6, \breve{\lambda}_7, \breve{\lambda}_8$, representing the SU(3) gauge group).

Furthermore, by assuming the group of seven spin-1/2 fermion fields (described by field Equations (110-9)–(110-12)) as the source currents of spin-1 boson fields, it would be also concluded that the group of seven uniform spin-1 boson fields $Z^{(b_1)}_{\mu\nu}$ represented by $(W^-, \tilde{W}^+, \overleftarrow{Z})$ and $(\psi, W^+, \tilde{W}^-, \vec{Z})$ (describing by general covariant field Equation (114-4)), and the group of eight uniform spin-1 boson fields $G^{(b_2)}_{\mu\nu}$ (describing by general covariant field Equation (114-5)), hold certain properties (including the electrical and color charges, so on) compatible with the known properties of ordinary bosons $W^-, W^+, Z$ and photon, and also eight gluon fields (with their known definite properties, including the color charges represented by 'color octet' [6,7]). In addition, based on the group representation of three additional new bosons that are predicted uniquely by this new mathematical axiomatic approach, denoted by: $\tilde{W}^+, \tilde{W}^-, \vec{Z}$, these new bosons could have properties similar to the ordinary bosons $W^-, W^+, Z$; where in particular new boson $\vec{Z}$ (as the complementary right-handed particle of ordinary $Z$ boson), can mix with $Z$ boson.

Moreover, as mentioned in Section 3.6.1.1, by assuming (as a basic natural assumption) the seven types of spin-1/2 fermion fields describing by general covariant field Equation (110-9), as the source currents of the uniquely determined two groups of seven and eight spin-1 boson fields (describing respectively by general covariant field Equations (114-4) and (114-5)), it would be concluded that there should be, in total, four specific groups of seven spin-1/2 fermion fields (each) with certain properties, corresponding to "1 + 3" generations of four fermions, including two groups of four leptons each, and two groups of four quarks each. Moreover, based on this basic circumstances, two groups of leptons would be represented uniquely by: "[($\nu_\mu$, $e^-$, $\nu_\tau$), ($\bar{\nu}_\mu$, $e^+$, $\bar{\nu}_\tau$ $z_e$)] and [($\mu^-$, $\nu_e$, $\tau^-$), ($\mu^+$, $\bar{\nu}_e$, $\tau^+$, $z_n$)], respectively, where each group includes a new single right-handed charge-less lepton, represented by: $z_e$ and $z_n$"; and two groups of quarks would be also represented uniquely by: "[(s, u, b), ($\bar{s}, \bar{u}, \bar{b}$, $z_u$)] and [(c, d, t), ($\bar{c}, \bar{d}, \bar{t}$, $z_d$)], respectively, where similar to leptons, each group includes a new single right-handed charge-less quark, represented by: $z_u$ and $z_d$". In addition, emerging two right-handed charhe-less quarks $z_u$ and $z_d$ specifically in two subgroups with anti-quarks ($\bar{s}, \bar{u}, \bar{b}$, $z_u$) and ($\bar{c}, \bar{d}, \bar{t}$, $z_d$), could explain the baryon asymmetry, and subsequently, the asymmetry between matter and antimatter in the universe.

## 4. Conclusions

The main results obtained in this article, are mainly, the outcomes of the new algebraic axiom (17) (along with the basic assumptions (2)–(3) defined in Section 3.1). This new axiom as a definite generalized form of the ordinary axiom of "no zero divisors" of integral domains (including the domain of integers), has been formulated soley in terms of square matrices (with integer entries, appeared as primary objects for representing the integer elements in their corresponding algebraic axiomatic formalism). In Section 3 of this article, as a new mathematical approach to origin of the laws of nature, using a new basic algebraic axiomatic (matrix) formalism based on the ring theory and Clifford algebras (presented in Section 2), "it is shown that certain mathematical forms of fundamental laws of nature, including laws governing the fundamental forces of nature (represented by a set of two definite classes of general covariant massive field equations, with new matrix formalisms), are derived uniquely from only a very few axioms"; where as a basic additional assumption (that is the assumption (2) in Section 3.1), in agreement with the rational Lorentz symmetry group, it has been also assumed that the components of relativistic energy-momentum (*D*-momentum) can only take the rational values. Concerning the basic assumption of rationality of relativistic energy-momentum, it is necessary to add (as mentioned in Section 3.1) that the rational Lorentz symmetry group is not only dense in the general form of Lorentz group, but also is compatible with the necessary conditions required basically for the formalism of a consistent relativistic quantum theory [15]. In essence, the main scheme of the new mathematical axiomatic approach to fundamental laws of nature presented in Section 3, is as follows. First in Section 3.1, based on the assumption of rationality of *D*-momentum, by linearization (along with a parameterization procedure) of the Lorentz invariant energy-momentum quadratic relation, a unique set of Lorentz invariant systems of homogeneous linear Equations (with matrix formalisms compatible with certain Clifford, and symmetric algebras) has been derived. Then in Section 3.4, by first quantization (followed by a basic procedure of minimal coupling to space-time geometry) of these determined systems of linear equations, a set of two classes of general covariant massive (tensor) field Equations (with matrix formalisms compatible with certain Clifford, and Weyl algebras) has been derived uniquely as well. Each class of the derived general covariant field equations also includes a definite form of torsion field appeared as generator of the corresponding field' invariant mass. In addition, in Sections 3.4–3.11, it has been shown that the (1 + 3)-dimensional cases of two classes of derived field equations represent a new general covariant massive formalism of bispinor fields of spin-2, and spin-1 particles, respectively. In fact, these uniquely determined bispinor fields represent a unique set of new generalized massive forms of the laws governing the fundamental forces of nature, including the Einstein (gravitational), Maxwell (electromagnetic) and Yang-Mills (nuclear) field equations. Moreover, it has been also shown that the (1 + 2)-dimensional cases of two classes of these field equations represent (asymptotically) a new general covariant massive formalism of bispinor fields of spin-3/2 and spin-1/2 particles, respectively, corresponding to the Dirac and Rarita–Schwinger equations.

As a particular consequence, in Section 3.6, it has been shown that a certain massive formalism of general relativity–with a definite form of torsion field appeared originally as the generator of gravitational field's invariant mass–is obtained only by first quantization (followed by a basic procedure of minimal coupling to space-time geometry) of a certain set of special relativistic algebraic matrix equations. In Section 3.9, it has been also proved that Lagrangian densities specified for the originally derived new massive forms of the Maxwell, Yang-Mills and Dirac field equations, are also gauge invariant, where the invariant mass of each field is generated solely by the corresponding torsion field. In addition, in Section 3.10, in agreement with recent astronomical data, a new particular form of massive boson has been identified (corresponding to U(1) gauge group) with invariant mass: $m_\gamma \approx 4.90571 \times 10^{-50}$ kg, which is specially generated by a coupled torsion field of the background space-time geometry.

Moreover, in Section 3.12, based on the definite mathematical formalism of this new axiomatic approach, along with the C, P and T symmetries (represented basically by the corresponding quantum matrix operators) of uniquely derived two fundamental classes of general covariant field equations, it has been concluded that the universe could be realized solely with the (1 + 2) and (1 + 3)-dimensional space-times (where this conclusion, in particular, is based on the time-reversal symmetry). In Sections 3.13 and 3.14, it has been proved that 'CPT' is the only (unique) combination of C, P, and T symmetries that could be defined as a symmetry for interacting fields. In addition, in Section 3.14, on the basis of these discrete symmetries of derived field equations, it has been also shown that only left-handed particle fields (along with their complementary right-handed fields) could be coupled to the corresponding (any) source currents. Furthermore, in Section 3.15, it has been shown that metric of the background space-time is diagonalized for the uniquely derived fermion field Equations (defined and expressed solely in (1 + 2)-dimensional space-time), where this property generates a certain set of additional symmetries corresponding uniquely to the $SU(2)_L \otimes U(2)_R$ symmetry group for spin-1/2 fermion fields (representing "1 + 3" generations of four fermions, including a group of eight leptons and a group of eight quarks), and also the $SU(2)_L \otimes U(2)_R$ and SU(3) gauge symmetry groups for spin-1 boson fields coupled to the spin-1/2 fermionic source currents. Hence, along with the known elementary particles, eight new elementary particles, including: four new charge-less right-handed spin-1/2 fermions (two leptons and two quarks, represented by "$z_e$ , $z_n$ and $z_u$ , $z_d$"), a spin-3/2 fermion, and also three new spin-1 massive bosons (represented by $\tilde{W}^+, \tilde{W}^-, \vec{Z}$, where in particular, the new boson $\vec{Z}$ is complementary right-handed particle of ordinary $Z$ boson), have been predicted uniquely by this fundamental axiomatic approach. As a particular result, in Section 3.6, based on the definite and unique formulation of the derived Maxwell's Equations (and also determined Yang-Mills equations, represented uniquely with two specific forms of gauge symmetries), it has been also concluded generally that magnetic monopoles could not exist in nature.

The new results obtained in this article, which are connecting with a number of longstanding essential issues in science and philosophy, demonstrate the wide efficiency of a new fundamental algebraic-axiomatic formalism presented in Section 2 of this article.

**Acknowledgments:** Special thanks are extended to Vitaly L. Ginzburg (Russia), Dmitry V. Shirkov (Russia), Leonid A. Shelepin (Russia), Vladimir Ya. Fainberg (Russia), Wolfgang Rindler (USA), Roman W. Jackiw (USA), Roger Penrose (UK), Steven Weinberg (USA), Ezra T. Newman (USA), Graham Jameson (UK), Sergey A. Reshetnjak (Russia), Sir Michael Atiyah (UK) (who, in particular, kindly encouraged me to continue this work as a new unorthodox mathematical approach to fundamental physics), and many others for their support and valuable guidance during my studies and research.

**Conflicts of Interest:** The author declares no conflict of interest.

## Appendix A

The matrix Equation (64) in Minkowski flat space-time (with metric signature (+ – ... –)) would be represented simply by:

$$(\alpha^{\mu} p_{\mu} - m_0 I)S = 0 \tag{A1}$$

where $I$ is the identity matrix, and column matrix $S$ is defined uniquely by formulas (66)–(70), … in (1 + 1), (1 + 2), (1 + 3), (1 + 4), (1 + 5), … space-time dimensions. The general contravariant forms of real matrices $\alpha^{\mu}$ that generate the Clifford algebra $C\ell_{1,N}$ (for $N \geq 2$) in (1 + $N$)-dimensional space-time, are (as mentioned in Section 3.3), are expressed by formulas (66)–(70),… in various space-times dimensions. Moreover, following the axiomatic approach of derivation of matrix Equation (64), matrices $\alpha^{\mu}$ in Minkowski flat space-time also hold the Hermiticity and anti-Hermiticity properties such that: $\alpha^0 = (\alpha^0)^*$ (compatible with $(\alpha^0)^2 = 1$), and $\alpha^{\mu} = -(\alpha^{\mu})^*$ (compatible with $(\alpha^{\mu})^2 = -1$, for $\mu$ = 1, 2, 3, …).

These matrices in the (1 + 1), (1 + 2), (1 + 3) and (1 + 4)-dimensional Minkowski space-time (as special cases of their general contravariant forms (65)–(69),…), have the following representations, respectively:

For (1 + 1)-dimensional space-time we have:

$$\alpha^0 = \begin{bmatrix} 1 & 0 \\ 0 & -1 \end{bmatrix}, \quad \alpha^1 = \begin{bmatrix} 0 & 1 \\ -1 & 0 \end{bmatrix} \tag{A2}$$

For (1 + 2)-dimensional case we get:

$$\begin{gathered}
\alpha^0 = \begin{bmatrix} \sigma^0 + \sigma^1 & 0 \\ 0 & -(\sigma^0 + \sigma^1) \end{bmatrix} = \begin{bmatrix} 1 & 0 & 0 & 0 \\ 0 & 1 & 0 & 0 \\ 0 & 0 & -1 & 0 \\ 0 & 0 & 0 & -1 \end{bmatrix}, \\
\alpha^1 = \begin{bmatrix} 0 & \sigma^2 - \sigma^3 \\ -\sigma^2 + \sigma^3 & 0 \end{bmatrix} = \begin{bmatrix} 0 & 0 & 0 & 1 \\ 0 & 0 & 1 & 0 \\ 0 & -1 & 0 & 0 \\ -1 & 0 & 0 & 0 \end{bmatrix}, \\
\alpha^2 = \begin{bmatrix} 0 & -\sigma^1 + \sigma^0 \\ \sigma^1 - \sigma^0 & 0 \end{bmatrix} = \begin{bmatrix} 0 & 0 & 1 & 0 \\ 0 & 0 & 0 & -1 \\ -1 & 0 & 0 & 0 \\ 0 & 1 & 0 & 0 \end{bmatrix}.
\end{gathered} \tag{A3}$$

In (1 + 3) dimensions, we have:

$$\alpha^0=\begin{bmatrix}\gamma^0+\gamma^1 & 0\\ 0 & -(\gamma^0+\gamma^1)\end{bmatrix}=\begin{bmatrix}1&0&0&0&0&0&0&0\\0&1&0&0&0&0&0&0\\0&0&1&0&0&0&0&0\\0&0&0&1&0&0&0&0\\0&0&0&0&-1&0&0&0\\0&0&0&0&0&-1&0&0\\0&0&0&0&0&0&-1&0\\0&0&0&0&0&0&0&-1\end{bmatrix},$$

$$\alpha^1=\begin{bmatrix}0 & \gamma^2-\gamma^3\\ \gamma^2-\gamma^3 & 0\end{bmatrix}=\begin{bmatrix}0&0&0&0&0&0&0&1\\0&0&0&0&0&0&1&0\\0&0&0&0&0&-1&0&0\\0&0&0&0&-1&0&0&0\\0&0&0&1&0&0&0&0\\0&0&1&0&0&0&0&0\\0&-1&0&0&0&0&0&0\\-1&0&0&0&0&0&0&0\end{bmatrix},$$

$$\alpha^2=\begin{bmatrix}0 & \gamma^4+\gamma^5\\ \gamma^4+\gamma^5 & 0\end{bmatrix}=\begin{bmatrix}0&0&0&0&0&0&-1&0\\0&0&0&0&0&0&0&1\\0&0&0&0&1&0&0&0\\0&0&0&0&0&-1&0&0\\0&0&-1&0&0&0&0&0\\0&0&0&1&0&0&0&0\\1&0&0&0&0&0&0&0\\0&-1&0&0&0&0&0&0\end{bmatrix},\quad \text{(A4)}$$

$$\alpha^3=\begin{bmatrix}0 & \gamma^6-\gamma^7\\ \gamma^6-\gamma^7 & 0\end{bmatrix}=\begin{bmatrix}0&0&0&0&0&1&0&0\\0&0&0&0&-1&0&0&0\\0&0&0&0&0&0&0&1\\0&0&0&0&0&0&-1&0\\0&1&0&0&0&0&0&0\\-1&0&0&0&0&0&0&0\\0&0&0&1&0&0&0&0\\0&0&-1&0&0&0&0&0\end{bmatrix}.$$

This article has been invited and presented at the following international conferences:

- The 2016 SIAM International Conference on Mathematical Aspects of Materials Science, Philadelphia, PA, USA, 2016. (https://www.siam.org/meetings/ms16).
- The 17th International Conference on Quantum Foundations: Quantum and Beyond, International Centre for Mathematical Modeling in Physics, Engineering and Cognitive Sciences (ICMM), Linnaeus University, Sweden, 2016. (https://lnu.se/en/qb/).
- The 4th International Conference on New Frontiers in Physics, CERN Organized and Supported Conference (Europe), ICNFP-2015. (https://indico.cern.ch/e/icnfp2015/).

- The 2016 International Conference on Algebraic Geometry and Mathematical Physics, University of Tromsø, Norway, 2016. (https://site.uit.no/).
- The XXXVII Max Born International Symposium, International Conference on Non-commutative Geometry, Quantum Symmetries and Quantum Gravity (II), Wroclaw University, Poland, 2016. (http://ift.uni.wroc.pl/~mborn37/).
- The 2016 GRavitational-wave Astronomy International Conference in Paris, Institute d'Astrophysique de Paris (IAP), The University of Paris VI—Sorbonne University, CNRS, LERU, EUA, Paris, France, 'Supported by the European Union's 7th Framework Program: FP7/PEOPLE-2011-CIG,' 2016. (http://www.iap.fr/vie_scientifique/ateliers/GravitationalWave/2016/scripts/abstract.aspx).
- The 22nd Internnational Australian Institute of Physics Congress (AIP), University of Queensland, Brisbane, Australia, 2016. (http://appc-aip2016.org.au).
- The 21st International Conference on General Relativity and Gravitation, Columbia University, New York, NY, USA, 2016. (http://www.gr21.org).